\documentclass[a4paper,11pt]{article}
\usepackage{amsmath,amssymb,epsfig,natbib,xspace}

\newcommand\TPBstyle\relax

\newcommand\bs[1]{\boldsymbol{#1}}
\newcommand\mc[1]{\mathcal{#1}}
\newcommand\bd[1]{\mathbf{#1}}

\newcommand{\Id}{\bd{I}}
\newcommand{\RR}{\mathbb R}

\newcommand{\ZZ}{\mathbb Z}

\newcommand\lambdamax{\lambda_{\text{max}}}
\newcommand\Rmax{R_{\text{max}}}
\newcommand\rmax{r_{\text{max}}}

\newcommand\rs{r_{\text{s}}}
\newcommand\xmin{x_{\text{min}}}
\newcommand\xmax{x_{\text{max}}}

\newcommand\dH{d_{\text{H}}}
\newcommand\muc{\mu_{\text{c}}}
\newcommand\xjump{x_{\text{jump}}}
\newcommand\scal[1]{\langle #1 \rangle}
\newcommand\ccint[1]{\mathopen[ #1 \mathclose]}
\newcommand\ooint[1]{\mathopen] #1 \mathclose[}
\newcommand\coint[1]{\mathopen[ #1 \mathclose[}

\newcommand\openunitint{\ooint{0,1}}

\newcommand\tr{\operatorname{tr}}

\newcommand\Var{\operatorname{Var}}
\newcommand\Cov{\operatorname{Cov}}
\newcommand\citec[1]{\citeauthor{#1}, \citeyear{#1}}

\newcommand\rawthreegnuplot[1]{%
\epsfig{file=#1,bbllx=66.6,bblly=50,bburx=225.2,bbury=193,clip=}}
\newcommand\threegnuplots[3]{%
\centerline{\rawthreegnuplot{#1}\hspace{0.5\columnsep}\rawthreegnuplot{#2}%
\hspace{0.5\columnsep}\rawthreegnuplot{#3}}}

\newcommand\Eref[1]{Eq.~(\ref{#1})}
\newcommand\eref[1]{\eqref{#1}}
\newcommand\fref[1]{Fig.~\ref{#1}}
\newcommand\Fref[1]{Fig.~\ref{#1}}
\newcommand\tref[1]{Table \ref{#1}}

\newcommand\sref[1]{Section \ref{#1}}
\newcommand\Sref[1]{Section \ref{#1}}
\newcommand\aref[1]{Appendix \ref{#1}}
\newcommand\Aref[1]{Appendix \ref{#1}}

\makeatletter
\long\def\@makecaption#1#2{%
  \vskip\abovecaptionskip
  \sbox\@tempboxa{\small\textbf{#1.} #2}
  \ifdim \wd\@tempboxa >\hsize
    \small\textbf{#1.} #2\par
  \else
    \global \@minipagefalse
    \hb@xt@\hsize{\hfil\box\@tempboxa\hfil}%
  \fi
  \vskip\belowcaptionskip}

\renewcommand\thetable{\@Roman\c@table}
\def\tabcaption{%
   \ifx\@captype\@undefined
     \@latex@error{\noexpand\caption outside float}\@ehd
     \expandafter\@gobble
   \else
     \refstepcounter\@captype
     \expandafter\@firstofone
   \fi
   {\@dblarg{\@tabcaption\@captype}}%
}
\long\def\@tabcaption#1[#2]#3{%
  \par
  \addcontentsline{\csname ext@#1\endcsname}{#1}%
    {\protect\numberline{\csname the#1\endcsname}{\ignorespaces #2}}%
  \begingroup
    \@parboxrestore
    \if@minipage
      \@setminipage
    \fi
    \normalsize
    \@maketabcaption{\csname fnum@#1\endcsname}{\ignorespaces #3}\par
  \endgroup}
\long\def\@maketabcaption#1#2{%
  \vskip\abovecaptionskip
  \sbox\@tempboxa{#1: #2}%
  \ifdim \wd\@tempboxa >\hsize
    \textbf{#1.} #2\par
  \else
    \global \@minipagefalse
    \hb@xt@\hsize{\hfil\box\@tempboxa\hfil}%
  \fi
  \vskip\belowcaptionskip}
\makeatother

\makeatletter
\def\narrowenumerate{%
  \ifnum \@enumdepth >\thr@@\@toodeep\else
    \advance\@enumdepth\@ne
    \edef\@enumctr{enum\romannumeral\the\@enumdepth}%
      \expandafter
      \list
        \csname label\@enumctr\endcsname
        {\itemsep0cm \topsep0cm \partopsep0cm \parsep0cm%
         \usecounter\@enumctr\def\makelabel##1{\hss\llap{##1}}}%
  \fi}

\def\itemlabel#1{\@bsphack
  \protected@write\@auxout{}%
         {\string\newlabel{#1}{{\@itemlabel}{\thepage}}}%
  \@esphack}
\makeatother

\begin{document}

\bibliographystyle{tpb}

\title{Mutation--Selection Balance:\\ Ancestry, Load, and Maximum Principle}
\author{{\sc Joachim Hermisson$^{1}$, Oliver Redner$^{2}$,}\\  
{\sc Holger Wagner$^{3}$, and Ellen Baake$^{2}$} 
\\[2mm]
${}^{1}$Department of Ecology and Evolutionary Biology, Yale
University,\\ New Haven, CT 06520, USA\\
${}^{2}$Institut f\"ur Mathematik und Informatik, Universit\"at
Greifswald,\\ Friedrich-Ludwig-Jahn-Str.\ 15a, 17487 Greifswald, Germany\\
${}^{3}$Technische Fakult\"at,  Universit\"at Bielefeld, \\ Postfach 100131,
33501 Bielefeld,
Germany}
\maketitle
\begin{quotation}
  \newlength\tempw \setlength\tempw\linewidth
  \addtolength\tempw{-2\fboxsep} \small\noindent
  \framebox{\parbox{\tempw}{\textbf{Note added for \texttt{cond-mat}:}
      We show how various concepts from statistical physics, such as
      order parameter, thermodynamic limit, and quantum phase
      transition, translate into corresponding biological concepts in
      mutation--selection models for sequence evolution and can be
      used in this context.  The article takes a biological point of
      view and works in a population genetics framework, but contains
      an appendix especially written for physicists, which makes this
      correspondence clear.}}
\end{quotation}
\begin{abstract}
  We analyze the equilibrium behavior of deterministic haploid
  mutation--selection models.  To this end, both the forward and the
  time-reversed evolution processes are considered. The stationary
  state of the latter is called the ancestral distribution, which
  turns out as a key for the study of mutation--selection balance.  We
  find that the ancestral genotype frequencies determine the
  sensitivity of the equilibrium mean fitness to changes in the
  corresponding fitness values and discuss implications for the
  evolution of mutational robustness.  We further show that the
  difference between the ancestral and the population mean fitness,
  termed mutational loss, provides a measure for the sensitivity of
  the equilibrium mean fitness to changes in the mutation rate.  The
  interrelation of the loss and the mutation load is discussed.  For a
  class of models in which the number of mutations in an individual is
  taken as the trait value, and fitness is a function of the trait, we
  use the ancestor formulation to derive a simple maximum principle,
  from which the mean and variance of fitness and the trait may be
  derived; the results are exact for a number of limiting cases, and
  otherwise yield approximations which are accurate for a wide range
  of parameters.  These results are applied to threshold phenomena
  caused by the interplay of selection and mutation (known as error
  thresholds).  They lead to a clarification of concepts, as well as
  criteria for the existence of error thresholds.
\end{abstract}

\TPBstyle

\section{Introduction}

A lot of research in theoretical population genetics has been directed
towards mutation--selection models in multilocus systems and infinite
populations. One is usually interested in statistical properties of
the equilibrium distribution of genotypes, like the means and
variances of fitness and trait(s).  The standard approach to determine
these starts out from the equilibrium condition for the genotype
frequencies (which takes the form of an eigenvalue equation if the
population is haploid).  On this basis, a wealth of methods and
results has been created; for a comprehensive and up-to-date account,
see \citet{Bue00}.

In this article, we present an alternative route, which relies on a
time-reversed version of the mutation--selection process and its
stationary distribution -- to be called the \emph{ancestral
  distribution}, as opposed to the equilibrium distribution of the
forward process.  We will apply this approach to tackle a rather
general class of models for haploids, or diploids without dominance.
It is only assumed that fitness is a function of a trait, and
genotypes with equal trait values have equivalent mutation patterns.
In fact, this is a standard assumption, and is often implied without
special mention. It applies, for example, if (geno)types are
identified with a collection of loci with two alleles each (wildtype
and mutant), which mutate independently and according to the same
process, and the number of (deleterious) mutations plays the role of
the trait.  The assumption of permutation invariance (with respect to
the loci) is certainly a distortion of biological reality, but, even
in this simplified setting, general answers have previously been
considered impossible \citep{Char90}, and researchers have resorted to
more specific choices of the fitness function and the mutation model
(e.g.\ quadratic fitness functions and unidirectional mutation).

With the help of the ancestral distribution, we will be able to tackle
general fitness functions (with arbitrary epistasis), as well as
general mutation schemes (including arbitrary amounts of back
mutation), from the permutation invariant class.  The
mutation--selection equilibrium will be characterized through a
maximum principle which relates the equilibrium population to the
ancestral one, and may be evaluated explicitly to yield expressions
for the mean fitness and the mean trait value, as well as the
variances of these quantities. The results are exact for a number of
limiting cases, and otherwise yield approximations which are accurate
for a wide range of parameters.

The results will then be used to settle the long-standing issue of
characterization and classification of error threshold phenomena in
this model class.  An error threshold may be generally, but vaguely,
circumscribed as a critical mutation rate beyond which mutation can no
longer be controlled by selection and leads to genetic degeneration;
for review, see \citet{EMcCS89}. Some, but by no means all,
mutation--selection models display such behavior.  It turns out that a
consistent definition of an error threshold is rather subtle and must
be sorted out first.  On this basis, we will classify
mutation--selection models according to their threshold behavior (if
any).

Since the article treats quite a number of topics, we start out with a
brief reader's guide to the main results here and give hints on what
parts can be skipped at a first reading.  Let us also mention that
\tref{tab:symbols} contains a glossary of repeatedly used notation.

The scene is set in \sref{sec:modelsetup}, where we will introduce the
model (the continuous-time mutation--selection model) and establish
its relationship with a multitype branching process.  Two concepts
that are central to this paper, the \emph{ancestor distribution} and
the \emph{mutation class limit}, are developed in this section.
\Sref{sec:branching} introduces the ancestor distribution as the
stationary distribution of the time-reversed branching process and
links the algebraic properties of the model to a probabilistic picture
that also helps to shape biological intuition.  In order to allow
quick progress to the results in the remainder of the article,
however, we have summarized the main points in \sref{sec:equilanc}.
In \ref{sec:averages}, the means and variances of the trait and of
fitness with respect to the equilibrium population and with respect to
the ancestors are introduced. In \ref{sec:response}, the difference
between the ancestral and the population mean fitness, termed
mutational loss, is shown to provide a measure for the sensitivity of
the equilibrium mean fitness to changes in the mutation rate. This
result is used and expanded in some of the applications in Sections
\ref{sec:interpret} and \ref{sec:thresholds}, but can be skipped at
first reading.  Sections \ref{sec:scaling} and \ref{sec:limit} are
mainly concerned with the mutation class limit, along with the proper
scaling of fitness functions and mutation schemes. Like the well-known
infinite-sites limit, this limit assumes an infinite number of types,
but uses a different scaling. As a consequence, it is valid if the
total mutation rate is large relative to typical fitness differences
of types. In this paper, the mutation class limit is used to derive
analytic expressions for means and variances of fitness and the trait
for the general case with back mutations and a non-linear fitness
function. It is also crucial for our discussion of threshold behavior
in \sref{sec:thresholds}.

\Sref{sec:results} is a condensed summary of the main results related
to the maximum principle.  The mean fitness at mutation--selection
balance equals the maximum of the difference between the fitness
function and a so-called \emph{mutational loss function}, where the
position of the maximum determines the mean ancestral trait.  Once
these means are known, explicit expressions are available for the mean
trait and the variances of fitness and trait.  The derivations are
postponed to \sref{sec:deriv}, which may be skipped at first reading.

The following two sections are devoted to applications.  Both are, to
a large extent, independent of each other and rely only on the matter
introduced in Sections \ref{sec:modelsetup} and \ref{sec:results}.  In
\sref{sec:interpret}, we first discuss the evolutionary significance
of the mutational loss, and then turn to the mutation load.  Explicit
expressions are derived for small (back) mutation rates; but arbitrary
mutation rates are covered by the maximum principle, which may be
interpreted as a generalized version of Haldane's principle.
Consequences for the evolution of mutational effects and for
mutational robustness are discussed.  Finally a note is added as to
the accuracy of the expressions for the means and variances.

In \sref{sec:thresholds}, we first analyze the definitions available
for the error threshold. It will turn out that various notions must be
distinguished, which coincide only in special cases.  For each of
these notions, a criterion for the existence of an error threshold is
derived from the maximum principle.  Furthermore, the phenomena are
illustrated by means of examples and discussed with respect to their
biological implications.  \Sref{sec:summary} provides a summary and an
outlook.

\Aref{app:physics} describes the connection between our
mutation--selection model and a system of quantum-statistical
mechanics, which had been used previously \citep[e.g.][]{BBW97,BaWa01}
to solve a more restricted model class, and which also served as the
source of concepts and methods for the current article. However, the
body of the paper does not require any knowledge of physics and
remains entirely within the framework of population genetics and
classical probability theory.  Appendices \ref{app:proofsderiv} and
\ref{app:proofsthr}, finally, contain the proofs from Sections
\ref{sec:deriv} and \ref{sec:thresholds}, respectively.

\section{Model setup}
\label{sec:modelsetup}

\begin{table}[t]
\begin{center}
\tabcaption{Glossary of repeatedly used notation.  Symbols are given
  together with the section or equation in which they are defined.
  Symbols whose scope is only a single section are not shown.}
\label{tab:symbols}
\begin{tabular}{lll}
\hline
$a$ & ancestor frequencies & \ref{sec:branching} \\
$G,g$ & mutational loss & \eref{eq:linres4} \\
$g$ & mutational loss function & \eref{eq:g} \\
$\bd{H}$ & evolution matrix & \eref{eq:H} \\
$\Id$ & identity matrix & \ref{sec:model} \\
$i,j,k,\ell$ & genotype/class labels & \ref{sec:model} \\
$L,l$ & mutation load & \ref{sec:averages} \\
$m$ & mutation rates & \ref{sec:model} \\
$\bd{M}$ & mutation matrix & \ref{sec:model} \\
$N$ & number of mutation classes / \\
    & sequence length & \ref{sec:model} \\
$p$ & population frequencies & \ref{sec:model} \\
$\bd{Q}$ & generator of reversed process & \ref{sec:branching} \\
$R,r$ & reproduction rates & \ref{sec:model} \\
$r$ & fitness function & \eref{eq:defrupm} \\
$\bd{R}$ & reproduction matrix & \ref{sec:model} \\
$s^\pm$ & mutational effects & \ref{sec:averages} \\
$\bs{s}$ & (binary) sequence & \ref{sec:model} \\
$\bd{T}$ & time evolution matrix & \ref{sec:model} \\
$t$ & time & \ref{sec:model} \\
$U^\pm,u^\pm$ & genomic mutation rates & \ref{sec:model} \\
$u^\pm$ &  mutation functions & \eref{eq:defrupm} \\
$V,v$ & variances & \ref{sec:averages} \\
$X,x$ & mutational distance & \ref{sec:averages} \\
$\mc{X}$ & mutation classes & \ref{sec:model} \\
$Y,y$ & arbitrary trait & \ref{sec:limit} \\
$z$ & relative reproductive success & \ref{sec:branching} \\
$\gamma$ & overall factor for reproduction \\
         & rates & \ref{sec:accuracy} \\
$\kappa$ & biallelic mutation asymmetry \\
         & parameter & \ref{sec:model} \\
$\lambdamax$ & largest eigenvalue of $\bd{H}$ & \ref{sec:model} \\
$\mu$ & overall mutation rate & \ref{sec:model}, \eref{eq:defmu} \\
\hline
\end{tabular}
\end{center}
\end{table}

\subsection{The model}
\label{sec:model}

We consider the evolution of an infinite population of haploid
individuals (or diploids without dominance) under mutation and
selection.  Disregarding environmental effects, we take individuals to
be fully described by their genotypes, which are labeled by the
elements of $\{1, \dots, M\}$.  Let $p_i(t)$ be the relative frequency
of type $i$ at time $t$, so that $\sum_i p_i(t) = 1$, and let
$\bs{p}(t) = (p_1(t), \dots, p_M(t))^T$ with $T$ denoting
transposition.  Throughout this article we will use the formalism for
overlapping generations, which works in continuous time, and only
comment on extensions to the analogous model for discrete generations.
The standard system of differential equations which describes the
evolution of the vector $\bs{p}(t)$ is (\citec{CK70}; \citec{Hof85};
see also \citec{Bue00}):
\begin{equation} \label{eq:paramuse}
\dot{p}_i(t) = [R_i - \bar{R}(t)] p_i(t) + 
\sum_j [m_{ij} p_j(t) - m_{ji} p_i(t)] \,.
\end{equation}
Here, $R_i$ is the Malthusian fitness of type $i$, which is connected
to the respective birth and death rates as $R_i = B_i - D_i$, and
$\bar{R}(t) = \sum_i R_i p_i(t)$ designates the mean fitness. Further,
$m_{ij}$ is the rate at which a $j$ individual mutates to $i$, and the
dot denotes the time derivative. In this model, mutation and selection
are assumed to be independent processes which go on in parallel.
However, mutation may also be viewed as occuring during reproduction.
In this case, the mutation rate is given by $m_{ij} := v_{ij}B_j$,
where $v_{ij}$ is the respective mutation \emph{probability} during a
reproduction event.  Since, formally, this leads to the same model, it
need not be discussed further.

For some of the main results of this article, further assumptions on
the mutation scheme are required.  To this end, we collect genotypes
into classes $\mc{X}_k$ of equal fitness, $0 \le k \le N$, and assume
mutations only to occur between neighboring classes.  Let $R_k$ denote
the fitness of class $k$ and $U^\pm_k$ the mutation rate from class
$\mc{X}_k$ to $\mc{X}_{k\pm1}$ (i.e.\ the total rate for each genotype
in $\mc{X}_k$ to mutate to some genotype in class $\mc{X}_{k\pm1}$),
with the convention $U_0^- = U_N^+ = 0$.  Thus, we obtain a variant of
the so-called single-step mutation model:
\begin{equation} \label{eq:step}
\dot{p}_k^{}(t) = [R_k^{} - \bar{R}(t) - U_k^+ - U_k^-] p_k^{}(t) 
+ U_{k-1}^+ p_{k-1}^{}(t) + U_{k+1}^- p_{k+1}^{}(t) \,.
\end{equation}
(Here, the convention $p_{-1}(t) = p^{}_{N+1}(t) = 0$ is used.)  We
can, for example, think of $\mc{X}_0$ as the \emph{wildtype} class
with maximum fitness and fitness only depending on the number of
mutations carried by an individual.  If, further, mutation is modeled
as a continuous point process (or if multiple mutations during
reproduction can be ignored), \Eref{eq:paramuse} reduces to
\eref{eq:step}, with an appropriate choice of mutation classes.
Depending on the realization one has in mind, the $U_k$ then describe
the total mutation rate affecting the whole genome or just some trait
or function.

\begin{figure}
\centerline{\epsfig{file=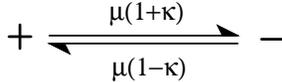,width=38mm}}
\caption{Rates for mutations and back mutations at each site or locus
  of a biallelic sequence.}
\label{fig:mutscheme}
\end{figure}

In most of our examples, we will use the Hamming graph as our genotype
space.  Here, genotypes are represented as binary sequences $\bs{s} =
s_1 s_2 \dots s_N \in \{+,-\}^N$, thus $M = 2^N$.  The two possible
values at each site, $+$ and $-$, may be understood either in a
molecular context as nucleotides (purines and pyrimidines) or, on a
coarser level, as wildtype and mutant alleles of a biallelic
multilocus model.  We will assume equal mutation rates at all sites,
but allow for different rates, $\mu(1+\kappa)$ and $\mu(1-\kappa)$,
for mutations from $+$ to $-$ and for back mutations, respectively,
according to the scheme depicted in \fref{fig:mutscheme}.

Clearly, the biallelic model reduces to a single-step mutation model
(with the same $N$) if the fitness landscape\footnote{We use the
  notion of a \emph{fitness landscape} \citep{KaLe87} as synonymous
  with \emph{fitness function} for the mapping from genotypes to
  individual fitness values.} is invariant under permutation of sites.
To this end, we distinguish a reference genotype $\bs{s}_+ =
++\ldots+$, in most cases the wildtype or master sequence, and assume
that the fitness $R_{\bs{s}}$ of sequence $\bs{s}$ depends only on the
Hamming distance $k = \dH(\bs{s},\bs{s}_+)$ to $\bs{s}_+$ (i.e.\ the
number of mutations, or `$-$' signs in the sequence).  The resulting
total mutation rates between the Hamming classes $\mc{X}_k$ and
$\mc{X}_{k\pm1}$ read
\begin{equation}
U_k^+ = \mu(1+\kappa)(N-k)
\quad\text{and}\quad
U_k^- = \mu(1-\kappa)k
\end{equation}
if mutation is assumed to be an independent point process at all
sites.  We usually have the situation in mind in which fitness
decreases with $k$ and will therefore speak of $U_k^+$ and $U_k^-$ as
the \emph{deleterious} and \emph{advantageous} mutation rates.
However, monotonic fitness is never \emph{assumed}, unless this is
stated explicitly.

In much of the following, we will treat the general model
\eref{eq:paramuse}, which builds on single genotypes, and the
single-step mutation model \eref{eq:step}, in which the units are
genotype classes, with the help of a common formalism.  To this end
note that both models can be recast into the following general form
using matrices of dimension $M$, respectively $N+1$:
\begin{equation} \label{eq:H}
\dot{\bs{p}}(t) = \big ( \bd{H}- \bar R(t) \Id \big ) \bs{p}(t) \,.
\end{equation}
Here, $\Id$ is the identity.  The evolution matrix $\bd{H} = \bd{R} +
\bd{M}$ is composed of a diagonal matrix $\bd{R}$ that holds the
Malthusian fitness values, and the mutation matrix $\bd{M} = (M_{ij})$
with either off-diagonal entries reading $m_{ij}$, or with $U_k^\pm$
on the secondary diagonals.  The diagonal elements in each case are
$M_{ii} = - \sum_{j \neq i} M_{ji}$, hence the column sums vanish.
Where the more restrictive form of the single-step model is needed,
this will be stated explicitly.  Unless we talk about unidirectional
mutation ($U_k^-\equiv 0$ for the single-step mutation model), we will
always assume that $\bd{M}$ is irreducible (i.e.\ each entry is
non-zero for a suitable power of $\bd{M}$).
  
Let now $\bd{T}(t) := \exp(t \bd{H})$, with matrix elements
$T_{ij}(t)$.  Then, the solution of \eref{eq:H} is given by
\citep[see, e.g.,][Ch.\ III.1]{Bue00}
\begin{equation}\label{eq:sol}
\bs{p}(t) = \frac{\bd{T}(t) \bs{p}(0)}{\sum_{i,j} T_{ij}(t) p_j(0)} \,,
\end{equation}
as can easily be established by differentiating and using $\sum_{i,j}
H_{ij} p_j(t) = \sum_i R_i p_i(t) = \bar R(t)$.  Due to
irreducibility, the population vector converges to a unique, globally
stable equilibrium distribution $\bs{p} := \lim_{t\to \infty}
\bs{p}(t)$ with $p_i > 0$ for all $i$, which describes
mutation--selection balance.  By the Perron--Frobenius theorem,
$\bs{p}$ is the (right) eigenvector corresponding to the largest
eigenvalue, $\lambdamax$, of $\bd{H}$.

\subsection{The branching process -- forward and backward in time}
\label{sec:branching}

Our approach will heavily rely on genealogical relationships, which
contain more detailed information than the time course of the relative
frequencies \eref{eq:sol} alone.  Let us, therefore, reconsider the
mutation--selection model as a branching process.  Branching processes
have been a classical tool in population genetics to approximate the
fixation rates of a single mutant type in a finite population. This
approach goes back to \citet{Hal27} \citep[see also][]{CK70}, and has
been used in many recent applications as well
\citep[e.g.][]{Bart95,OtBa97}.

We pursue a different route here by considering the process of
mutation, reproduction and death as a (continuous-time)
\emph{multitype} branching process, as described previously for the
quasispecies model (\citeauthor{DSS85}, \citeyear{DSS85};
\citeauthor{HoSi88} \citeyear{HoSi88}, Ch.\ 11.5).  Let us start with
a \emph{finite} population of individuals, which reproduce (at rates
$B_i$), die (at rates $D_i$), or change type (at rates $M_{ij}$)
independently of each other, \emph{without any restriction on
  population size}.  Let $Y_i(t)$ be the random variable denoting the
number of individuals of type $i$ at time $t$, and $n_i(t)$ the
corresponding realization; collect the components into vectors
$\bs{Y}$ and $\bs{n}$, and let $\bs{e}_i$ be the $i$-th unit vector.
The transition probabilities for the joint distribution, $\Pr
\bigl(\bs{Y}(t)=\bs{n}(t) \mathbin| \bs{Y}(0) = \bs{n}(0) \bigr)$,
which we will abbreviate as $\Pr \bigl(\bs{n}(t) \mathbin| \bs{n}(0)
\bigr)$ by abuse of notation, are governed by the differential
equation\footnote{Note that differentiability of the transition
  \emph{probabilities} is guaranteed in a finite-state,
  continuous-time Markov chain, provided the transition \emph{rates}
  are finite, cf \citet[Ch.\ 4]{KT75} and \citet[Ch.\ 14]{KT81}.}
\begin{equation}
\begin{split}
\frac{d}{dt} \Pr \bigl( \bs{n}(t) \mathbin| \bs{n}(0) \bigr) = & 
 - \bigl( \sum_i (B_i + D_i + \sum_{j \neq i} M_{ji}) n_i(t) \bigr) 
   \Pr \bigl( \bs{n}(t) \mathbin| \bs{n}(0) \bigr) \\
&  + \sum_i B_i \bigl( n_i(t)-1 \bigr) 
    \Pr \bigl( \bs{n}(t)-\bs{e}_i \mathbin| \bs{n}(0) \bigr) \\
&  + \sum_i D_i \bigl(n_i(t)+1 \bigr) 
\Pr \bigl(\bs{n}(t)+\bs{e}_i \mathbin| \bs{n}(0) \bigr) \\ 
&  + \sum_{\substack{i,j \\ i \neq j}} M_{ij} \bigl( n_j(t)+1 \bigr) 
     \Pr \bigl( \bs{n}(t)-\bs{e}_i + \bs{e}_j \mathbin| \bs{n}(0) \bigr) \,. 
\end{split}
\end{equation}

The connection of this stochastic process with the deterministic model
described in \Sref{sec:model} is twofold.  Firstly, in the limit of an
infinite number of individuals ($n:= \sum_i n_i(0) \to \infty$), the
sequence of random variables $\bs{Y}^{(n)}(t)/n$ converges almost
surely to the solution $\bs{y}(t)$ of $\dot{\bs{y}} = \bd H \bs{y}$
with initial condition $\bs{y}(0)= \bs{n}(0)/n$ \citep[][Ch.\ 11,
Thm.\ 2.1]{EtKu86}.  That is, $\Pr \bigl (\lim_{n \to \infty}
\bs{Y}^{(n)}(t)/n = \bs{y}(t) \bigr) = 1$, and the superscript $(n)$
denotes the dependence on the number of individuals.  The connection
is now clear since $\bs{p}(t) := \bs{y}(t) / \sum_i y_i(t)$ solves the
mutation--selection equation \eref{eq:paramuse}.

\begin{figure}
\centerline{\epsfig{file=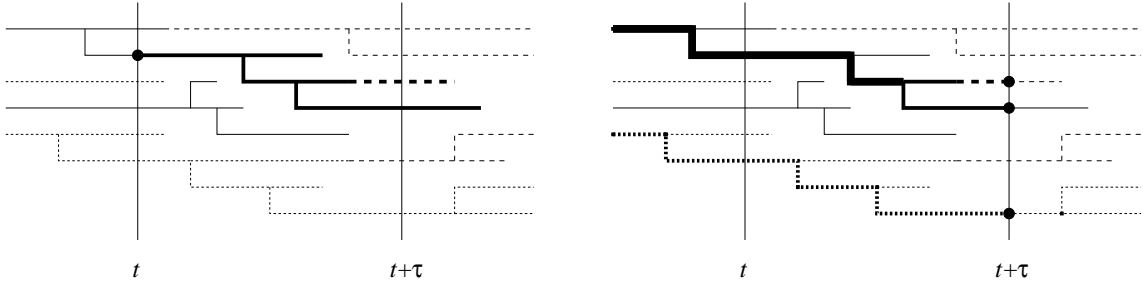}}
\caption{The multitype branching process. Individuals reproduce
  (branching lines), die (ending lines), or mutate (lines changing
  type) independently of each other; the various types are indicated
  by different line styles.  Left: The fat lines mark the clone
  founded by a single individual (bullet) at time $t$.  Right: The fat
  lines mark the lines of descent defined by three individuals
  (bullets) at time $t + \tau$.  After coalescence of two lines, their
  ancestor receives twice the `weight', as indicated by extra fat
  lines.}
\label{fig:branching}
\end{figure}
Secondly, taking expectations of $Y_i$ and marginalizing over all
other variables, one obtains the differential equation for the
conditional expectations
\begin{equation}
\begin{split}\label{eq:condexp}
 \frac{d}{dt} E \bigl(Y_i(t) \mathbin| \bs{n}(0) \bigr) = 
   & (B_i-D_i) E \bigl(Y_i(t) \mathbin| \bs{n}(0) \bigr) \\
 & + \sum_j  \bigl[ M_{ij} E \bigl(Y_j(t) \mathbin| \bs{n}(0) \bigr)
          - M_{ji} E \bigl(Y_i(t) \mathbin| \bs{n}(0) \bigr) \bigr] \,.
\end{split}
\end{equation} 
Clearly, our evolution matrix $\bd{H}$ appears as the infinitesimal
generator here, and the solution is given by $\bd{T}(t) \bs{n}(0)$,
where $\bd{T}(t):=\exp(t\bd{H})$ \citep[see also][Ch.\ 11.5]{HoSi88}.
In particular, we have $E \bigl(Y_i(t)\mathbin|\bs{e}_j
\bigr)=T_{ij}(t)$ for the expected number of $i$-individuals at time
$t$, in a population started by a single $j$-individual at time $0$ (a
`$j$-clone').  In the same way, $T_{ij}(\tau)$ is the expected number
of descendants of type $i$ at time $t+\tau$ in a $j$-clone started at
an arbitrary time $t$, cf left panel of \fref{fig:branching}.  (Note
that, due to the independence of individuals and the Markov property,
the progeny distribution depends only on the age of the clone, and on
the founder type.)  Further, the expected total size of a $j$-clone of
age $\tau$, irrespective of the descendants' types, is $ \sum_i
T_{ij}(\tau)$.
 
Initial conditions come into play if we consider the reproductive
success of a clone \emph{relative to the whole population}.  A
population of independent individuals, with initial composition
$\bs{p}(t)$, has expected mean clone size $\sum_{i,j} T_{ij}(\tau)
p_j(t)$ at time $t+\tau$ (note that $t$ always means `absolute' time,
whereas $\tau$ denotes a time increment).  The expected size of a
single $j$-clone at time $t+\tau$, relative to the expected mean clone
size of the whole population, then is
\begin{equation}\label{eq:zi}
z_j(\tau,t):= 
  \sum_i T_{ij} (\tau)/\sum_{k,\ell}  T_{k\ell}(\tau) p_{\ell}(t)\,.
\end{equation}

The $z_j$ express the expected relative success of a type after
evolution for a time interval $\tau$, in the sense that, if
$z_j(\tau,t) > 1$ ($<1$), we can expect the clone to flourish more
(less) than average (this does in general not mean that type $j$ is
expected to increase (decrease) in abundance relative to the initial
population).  Clearly, the values of the $z_j$ depend on the fitness
of type $j$, but also on its mutation rate and the fitness of its
(mutated) offspring.  (If there is only mutation, but no reproduction
or death, one has a Markov chain and $ z_j(\tau,t)\equiv 1$.)

We now consider \emph{lines of descent}, as in the right panel of
\fref{fig:branching}.  To this end, we randomly pick an individual
alive at time $t+\tau$, and trace its ancestry back in time; this
results in an unbranched line (in contrast to the lineage forward in
time).  Let $Z_{t+\tau}(t)$ denote the type found at time $t \le
t+\tau$, where we will drop the index for easier readability.  We seek
its probability distribution $\Pr\bigl(Z(t)=j\bigr)$.  Since the
(relative) clone size $z_j(\tau,t)$ also determines the expected
(relative) frequency of lines present at time $t+\tau$ that contain a
$j$-type ancestor at time $t$, we have:
\begin{equation} \label{eq:ancestors}
\Pr\bigl(Z(t)=j\bigr) = z_j(\tau,t) p_j(t) =: a_j(\tau,t) \,.
\end{equation}
The $a_j(\tau,t)$ define a probability distribution ($\sum_j
a_j(\tau,t) \equiv 1$), which will be of major importance, and may be
interpreted in two ways.  Forward in time, $a_j(\tau,t)$ is the
frequency of $j$-individuals at time $t$, \emph{weighted by their
  relative number of descendants} after evolution for some time
$\tau$.  Looking backward in time, $a_j(\tau,t)$ is the fraction of
the ($\bs{p}$-distributed) population at time $t+\tau$ whose ancestor
at time $t$ is of type $j$.  We shall therefore refer to
$\bs{a}(\tau,t)$ as the \emph{ancestral distribution} at the earlier
time, $t$.

Let us, at this point, expand a little further on this backward
picture by explicitly constructing the time-reversed process.  This is
done in the usual way, by writing the joint distribution of
parent--offspring pairs (i.e.\ pairs $Z(t)$ and $Z(t+\tau)$) in terms
of forward and backward transition probabilities.  On the one hand,
\begin{equation}\label{eq:cond_forward}
\begin{split}
\Pr\bigl(Z(t+\tau)=i,Z(t)=j \bigr) &=   
\Pr\bigl(Z(t+\tau)= i \mathbin| Z(t)= j \bigr) \Pr\bigl(Z(t)= j \bigr)\\
  &=  P_{ij}(\tau) a_j(\tau,t)\,. 
\end{split}
\end{equation}
Here, the $P_{ij}(\tau) := \Pr \bigl( Z(t+\tau)=i \mathbin| Z(t)=j
\bigr)$ may be obtained by rewriting the (conditional) expectations
defining the (forward) branching process as $T_{ij}(\tau) =
P_{ij}(\tau) \sum_k T_{kj}(\tau)$, which gives
\begin{equation}\label{eq:preverse}
   P_{ij}(\tau) = T_{ij}(\tau) / \sum_k T_{kj}(\tau).
\end{equation}
On the other hand,
\begin{equation}\label{eq:cond_back}
\begin{split}
\Pr\bigl( Z(t+\tau)=i,Z(t)=j \bigr) & =  
 \Pr\bigl(Z(t)=j \mathbin| Z(t+\tau)=i \bigr) \Pr\bigl(Z(t+\tau)= i\bigr) \\
 & =  \tilde P_{ji}(\tau,t) p_i(t+\tau) \,, 
\end{split}
\end{equation}
where $\tilde P_{ji}(\tau,t) := \Pr \bigl( Z(t)=j \mathbin|
Z(t+\tau)=i \bigr)$ is the transition probability of the
\emph{time-reversed} process and is obtained from
\eref{eq:cond_forward} and \eref{eq:cond_back} as $ \tilde
P_{ji}(\tau,t) = a_j(\tau,t) P_{ij}(\tau) \bigl( p_i(t+\tau)
\bigr)^{-1}$.  With Eqs.\ \eref{eq:zi}, \eref{eq:ancestors}, and
\eref{eq:preverse}, one therefore obtains the elements of the backward
transition matrix $\tilde{\bd{P}}$ as
\begin{equation}
   \tilde P_{ji}(\tau,t) = 
    p_j(t) \frac{ T_{ij}(\tau)}{\sum_{k,\ell} T_{k\ell}(\tau) p_\ell(t)} 
    \bigl( p_i(t+\tau) \bigr)^{-1}.
\end{equation}
By differentiating $\tilde{\bd{P}}(\tau,t)$ with respect to $\tau$ and
evaluating it at $\tau=0$, one obtains the matrix $\bd{Q}(t)$
governing the corresponding backward process in continuous time.  Its
elements read $Q_{ji}(t) = \frac{d}{d\tau}
\tilde{P}_{ji}(\tau,t)\bigr|_{\tau=0} = p_j(t) \bigl( H_{ij} -
\delta_{ij} \bar{R}(t) \bigr) \bigl(p_i(t) \bigr)^{-1} - \delta_{ij}
\dot{p}_i(t)/p_i(t)$.  Using \eref{eq:H} this simplifies to:
\begin{equation}
Q_{ji}(t)=\begin{cases} p_j(t)  H_{ij} \bigl(p_i(t) \bigr)^{-1}, & i \neq j \\
            - \sum_{k\neq i}
              p_k(t)  H_{ik} \bigl(p_i(t) \bigr)^{-1}, & i = j\,.
          \end{cases}
\end{equation}
Note that the backward process is, in general, state-dependent (it
does not generate a Markov chain).  Note also that time reversal works
in the same way if sets of types $\mc{X}_k$ instead of single types
are considered, as long as mutation and reproduction rates are the
same within classes.  Furthermore, an analogous treatment is possible
both  for mutation coupled to reproduction, as well as for discrete
generations.

As to the asymptotic behavior of our branching process, it is
well-known that, for irreducible $\bd{H}$ and $t \to \infty$, the time
evolution matrix $\exp \bigl(t(\bd{H} - \lambdamax \Id) \bigr)$
becomes a projector onto the equilibrium distribution $\bs{p}$, with
matrix elements $p_i z_j$ \citep[e.g.][Appendix]{KT81}.  Here,
$\bs{z}$ is the Perron--Frobenius (PF) left eigenvector of $\bd{H}$,
normalized such that $\sum_i z_i p_i = 1$.  As suggested by our
notation, one also has
\begin{equation}
   \lim_{t,\tau \to \infty} \bs{z}(\tau,t) = \bs{z} \,,
\end{equation}
which may be seen from \eref{eq:zi}.\footnote{Both $\bs{z}$ and
  $\bs{p}$ also admit a more stochastic interpretation.  If the
  population does not go to extinction, one has $\lim_{t \to \infty}
  Y_i^{}(t) / \bigl(\sum_j Y_j^{}(t) \bigr) = p_i^{}$ almost surely,
  i.e.\ the stochasticity is in the population size, not in the
  relative frequencies \citep[][Ch.\ 11.5]{KeSt66,HoSi88}.  Further,
  for the \emph{critical process} generated by $\bd{H} - \lambdamax
  \Id$, one has $ \lim_{t \to \infty} t \Pr \bigl(\bd{Y}(t) \neq
  \bs{0} \mathbin| \bd{Y}(0)=\bs{e}_j \bigr) =z_j/C$ and $ \lim_{t \to
    \infty} \frac{1}{t} \bigl( E (Y_i(t) \mathbin| \bs{Y}(0)=\bs{e}_j,
  \bd{Y}(t) \neq \bs{0} \bigr)=C p_i$, where $C$ is a constant; this
  is the continuous-time analog of a result by \citet[][p.\ 
  94]{Jag75}.  Note that, in the long run, the expected offspring
  depend on the founder type only through the probability of
  nonextinction of its progeny.}  We therefore term $z_i$ the
\emph{relative reproductive success} of type $i$.

At stationarity, the matrix governing the backward process simplifies
to $Q_{ji}^{} = p_j^{} \bigl( H_{ij} - \delta_{ij}^{} \lambdamax^{}
\bigr) p_i^{-1}$, which can now be interpreted as a Markov generator.
Further, the (asymptotic) ancestor distribution, given by $a_i = z_i
p_i$, turns out to be the stationary distribution of the backward
process, since $\sum_i Q_{ji}^{} a_i^{} = \sum_i p_j^{} (H_{ij}^{} -
\delta_{ij}^{} \lambdamax^{}) p_i^{-1} z_i^{} p_i^{} = \sum_i p_j^{}
z_i^{} (H_{ij}^{} - \delta_{ij}^{} \lambdamax^{}) = 0$.  Due to
ergodicity of the backward process ($\bd{Q}$ is irreducible if
$\bd{H}$ is), $\bs{a}$ is, at the same time, the distribution of types
along each line of descent (with probability 1).

\subsection{The equilibrium ancestor distribution}
\label{sec:equilanc}

As we saw in the last subsection, there is a simple link between the
algebraic properties of $\bd{H}$ and the probabilistic structure of
the mutation--selection process at equilibrium, which may be
summarized as follows.  The PF right eigenvector $\bs{p}$ (with
$\sum_i p_i=1$) determines the composition of the population at
mutation--selection balance; the corresponding left eigenvector
$\bs{z}$ (normalized so that $\sum_i z_i p_i=1$) contains the
asymptotic offspring expectation (or relative reproductive success) of
the various types; and the ancestral distribution, defined by $a_i=p_i
z_i$, gives the asymptotic distribution of types that are met when
lines of descent are followed backward in time (cf
\fref{fig:branching}).
\begin{figure}
\centerline{\epsfig{file=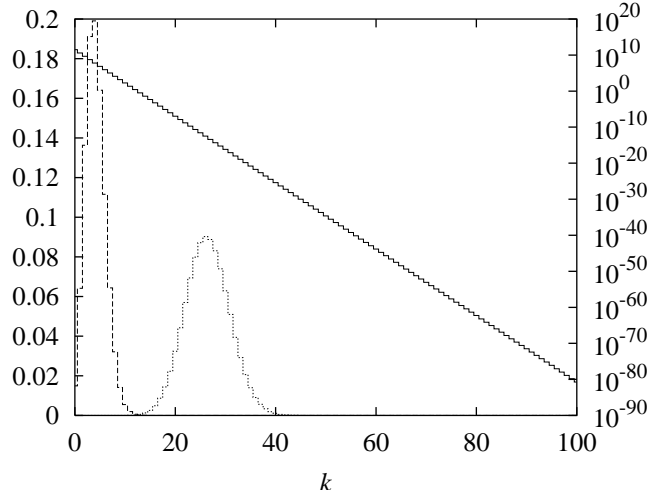,bbllx=67,bblly=52,bburx=308,bbury=236,clip=}}
\caption{Equilibrium values of population frequencies $p_k$ (dotted
  line), ancestor frequencies $a_k$ (dashed line), and relative
  reproductive success $z_k$ (solid line) for the biallelic model
  with additive fitness $R_k = \gamma\,(N-k)$ (where $\gamma$ is the
  loss in reproduction rate due to a single mutation), point mutation
  rate $\mu=0.2\gamma$, mutation asymmetry parameter $\kappa=\frac12$,
  and sequence length $N=100$.  The logarithmic right axis refers to
  the $z_k$ only.}
\label{fig:paz}
\end{figure}
\Fref{fig:paz} shows $\bs{p}$, $\bs{a}$, and $\bs{z}$ for a
single-step mutation model with a linear fitness function.  One sees
that $\bs{z}$ decreases exponentially.

For the single-step mutation model, we may directly transform the
eigenvalue equation $\bd{H}\bs{p} = \lambdamax\bs{p}$ into an equation
for $\bs{a}$.  To this end, we define a diagonal transformation matrix
$\bd{S}$ with non-zero elements $S_{kk} = \prod_{\ell=1}^k
\sqrt{U^-_\ell/U^+_{\ell-1}}$ and obtain a symmetric matrix by
$\tilde{\bd{H}} := \bd{SHS}^{-1}$.  The corresponding PF right and
left eigenvectors are given by $\tilde{\bs{p}} = \bd{S}\bs{p}$ and
$\tilde{\bs{z}} = \bd{S}^{-1}\bs{z}$.  But now, as $\tilde{\bd{H}}$ is
symmetric, we have $\tilde{\bs{z}} \sim \tilde{\bs{p}}$ (where $\sim$
means proportional to).  Hence, due to $a_k = z_k p_k = \tilde{z}_k
\tilde{p}_k \sim \tilde{p}_k^2$, one has $\tilde{p}_k \sim
\sqrt{a_k}$.  Thus, we obtain the following explicit form of the
eigenvalue equation for $\tilde{\bd{H}}$, which will be crucial for
the derivation of our main results:
\begin{equation}
\label{eq:evsymm}
[R_k^{} - U^+_k - U^-_k] \sqrt{a_k^{}} +
\sqrt{U^+_{k-1} U^-_k} \sqrt{a_{k-1}^{}} + \sqrt{U^+_k U^-_{k+1}} 
\sqrt{a_{k+1}^{}} = \lambdamax^{} \sqrt{a_k^{}} \,.
\end{equation}
Note that \Eref{eq:evsymm} relates the mean fitness of the
\emph{equilibrium} population ($\bar{R} = \lambdamax$) to the
\emph{ancestor} frequencies $a_k$.

\subsection{Observables and averages}
\label{sec:averages}

In this subsection we define the observables, i.e.\ measurable
quantities, that are used to describe the population on its
evolutionary course.  Besides the usual population mean, we shall also
introduce the mean with respect to the ancestor distribution (see
\sref{sec:equilanc}).

We shall consider means and variances of two observables in the
following.  To this end, we characterize each type (or class) $i$ by
its \emph{fitness value} $R_i$ and its \emph{mutational distance}
$X_i$ from the reference genotype (or the class $\mc{X}_0$).  For the
biallelic model in particular, mutational distance corresponds to the
Hamming distance to $\bs{s}_+$.  If, in addition, this is the fittest
type, $X_i$ just gives the number of deleterious mutations.  But in
general it can also be used to describe the value of any additive
trait with equal contributions of sites or loci.  Similarly, for
single-step mutation, we define $X_k$ to be the distance from the
class $\mc{X}_0$, thus $X_k = k$ for class $\mc{X}_k$.  Again, $X_k$
may be viewed as (the genetic contribution to) any character with
discrete values that depends linearly on the mutation classes.

\paragraph{Population average.}

Representing an arbitrary observable as $(O_i)$, such as $(R_i)$ or
$(X_i)$, we will denote its \emph{population average} as
\begin{equation}
\bar{O}(t) := \sum_i O_i p_i(t)\,.
\end{equation}
By omission of the time dependence we will indicate the corresponding
equilibrium average.

As to mean fitness, $\bar{R}(t)$ determines the mutation load,
$L(t):=\Rmax-\bar{R}(t)$.  Here, $\Rmax = \max_i R_i$ is the fitness
of the fittest genotype, in line with the usual convention \citep[see,
e.g.,][]{Ewe79,Bue00}.  It is well-known that the equilibrium value
$\bar{R} := \lim_{t\to\infty}\bar{R}(t)$ is given by the largest
eigenvalue, $\lambdamax$, of the evolution matrix $\bd{H}$.

For the variance of fitness, $V_R^{}(t) = \sum_i (R_i^{} -
\bar{R}(t))^2 p_i^{}(t)$, we differentiate $\bar{R}(t)$ according to
\eqref{eq:paramuse}, i.e.\ $\frac{d}{dt} \bar{R}(t) = \sum_i R_i \dot
p_i(t) = V_R(t) + \sum_{i,j} R_i M_{ij} p_j(t)$, and hence
\begin{equation} \label{eq:varfin}
V_R^{}(t) = \frac{d}{dt} \bar{R}(t) - \sum_{i,j} R_i M_{ij} p_j(t)
= \frac{d}{dt} \bar{R}(t) + \sum_j \Big({\textstyle\sum\limits_{i}}(R_j-
R_i)M_{ij}\Big)p_j(t) \,.
\end{equation}
The interpretation of this completely general formula is as follows:
In absence of mutation, \Eref{eq:varfin} just reproduces Fisher's
Fundamental Theorem, i.e.\ the variance in fitness equals the change
in mean fitness, as long as there is no dominance \citep[see,
e.g.,][]{Ewe79}.  If mutation is present, however, a second component
emerges, which is given by the population mean of the mutational
effects on fitness, weighted by the corresponding rates.  It may be
understood as the \emph{rate of change in mean fitness due to mutation
  alone}.  At mutation--selection balance, this second term is
obviously the only contribution to variance in fitness.

For the single-step mutation model in particular, we can define
\emph{deleterious} and \emph{advantageous mutational effects}
separately as $s^+_k = R_k^{}-R_{k+1}^{}$ and $s^-_k =
R_{k-1}^{}-R_k^{}$, respectively.  For decreasing fitness values
(which is the usual case, but not strictly presupposed here) these are
positive.  This way we obtain
\begin{equation}
\label{eq:varmuteff}
V_R^{} =  \overline{s^+ U^+ - s^- U^-} 
= \overline{s^+}\; \overline{U^+}  - \overline{s^-} \;\overline{U^-} 
+ \Cov(s^+,U^+) - \Cov(s^-,U^-)
\end{equation}
for the equilibrium variance, a result we will rely on in the
following.

Just as for the fitness distribution, we define the population mean,
$\bar{X}(t) = \sum_{i=0}^N X_i^{} p_i^{}(t)$, and variance, $V_X^{}(t)
= \sum_i (X_i^{} - \bar{X}(t))^2 p_i^{}(t)$, of the mutational
distance.

\paragraph{Ancestral average.}

We will also need the ancestral average of our observables, that is,
the average with respect to the ancestral distribution defined in
\Eref{eq:ancestors}: $\hat{O}(\tau,t) := \sum_i O_i a_i(\tau,t) =
\sum_i z_i^{}(\tau,t) O_i^{} p_i^{}(t)$.  In the following, we will
only be concerned with the ancestral distribution in equilibrium,
i.e.\ with both $t$ and $\tau$ going to infinity.  We obtain the
ancestral average of any observable $(O_i)$ in this limit as
\begin{equation}
\hat{O} := \sum_i O_i^{} a_i^{} = \sum_i z_i^{} O_i^{} p_i^{}  \,.
\end{equation}

These averages may be read forward in time (corresponding to a
weighting of the current population with expected offspring numbers),
and backward in time (corresponding to an averaging w.r.t.\ the
distribution of the ancestors).  A third interpretation is available
if the mutation matrix is irreducible, which entails that the
equilibrium backward process defined by $\bd{Q}$ is ergodic.  Then,
with probability 1, the equilibrium ancestral average also coincides
with the average of the observable over a lineage backwards in time.
Note that the information so obtained is not contained in the
population average, which is merely a `time-slice' average.  The
ancestral mean adds a time component to the averaging procedure, which
provides extra information on the evolutionary dynamics.  In
\aref{app:physics}, we shall show that our ancestral averaging
coincides with the way observables are evaluated in a system of
quantum statistical mechanics.

\subsection{Linear response and mutational loss}
\label{sec:response}

We now come to another interpretation of the equilibrium ancestor
frequencies introduced in \sref{sec:branching}.  Consider the
derivative of the equilibrium mean fitness with respect to the $i$-th
fitness value in a general system of parallel mutation and selection
\eref{eq:paramuse}:
\begin{equation}\label{eq:linres1}
\frac{\partial\bar{R}}{\partial R_i}  
= \frac{\partial}{\partial R_i} \biggl[\sum_{j,k} z_j H_{jk} p_k \biggr]  
= a_i + \bar{R} \frac{\partial}{\partial R_i} \biggl[\sum_j z_j p_j
\biggr] = a_i \,,
\end{equation}
where we made use of the normalization condition $\sum_j z_j p_j =
\sum_j a_j \equiv 1$.  The ancestor frequency $a_i$ therefore measures
the \emph{linear response} (or \emph{sensitivity}) of the equilibrium
mean fitness to changes in the $i$-th fitness value.\footnote{If
  mutation is coupled to reproduction, the linear response to
  variations in the death rate $D_i$ is given by $-a_i$.}  A similar
calculation for the response to changes in the mutation rates results
in
\begin{equation}\label{eq:linres2}
\frac{\partial\bar{R}}{\partial M_{ij}} = (z_i - z_j)p_j \,. 
\end{equation}
Using \eref{eq:linres1} and \eref{eq:linres2}, we can express the
equilibrium mean fitness as follows:
\begin{equation}\label{eq:linres3}
\bar{R} = \hat{R} + \sum_{i,j} z_i M_{ij} p_j =
\sum_i R_i \frac{\partial \bar{R}}{\partial R_i} + 
\sum_{i,j} M_{ij} \frac{\partial \bar{R}}{\partial M_{ij}}\,.
\end{equation}

Let us give a variational interpretation for the ancestor mean fitness
as well. To this end, we define the \emph{mutational loss} $G$ of the
system as the difference between ancestor and population mean fitness
in equilibrium. Assume now that we change all mutation rates $M_{ij}$
by variations in a common factor $\mu$.  From \eref{eq:linres3} and
\eref{eq:linres1} we then find that the mutational loss relates to the
linear response of the equilibrium mean fitness to changes in the
mutation rates as:
\begin{equation} \label{eq:linres4}
G := \hat{R} - \bar{R} = - \mu\frac{\partial \bar{R}}{\partial \mu} \,.
\end{equation}
Actually, this relation holds for arbitrary (clonal)
mutation--selection systems, in particular also if mutation and
reproduction are coupled (in which case the mutation \emph{rates} are
replaced by mutation \emph{probabilities}).

Eqs.\ \eref{eq:linres1} and \eref{eq:linres4} may also be used to
determine the change in mean fitness if $\bd{H}$ changes to $\bd{H} +
\bd{\Delta H}$, to linear order in $\bd{\Delta H}$. (Small changes in
the fitness values, or mutation rates, may be due to environmental
changes, or changes in the genetic background.)  Clearly, $\bd{H} +
\bd{\Delta H}$ has $\bar{R} + \Delta\bar{R}$ as largest eigenvalue,
with $\Delta\bar{R} \simeq \sum_i \Delta R_i (\partial \bar{R}/
\partial R_i) + \sum_{i,j} \Delta M_{ij} (\partial \bar{R} / \partial
M_{ij})$ to linear order in $\Delta R_i$ and $\Delta M_{ij}$. If only
fitness values are affected, \eref{eq:linres1} yields
\begin{equation}\label{eq:linresr}
\Delta \bar{R} \simeq \sum_i \Delta R_i \, a_i,
\end{equation}
where the $a_i$ belong to the original system. If only the mutation
rates change by variations in a common factor $\mu$, \eref{eq:linres4}
leads to
\begin{equation}\label{eq:linresmu}
\Delta\bar{R} \simeq -\frac{\Delta\mu}{\mu} G\,.
\end{equation}
We will come to further interpretation and discussion of the
mutational loss and the response relations in \sref{sec:mutflow}.

\subsection{Fitness functions and mutation models}
\label{sec:fitmut}

For many of the results and all of our examples, we will restrict our
treatment to the case of the single-step mutation model as described
by \Eref{eq:step}.  Although most of our results do not depend on this
particular choice we will, for simplicity, concentrate on this scheme
here, and only briefly discuss possible extensions.  We will start out
with a discussion of fitness functions and mutation schemes in this
context.  Depending on whether the phenotype or the genotype is
considered the primary quantity for the model, the inherent
approximation mainly concerns the mutation or the fitness part,
respectively.

If the $X_k^{}$ ($0 \le k \le N$) are the values of a quantitative
trait on which selection acts, fitness may be taken as an arbitrary
function of it.  The essential assumption, in this case, is that
genotypes with equal trait values have equivalent mutation patterns,
with mutation in single steps as an additional simplification.  This
is the original view in which this assumption first appeared, with
$X_k^{}$ as the electric charge of proteins \citep{OhKi73}.  The
numerous papers to follow have been reviewed by \citet{Bue98,Bue00}.

If, on the other hand, $X_k^{}$ is the number of mutations with
respect to the wildtype (i.e.\ $X_k=k$ as in the biallelic model),
single-step mutation is a natural approximation and directly emerges
if mutation and reproduction are modeled as independent processes.
The essential simplification, in this case, consists in the choice of
the genotype fitness values, which depend only on $k$.  This way, only
the average epistatic effect is included in the model, whereas any
variance among epistatically interacting mutations is disregarded.
Fitness functions of this kind, although undoubtedly lacking much of
the biological complexity, have been used as standard landscapes
throughout population genetics literature.  While the principal reason
for this seems to lie in the large simplifications due to permutation
invariance, they already take full account of the limited information
on fitness provided by mutation accumulation experiments (e.g.\ 
\citec{CrSi83}; but see also the discussion by \citec{POW00}).
Further, they include a broad range of examples with vastly diverging
properties, ranging from simple additive fitness over \emph{quadratic}
-- or otherwise polynomial or exponential -- landscapes with smoothly
varying fitness values \citep[e.g.][]{Char90} to \emph{truncation
  selection} \citep[e.g.][]{Kond88} and Eigen's \emph{sharply peaked}
landscape \citep{EMcCS89}.

\label{sec:scaling}
For a consistent treatment of our model in the mutation class limit
$N\to\infty$ (to be defined in the next subsection), it will be
advantageous to think of the fitness values and mutation rates as
being determined by the mutational distance \emph{per class} (or
site), $x_k := X_k/N \in [0,1]$:
\begin{equation}
\label{eq:defrupm}
R_k^{} = N r_k^{} = N r(x_k^{})\,, \qquad 
U^\pm_k = N u^\pm_k = N u^\pm(x_k^{}) \,.
\end{equation}
Here, also $r_k^{}$ and $u^\pm_k$ are introduced as fitness and total
mutation rates per class.  They can now be thought of as being
defined, without loss of generality, by three functions $r$ and
$u^\pm$ on the compact interval $[0,1]$.  We will refer to $r$ as the
\emph{fitness function}, and to $u^+$ and $u^-$ as the (deleterious
and advantageous) \emph{mutation functions} of the model.  Both $u^+$
and $u^-$ are assumed to be continuous and positive, with boundary
conditions $u^-(0) = u^+(1) = 0$, and $r$ to have at most finitely
many discontinuities, being either left or right continuous at each
discontinuity in $\openunitint$.  This should include all biologically
relevant examples.  For the biallelic model, the mutation functions
are simple linear functions of $x$:
\begin{equation}
\label{eq:mutbiall}
u^+(x) = \mu (1 + \kappa)(1-x) \,, \qquad
u^-(x) = \mu (1 - \kappa)x \,.
\end{equation}
Note that the classical stepwise mutation model \citep{OhKi73} is not
covered by this framework, since its genotype space $\ZZ$ is inherently
non-compact.

\subsection{Three limiting cases}
\label{sec:limit}

Our primary aim in the following sections is to establish simple
relations for the equilibrium means and variances of mutational
distance and fitness that lend themselves to biological
interpretation.  Whereas these relations are approximations in the
general case, they rest on three limiting cases as pillars, for which
they hold as exact identities.  All three are biologically meaningful
by themselves, two of them are well studied, and we will show that the
formulas reduce to well-known results there.

The first case is the \emph{limit of vanishing back mutations},
defined by $U^-_k \equiv 0$ in our model.  The second one is a
\emph{limit of linearity}, in which fitness and mutation rates depend
linearly on some trait $Y_k = N y_k = N y(x_k)$ with $Y_0 = 0$ and
$Y_N = N$, such as
\begin{equation}\label{eq:lincond}
r(x) = r_0 - \alpha y(x) \,, \qquad 
u^+(x) = \beta^+(1-y(x)) \,, \qquad  u^-(x) = \beta^-y(x) \,.
\end{equation}
Note that, if $Y_k$ is proportional to the mutational distance $X_k =
k$, the fitness function is linear whereas the mutation functions
$u^\pm$ reproduce the mutation scheme of the biallelic model if
$\beta^\pm = \mu(1\pm\kappa)$.  This limit can be understood as the
limit of vanishing epistasis, in which the system is known as the
Fujiyama model in the sequence space literature \citep[cf][]{Kauf93}.

The third case is the limit of an infinite number of mutation classes,
$N\to\infty$, which we will call \emph{mutation class limit} for
short.  In the case of the biallelic multilocus model, this limit has
been used and discussed in a recent publication \citep[]{BaWa01}. It
addresses the situation of weak or almost neutral mutations, where the
average mutational effect (over the mutation classes) is small
compared to the mean total mutation rate, $U \gg s$.  The limit
further assumes that differences in mutation rate between neighboring
(pairs of) classes are small compared to the mean rate itself.  In
this case, genetic change by mutation proceeds in many steps of small
average effect and the model is a genuine multi-class model in the
sense that typically a large number of classes are relevant in
mutation--selection equilibrium.  Note that only the \emph{average}
mutational effect must be small; this includes the possibility of
single steps with much larger effect (such as in truncation
selection).

Technically, the limit $N \to \infty$ is performed such that the
mutational effects $s^\pm$ and the fitness values and mutation rates
\emph{per class}, $r$ and $u^\pm$, remain constant.  If fitness values
and mutation rates are defined by the three functions $r$ and $u^\pm$
as described above \eref{eq:defrupm}, increasing $N$ simply leads to
finer `sampling' of the functions.

With this kind of scaling, the means and variances \emph{per class} of
the observables defined in \sref{sec:averages} approach well defined
limits, which then serve as approximations for the original model with
finite $N$.  We will denote them by the corresponding lower case
letters, i.e.\ $\hat{r} := \hat{R}/N$, $v_X := V_X/N$, etc.; an
additional subscript will indicate the limit value, e.g.\ 
$\bar{x}_\infty := \lim_{N\to\infty} \bar{x}$.  Note that it is, in
general, the \emph{variance per class} of a given quantity that is
meaningful in this limit, not the variance of the \emph{quantity per
  class} (e.g.\ $\Var(X/N)$), which tends to zero (cf
\sref{sec:derivrest}).  The described limit is the biological analog
of the \emph{thermodynamic limit} in statistical physics.  We will
further discuss this issue for physically interested readers in
\aref{app:physics}.

Let us finally compare the mutation class limit with the more familiar
\emph{infinite-sites limit}, which, when applied to the biallelic
model, also leads to a stepwise mutation model with an infinite number
of classes \citep[as found, e.g., in][]{Bue00}.  Both limits, however,
approximate an original situation with a large, but finite number of
types in quite different ways.  In the infinite-sites limit, the
original model is \emph{extrapolated} to an infinite one by adding new
states at the boundaries, where the population distribution is
(assumed to be) small.  In contrast, the present approach arrives at
the limit by \emph{interpolation} of the types of the finite model.
Mathematically, this leads to a non-compact state space (such as
$\ZZ$) in the infinite-sites limit, whereas the state space in the
mutation class limit is a compact interval (bounded by the extreme
types of the original model).  To approximate biological observables
of the finite model in the limit, the approaches use different
scaling.  In the infinite-sites case, the range of Malthusian fitness
parameters $R$ usually diverges (depending on how the extrapolation is
done), while the total (`genomic') mutation rate $U$ is kept constant.
In the mutation class limit, both $R$ and $U$ diverge with $N$, but
the ratio $U/R$ is kept constant.  These differences in scaling result
in different ranges of validity of the two limits.  The mutation class
limit assumes $U \gg s$, it is accurate if the total mutation rate is
large or fitness differences are small, and allows a sizable fraction
of sites to be mutated ($\bar{x} = \bar{X}/N$ may approach a non-zero
limit).  In this article, we are mainly interested in this regime, in
particular in \sref{sec:thresholds}, where we discuss error
thresholds.  Infinite-sites models, on the other hand, typically
assume $U \ll s$.  Then back mutations can be neglected, and the bulk
of the population is concentrated on just a few classes with a finite
number of mutations.

\section{Results for observable means and variances}
\label{sec:results}

In this section, we want to give a short summary of our main findings
for the single-step mutation model.  Derivations and a more extended
discussion are postponed to Sections \ref{sec:deriv} and
\ref{sec:interpret}.

A key result of this article is the following estimate of the
equilibrium mean fitness, which states a maximum principle and holds
as an exact identity in the three limiting cases described in the
preceding section:
\begin{equation} \label{eq:maxprinc}
\bar{r} \simeq \bar{r}_\infty =  \sup_{x\in[0,1]} \!\big(r(x)-g(x)\big) \,.
\end{equation}
Here, the function $g$ is defined as twice the difference between the 
arithmetic and geometric mean of the mutation functions:
\begin{equation}
\label{eq:g}
g(x) = u^+(x) + u^-(x) - 2\sqrt{u^+(x)u^-(x)} \,.
\end{equation}
For reasons that will become clear in \sref{sec:mutflow}, we will call
it \emph{mutational loss function}.  For the biallelic model, it reads 
explicitly:
\begin{equation}
\label{eq:gbiall}
g(x) = \mu\left(1+\kappa-2\kappa x - 2\sqrt{(1-\kappa^2)x(1-x)}\right) \,.
\end{equation}
In general, $\bar{r}_\infty$ describes the equilibrium mean fitness
$\bar{r}$ to leading order in $1/N$ and to next to leading order in
$u^-$.  The approximation is indeed rather accurate already for
moderately large $N$ and/or weak back mutation rates, cf
\sref{sec:accuracy} and the examples in \sref{sec:thresholds}.

The maximum principle \eref{eq:maxprinc} is closely linked to the
ancestor distribution.  In particular, if the maximum is attained at a
unique value, this is precisely the ancestor mean $\hat{x}_\infty$:
\begin{equation}
\label{eq:hatx}
\bar{r}_\infty = r(\hat{x}_\infty) - g(\hat{x}_\infty) 
  = \hat{r}_\infty - g(\hat{x}_\infty) \,,
\end{equation}
where the relation $r(\hat{x}_\infty) = \hat{r}_\infty$ can be proved
for all three limiting cases.  A corresponding relation for the
population mean $\bar{x}_\infty$,
\begin{equation}
\label{eq:barr}
\bar{r}_\infty = r(\bar{x}_\infty) \,,
\end{equation}
holds in the mutation class limit and the linear case, if this
equation has a unique solution (e.g.\ for strictly monotonic $r$).

The variances per site of fitness and of distance from wildtype are
then given by
\begin{equation}
\label{eq:var}
v_{R,\infty}^{} = 
  -r'(\bar{x}_\infty) \bigl(u^+(\bar{x}_\infty)-u^-(\bar{x}_\infty)\bigr)
\quad\text{and}\quad
v_{X,\infty}^{} = \frac{v_{R,\infty}^{}}{\left(r'(\bar{x}_\infty)\right)^2} \,,
\end{equation}
provided $r$ is differentiable, in which case $-r'(\bar{x}_\infty)$ is
the population mean of the mutational effects.  For the biallelic
model this is explicitly:
\begin{equation}
\label{eq:varbiall}
v_{R,\infty}^{} = -r'(\bar{x}_\infty) \mu\,(1+\kappa - 2\bar{x}_\infty)
\quad\text{and}\quad
v_{X,\infty}^{} = 
  -\frac{\mu\,(1+\kappa - 2\bar{x}_\infty)}{r'(\bar{x}_\infty)} \,.
\end{equation}
If $r$ has a jump discontinuity at $\xjump$ from $r^+$ to $r^-$ and we
have $r^+ \le \bar{r}_\infty \le r^-$, then $\bar{x}_\infty = \xjump$
and $v_{R,\infty}$ diverges.  In this case, $V_{r,\infty} =
\lim_{N\to\infty} V_R/N^2$ is finite (cf the example of truncation
selection in \fref{fig:truncsel} and the one in \fref{fig:wtthr}):
\begin{equation}
\label{eq:varrjump}
V_{r,\infty}^{} = (r^+ - \bar{r}_\infty) (\bar{r}_\infty - r^-) \,.
\end{equation}

\begin{figure}[t]
\centerline{\epsfig{file=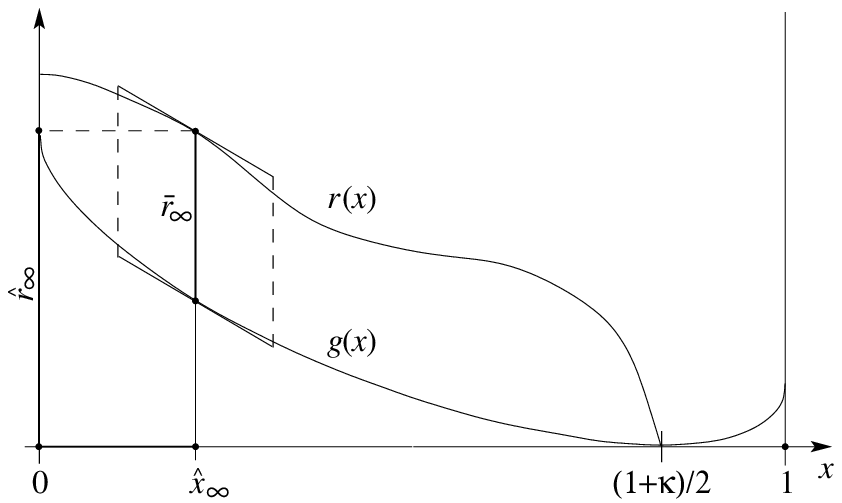,bbllx=0,bblly=0,bburx=248,bbury=143,clip=}}
\centerline{\epsfig{file=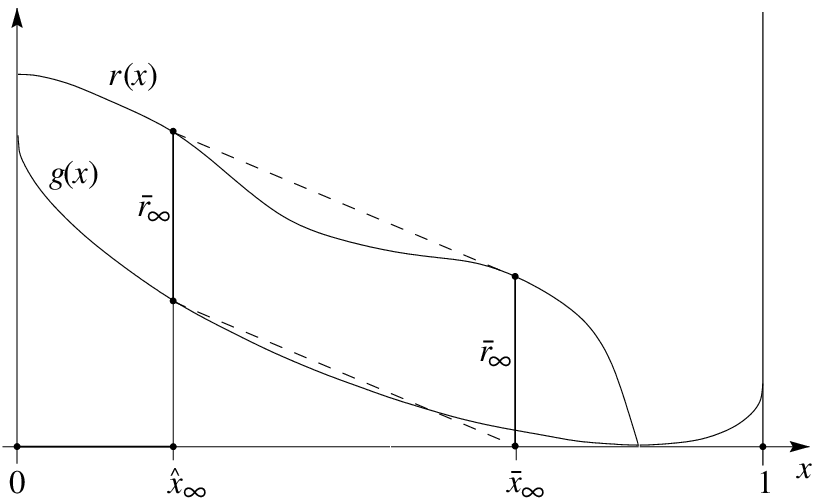}}
\caption{Graphical constructions for the observable means following
  the results in \sref{sec:results}.  Upper part: $\bar{r}_\infty$ is
  the maximal distance $r(x) - g(x)$, cf \eref{eq:maxprinc}.  This is
  attained at $x = \hat{x}_\infty$, cf \eref{eq:hatx}, where
  $r'(\hat{x}_\infty) = g'(\hat{x}_\infty)$.  Lower part:
  $\bar{x}_\infty$ is the solution of $\bar{r}_\infty =
  r(\bar{x}_\infty)$, cf \eref{eq:barr}.}
\label{fig:comics}
\end{figure}
The results presented here lead to simple graphical constructions of
the means as shown in \fref{fig:comics}.  This allows for an intuitive
overview over the dependence of these quantities on (the shape of) the
fitness function and mutation rates, without the need for explicit
calculations.

\section{Derivations}
\label{sec:deriv}

We now come to the proofs and some first interpretation of the results
presented in the previous section.  Our starting point is the
mutation--selection equilibrium of the single-step model \eref{eq:step}
for finite $N$, i.e.\ the eigenvalue equation
\begin{equation}
\label{eq:intens}
[r(x_k) - u^+(x_k) - u^-(x_k)] p_k 
+ u^+(x_{k-1}) p_{k-1} + u^-(x_{k+1}) p_{k+1} = \bar{r} p_k \,.
\end{equation}
For most of our calculations, we will use the equivalent equation for
the ancestor distribution, cf \eref{eq:evsymm},
\begin{multline}
\label{eq:symm}
[r(x_k) - u^+(x_k) - u^-(x_k)] \sqrt{a_k} \,+\\
\sqrt{u^+(x_{k-1}) u^-(x_k)} \sqrt{a_{k-1}} + \sqrt{u^+(x_k) u^-(x_{k+1})} 
\sqrt{a_{k+1}} = \bar{r} \sqrt{a_k} \,,
\end{multline}
which is the eigenvalue equation for the largest eigenvalue of the
symmetric matrix $\tilde{\bd{H}}$.  For the latter, Rayleigh's
principle is applicable, which is a general maximum principle
involving the full $(N+1)$-dimensional space: $\bar{r} = \sup_{\bs{y}}
\sum_{k,\ell} y_k^{} \tilde{H}_{k\ell}^{} y_\ell^{}/\sum_k y_k^2$,
with non-zero $\bs{y}$.  In the following subsections we will show,
for each of the three limiting cases (cf \sref{sec:limit}) separately,
how it boils down to the simple scalar maximum principle
\eref{eq:maxprinc} and the relation \eref{eq:hatx}, and give a
biological interpretation.  We will then come to the derivation of the
other identities.

\subsection{Unidirectional mutation}
\label{sec:derivuni}

We start with the limit of unidirectional mutation, since exclusion of
back mutations leads to a considerably simpler situation, and we can
show how our findings connect to well-known results.  To be specific,
we assume $u^-_k \equiv 0$ and $u^+_k > 0$ for $k<N$.  All results
then follow fairly directly from the equilibrium condition
\eref{eq:intens}.

Owing to $u^-_k \equiv 0$, the equilibrium distribution $\bs{p}$ in
general depends on initial conditions.  But $u^+_k > 0$ implies that
for any such $\bs{p}$, there exists a particular label $\hat{k}$, $0
\le \hat{k} \le N$, which divides all classes of genotypes into two
parts according to
\begin{equation}
p_k = 0 , \; k < \hat{k} , \qquad p_k > 0, \; k \ge \hat{k} \,. 
\end{equation}
Equivalently, we obtain for the corresponding left eigenvector
$\bs{z}$:
\begin{equation}
z_k = 0,  \; k > \hat{k} , \qquad  z_k > 0, \;  k \le \hat{k} \,.
\end{equation}
Since $a_k = p_k z_k$, this shows that the only non-zero element of
the ancestral distribution is $a_{\hat{k}} = 1$, and that $\hat{k}$ is
the equilibrium ancestor mean $\hat{X}$ of the mutational distance
from the reference class $\mc{X}_0$.  In line with this, the
mutational distance of every line of ancestors in equilibrium dynamics
converges to $\hat{k}$ (with probability 1).  For the classes with
non-vanishing frequency, the fitness and total mutation rate are thus
related according to
\begin{equation} \label{eq:rgrel}
\hat{r} - \bar{r} =
r(\hat{x}) - \bar{r} = u^+(\hat{x}) , \qquad 
r(x_k) - \bar{r} < u^+(x_k), \; k > \hat{k} \,,
\end{equation}
the first part of which corresponds to \eref{eq:hatx}.

Although the equilibrium distribution is not unique, \eref{eq:rgrel}
implies that the one with maximal mean fitness (which is the only
stable one and is automatically obtained in the limit of vanishing
back mutations, $u^- \to 0$, or by starting with a population with
$p_0(t=0) > 0$) is characterized by
\begin{equation} \label{eq:max1}
\bar{r} = r(\hat{x}) - u^+(\hat{x}) = 
  \max_k \left(r(x_k) - u^+(x_k) \right)
\end{equation}
for arbitrary choices of $r(x_k)$ and $u^+(x_k)$.  Obviously,
\eref{eq:max1} is the discrete version of the maximum principle given
in \Eref{eq:maxprinc}.

If the sequence $r(x_k)$ \emph{or} the sequence $u^+(x_k)$ is
monotonically decreasing (as in the biallelic model), $\hat{k}$ is
also the fittest class present in the equilibrium population:
\begin{equation} \label{eq:unimaxfit}
\hat{r} = r(\hat{x}) = 
  \max_k \left\{ r(x_k) \mathrel\big| p_k \neq 0\right\} \,.
\end{equation}
If additionally $\hat{k}$ coincides with the class of maximal fitness,
i.e.\ $\hat{r} = \rmax$, then \eref{eq:rgrel} is a special case of
Haldane's principle, which relates the mutation load $l$ to the
deleterious mutation rate of the fittest class \citep{KiMa66,Bue98}:
\begin{equation} \label{eq:haldane}
l = \rmax - \bar{r} = u^+(\hat{x}) \,.
\end{equation}
In derivations of (variants of) \Eref{eq:haldane}, it is often tacitly
assumed that the equilibrium frequency of the fittest class is
non-zero.  This, however, is in general not the case and must be made
explicit here since we are also interested in the change of the
equilibrium distribution with varying mutation rates.  This can lead
to a shift in $\hat{k}$ and hence in $\hat{r}$.

\subsection{The linear case}

If fitness values and mutation rates depend linearly on some trait
$Y$, as described in \eref{eq:lincond}, the maximum principle holds as
an exact identity.  This may be derived from \eref{eq:symm} by a short
direct calculation, which we present in \aref{app:prooflin}.

For an interpretation of this result, first consider a trait
proportional to the mutational distance $X$ from the reference class,
in which case the system coincides with the Fujiyama model.  Since
this is a model without epistasis, the means and variances are easily
obtained \citep{OBri85,BaWa01}.  In particular, they are independent
of the number of classes.  What is more, our derivation shows that
they only rely on a linear dependence of fitness and mutation
functions on some trait, as well as the boundary conditions for the
mutation functions.  This means that they remain unchanged if mutation
classes are permuted, or even subjoined or removed.

\subsection{Mutation class limit}
\label{sec:discinf}

Since the proof of the maximum principle \eref{eq:maxprinc} and the
relation \eref{eq:hatx} in the limit $N \to \infty$ is somewhat
technical we will just give a sketch here and defer the details to
\aref{app:proofinf}.  The main idea is to look at the system locally,
i.e.\ at some interval of mutation classes in \eref{eq:intens} and
\eref{eq:symm}.  This will provide us with upper and lower bounds for
the mean fitness of a system with finite $N$ (denoted by
$\bar{r}^{}_N$).  In the limit $N \to \infty$, they can then be shown
to converge to the same value $\bar{r}_\infty = \lim_{N\to\infty}
\bar{r}^{}_N$.

For a lower bound, let us consider submatrices of the evolution matrix
$\bd{H}$ that, for any class $\mc{X}_k$, consist of the rows (and
columns) corresponding to $\mc{X}_{k-m}$ through $\mc{X}_{k+n}$.  Each
of them describes the evolution process on a certain interval of
mutation classes at whose boundaries there is mutational flow out, but
none in.  Thus, each largest eigenvalue, $\bar{r}_{k,m,n}$,
corresponding to the local growth rate, is a lower bound for
$\bar{r}^{}_N$.  In order to estimate $\bar{r}_{k,m,n}$, it is
advantageous to use the formulation in ancestor form -- with the same
local growth rates as largest eigenvalues of the corresponding
symmetric submatrices of $\tilde{\bd{H}}$.  Here, lower bounds can be
found due to Rayleigh's principle, and follow from evaluating the
corresponding quadratic form for the vector $(1,1,\dots,1)^T$:
\begin{equation}
\label{eq:lowerfin}
\bar{r}^{}_N \ge \bar{r}_{k,m,n} \ge \frac{1}{n+m+1}
 \bigg[ \sum_{\ell=k-m}^{k+n}\Big( r_\ell - g^{}_{N,\ell} \Big) 
 - \sqrt{u^+_{k-m-1}u^-_{k-m}} - 
\sqrt{u^+_{k+n}u^-_{k+n+1}} \bigg] \,,
\end{equation}
where $g^{}_{N,\ell} = u^+_\ell + u^-_\ell - \sqrt{u^+_{\ell-1} u^-_\ell}
- \sqrt{u^+_\ell u^-_{\ell+1}}$.

For an upper bound, consider a local maximum of the ancestor
distribution, i.e.\ a $k^+$ such that $a_{k^+} \ge a_{k^+\pm1}$ (with
the convention $a_{N+1} = a_{-1} = 0$ such a maximum always exists).
Evaluating \eref{eq:symm} for this $k^+$ then yields the inequality
\begin{equation} \label{eq:upper}
\bar{r}^{}_N \le  r^{}_{k^+} - g^{}_{N,k^+} \le 
  \sup_k \big( r^{}_k - g^{}_{N,k} \big) \,.
\end{equation}

Let now $r_k = r(x_k)$ and $u^\pm_k = u^\pm(x_k^{})$ be given by
continuous functions as described in \Eref{eq:defrupm}, and
analogously $g^{}_{N,k} = g_N(x_k)$.  (The more general case with a
finite number of steps in $r$ is treated in \aref{app:proofinf}.)  For
an increasing number of mutation classes, fitness values and mutation
rates of neighboring classes will then become more and more similar on
the scale of the total range of values.  More generally, we can use
that $x_k - x_{k\pm i} = \pm\frac{i}{N} \to 0$ for any finite $i$ as
$N \to \infty$.  Defining, for each $x \in [0,1]$, an appropriate
sequence $(k_N) = (k_N(x))$, such that $x_{k_N} \to x$, we therefore
obtain $r(x_{k_N\pm i}) - g_N(x_{k_N\pm i}) \to r(x) - g(x)$, with
$g(x)$ as defined in \Eref{eq:g}.  Evaluating $\bar{r}_{k_N,m,n}$ for
increasing submatrix dimension $n+m \to \infty$ in that limit, we have
$\lim_{n+m\to\infty} \lim_{N\to\infty} \bar{r}_{k_N,m,n} = r(x) -
g(x)$ for each $x$.  Combining this with the upper bound
\eref{eq:upper}, in which $\sup_k (r^{}_k - g^{}_{N,k}) \le
\sup_{x\in[0,1]} (r(x)-g_N(x)) \to \sup_{x\in[0,1]} (r(x)-g(x))$ due
to the uniform convergence $g_N \to g$ (see \aref{app:proofinf}),
gives
\begin{equation} \label{eq:inequal}
\sup_{x\in[0,1]} \!\big( r(x) - g(x) \big) \le
\bar{r}_\infty \le 
\sup_{x\in[0,1]} \!\big( r(x) - g(x) \big) \,,
\end{equation}
which implies the maximum principle \eref{eq:maxprinc}.  As shown at
the end of \aref{app:proofinf}, the ancestral distribution is sharply
peaked around those $x$ at which $r(x)-g(x)$ is maximal.  Thus,
whenever the supremum is unique (which is the generic case),
\Eref{eq:hatx} follows.

\subsection{Mean mutational distance and the variances}
\label{sec:derivrest}

In this subsection, we derive and discuss the results for the mean
mutational distance and the variances, which hold in the linear case
and for $N \to \infty$.

If fitness is linear in an arbitrary trait $y_k = y(x_k)$, the
relation $\bar{r} = r(\bar{y})$ is immediate.  For the variance
formulas, we must additionally assume that fitness is linear in the
mutational distance, $r(x) = \rmax - \alpha x$, or, equivalently, that
all mutational effects are equal.  Thus, the covariances in the
general formula \eref{eq:varmuteff} vanish, and $v^{}_R = \alpha\,
(\bar{u}^+ - \bar{u}^-)$.  Due to linearity, this also determines the
variance in mutational distance as $v^{}_X = (\bar{u}^+ - \bar{u}^-)/
\alpha$.  These relations do not require that $u^\pm(x)$ are linear in
$x$; they reduce to \eref{eq:var} if this is the case.

In the mutation class limit, let us first assume $r$ to be
continuously differentiable on $[0,1]$ with derivative $r'$.
Expressing $v_{R,\infty}$ as the limit variance for increasing system
size $N$, and using \eref{eq:varmuteff} for the variance of each
finite system, $v_{R,N}$, we obtain
\begin{equation}\label{eq:vr1}
v_{R,\infty} 
= \lim_{N\to\infty}\sum_{k=0}^N \Big( \frac{r_k - r_{k+1}}{N^{-1}}\, u^+_k
-\frac{r_{k-1}-r_k}{N^{-1}}\,u^-_k\Big)p_k =
-\overline{r'\,(u^+ - u^-)}_\infty \,.
\end{equation}
Here, we made use of the fact that the mutational effects converge to
the corresponding values of $-r'$, i.e.\ the negative slope of the
fitness function.

Since $r'$ is bounded, \eref{eq:vr1} in particular shows that
$v_{R,\infty}$ is finite, and hence
\begin{equation} \label{eq:vrzero}
V_{r,\infty} = \lim_{N\to\infty}
  \bigg[\sum_{k=0}^N r_k^2 p_k - \Big( \sum_{k=0}^N r_k p_k\Big)^2\bigg] =
  \lim_{N\to\infty} N^{-1} v_{R,N} = 0\,.
\end{equation}
For increasing $N$, the distribution of fitness values \emph{per
  class} therefore concentrates around $\bar{r}$.  In the limit, if
$r$ is invertible at $\bar{r}_\infty$, this fixes the mean mutational
distance at $\bar{x}_\infty = r^{-1}(\bar{r}_\infty)$, cf
\eref{eq:barr}, which approximates the mean distance $\bar{x}_N =
\bar{X}_N/N$ of a finite system to leading order in $N^{-1}$.

With this, we have $v_{R,\infty} = -r'(\bar{x}_\infty)
\bigl(u^+(\bar{x}_\infty) - u^-(\bar{x}_\infty)\bigr)$, cf
\eref{eq:var}, which approximates $v_{R,N} = V_{R,N}/N$.  Note that
the leading order term w.r.t.\ $N^{-1}$ is proportional to
$-r'(\bar{x}_\infty)$, which is the population mean of the mutational
effects in the limit: $\overline{s^\pm}_N \to \overline{s^+}_\infty =
\overline{s^-}_\infty = -r'(\bar{x}_\infty)$.  (The local curvature of
$r$ only gives rise to higher order corrections.)  Obviously, the
leading order depends only on the effective deleterious mutation rate,
$u^+(\bar{x}_\infty) - u^-(\bar{x}_\infty)$, if this does not vanish.
Otherwise, the dominant term is of higher order in $N^{-1}$.

The variance in $x$ can be obtained via the linear approximation $r(x)
\simeq r(\bar{x}_\infty) + r'(\bar{x}_\infty)(x-\bar{x}_\infty)$ as
$v_{X,\infty} = v_{R,\infty} / (r'(\bar{x}_\infty))^2$, cf
\eref{eq:var}.  In contrast to $v^{}_R$, $v^{}_X$ decreases with
increasing mutational effects at $\bar{x}$. Interestingly,
$\sqrt{v^{}_R/v^{}_X}$ can serve as an estimate for the mean
mutational effect (at least in our simple setup) -- a quantity which
is difficult to determine experimentally.  For our numerical examples
in Sections \ref{sec:accuracy} and \ref{sec:thresholds}, this works
reasonably well (not shown).

Comparing the results with those for the linear case above, we see
that, given $\bar{r}$, the infinite mutation class limit can be
interpreted as a local linear approximation.  This does not mean,
however, that nonlinearities (i.e.\ epistasis) are ignored.  They
enter indirectly through the mean fitness as determined by the maximum
principle.

For fitness functions with kinks, the derivation is analogous, as long
as the left- and right-sided limits of $r'$, and thus the mutational
effects in the limit $N \to \infty$, remain bounded.  If $r'$ diverges
at $\bar{x}_\infty$, or if there is even a jump in the fitness
function, $v^{}_R$ diverges according to the above relation.  In the
latter case, $V_{r,\infty}$ is finite and determined by the fraction
of the population below and above the jump, which yields
\eref{eq:varrjump}.

\section{Interlude: Applications and Discussion}
\label{sec:interpret}

\subsection{Mutational loss}
\label{sec:mutflow}

A central role in this article is played by the mutational loss $G$,
which was defined in \Eref{eq:linres4} as the difference between the
ancestor and population mean fitness in equilibrium.  Let us now add
some further interpretation to this quantity.  Recapitulating
relations \eref{eq:linres2}--\eref{eq:linres4}, we obtain for $g:=
G/N$ in the framework of the general mutation--selection model
\eref{eq:paramuse}:
\begin{equation} \label{eq:loss}
 g = \hat{r} - \bar{r} 
 = - \frac{1}{N} \sum_{i,j} z_i M_{ij} p_j 
 =  \frac{1}{N} \sum_{i,j} M_{ij} (z_j-z_i) p_j 
 = - \mu\frac{\partial \bar{r}}{\partial \mu}\,.
\end{equation}

It is instructive to compare $g$ with the mutation load $l = \rmax -
\bar{r}$. Both quantities describe the effect of mutation on the
equilibrium mean fitness.  But whereas the mutation load compares the
biological system with a fictitious system free of mutations, the loss
is essentially a response quantity: In analogy with
\eref{eq:linresmu}, we have $\delta\bar{r} \simeq - (\delta\mu/\mu)
g$.  Since mutation rates are usually not switched on or off in
nature, but may be subject to gradual change, the mutational loss
seems to be the quantity of more direct relevance for questions
connected with the evolution of mutation rates.

From the above, we see that the loss can be understood as the linear
component of the load. In particular, loss and load will coincide if
the latter is linear in $\mu$.  This holds for unidirectional mutation
as long as the wildtype has non-vanishing equilibrium frequency (and,
more generally, below a wildtype threshold, see \sref{sec:wtthr}),
where $\rmax = \hat{r}$. In general, however, also non-linear terms in
$\mu$ will contribute to the load and we find $l > g$.

The genetic load concept has often been criticized, since the
reference genotype (usually the one with maximum fitness) is often
extremely unlikely to be found in the population at all. This argument
is made precise by \citet[][Ch.\ 9.2]{Ewe79} and \citet[][Ch.\ 
6.2]{Gill91} for the substitution and the segregation load in finite
populations. An analogous point may be made against the mutation load,
even in infinite populations: The equilibrium frequency of the fittest
class is often close to zero (or may even vanish for unidirectional
mutation).  Therefore, measurements of $\rmax$ in real populations are
difficult, if not impossible, and the evolutionary significance of the
reference type seems questionable.

This problem is circumvented in the definition of the mutational loss.
As a response quantity, $g$ is well-defined as long as it is
meaningful to think of a system as in equilibrium.  Measurements of
$g$ could make use of marker techniques in (bacterial or viral) clones
in order to determine clone sizes (and thus $\bs{z}$) and ancestor
frequencies, or determine directly the response of $\bar{r}$ to
changes in mutation rates, e.g.\ by comparing strains with different
mutation repair efficiencies.
  
Up to this point, we have entirely concentrated on the mutational loss
as a response quantity. There is, however, a second line of
interpretation, which clarifies the role of $g$ in the equilibrium
dynamics and also sheds some light on the maximum principle.  If an
individual mutates from $j$ to $i$, its offspring expectation changes
by $z_j^{} - z_i^{}$, where the sign determines whether a loss ($+$)
or gain ($-$) is implied.  Since the mutational flow from $j$ to $i$
in equilibrium is $M_{ij}^{}\,p_j^{}$, the entire system loses
offspring at rate $\sum_{i,j} (z_j^{} - z_i^{})\, M_{ij}^{}\, p_j^{}$,
which is the same as $G$ (compare with \Eref{eq:linres3} or
\eref{eq:loss}).

The mutational loss does not include any information about the
destination of the `lost' offspring.  This, however, may easily be
found by recalling that, asymptotically, every ancestor of type $i$
leaves $z_i p_j$ descendants of type $j$ in the equilibrium
population.  Further, $p_i (z_i - 1) = a_i - p_i$ is the excess
offspring produced by an $i$-individual.  We thus come to a picture of
a constant flow of mutants from the ancestor to the equilibrium
population.

Let us now turn to the mutational loss function $g(x)$.  Recall that,
in the derivation of the maximum principle in the mutation class
limit, we obtained $r(x) - g(x)$ as the leading eigenvalue of a local
open subsystem around $x$; if $\bar r_{\infty}$ is the death rate due
to population regulation in the entire system, $r(x) - \bar r_{\infty}
- g(x)$ is the net growth rate of the subsystem at $x$.  Hence, $g(x)$
must describe the rate of mutational loss due to the flow out of the
local system.  This can be made more precise within the framework of
large-deviation theory, which will be presented in a future
publication.  If $r(x)-g(x)$ has a unique maximum, in which case the
ancestor distribution has a single peak, the maximum principle
\eref{eq:maxprinc} along with $\bar{r} \to \bar{r}_\infty$ and
$\hat{r} \to r(\hat{x}_\infty)$ as $N\to\infty$ implies that the
mutational loss $g = G/N$ converges to $g(\hat{x}_\infty)$.  Thus,
$g(\hat{x})$ can be taken as an approximation to the actual mutational
loss $g$ in this case.

Let us finally add a remark concerning the influence of epistasis on
the mutational loss (in the sense of a response quantity) in the
single-step model.  Following the suggestion of \citet{POW00}, we
speak of \emph{negative} (\emph{positive}) epistasis w.r.t.\ some
class $k$ if $R_{k+1}-R_k < (>)\; R_k-R_{k-1}$.  This entails
\emph{synergistic} (\emph{antagonistic}) interaction of deleterious
mutations.  This way, negative and positive epistasis are connected to
concavity and convexity of the fitness function, and thus to its
second derivative (if well-defined) being negative, respectively
positive.

Let us now keep the mutation rates fixed and compare fitness functions
with different degrees of epistasis.  Let $g$ be a decreasing loss
function, and $r$ and $\rs$ two decreasing fitness functions which are
either convex or concave, and only differ in an open subinterval of
$[0,1]$ that includes $\hat{x}$ (the ancestral mean trait under $r$).
Assume that $\rs - r$ is a concave function in our subinterval.  Then
$\rs$ describes more negative, or less positive, epistasis than $r$.
Under the above assumption, $\rs-r$ has a unique maximum, whose
position we denote by $x_0$.  As is most easily seen from the
graphical representation of the maximum principle (\fref{fig:comics}),
one then finds $\hat{x}_{\text{s}} > \hat{x} $ whenever $x_0 >
\hat{x}$ (and vice versa), where $\hat{x}_{\text{s}}$ is the ancestral
mean trait under the modified fitness function.  Since $g(x)$ is
decreasing, it follows that $g(\hat{x}_{\text{s}}) < g(\hat{x})$ if
$x_0 > \hat{x}$.  If $\hat{x}$ is small (as may be considered typical
of realistic examples), increased negative epistasis will reduce the
loss.  The opposite may be said of decreased negative or increased
positive epistasis, in line with the fact that the loss is maximal for
the sharply-peaked landscape, which displays extreme positive
epistasis.

\subsection{Haldane's principle and evolution of mutational effects}
\label{sec:haldane}

As we have seen in the discussion of the unidirectional case, the
maximum principle reduces to a well-known form of the Haldane--Muller
principle in that limit. Using the concept of the ancestor
distribution, we will now re-analyze this principle in the broader
context of models with back mutations. We will also discuss
consequences for the evolution of mutational effects and mutational
robustness.

For models without back mutations to the fittest genotype with
non-zero equilibrium frequency, Haldane's principle says that the
difference in fitness between this type, $\hat{i}$, and the population
mean is equal to the total mutation rate for $\hat{i}$.  For our
general model this reads $L = \Rmax - R_{\hat{i}} + \sum_{j\neq
  \hat{i}} M_{j \hat{i}}$, where $M_{j \hat{i}}$ is the mutation rate
from the fittest type (or class) $\hat{i}$ to some other type (or
class) $j\neq \hat{i}$.  Note in particular that the load is
independent of the mutant fitness values if the wildtype itself has
non-zero frequency in equilibrium, i.e.\ $\hat{i}=0$.  We will assume
$R_0 = \Rmax$ for simplicity in this section.

If back mutations to the fittest class are present, but mutation
rates, denoted by $u$, are small compared to the fitness advantage, $u
\ll s$, the relation for the load is modified by a correction term of
order $u^2/s$ \citep{BuHo94}. In the following, we will reproduce this
result in our setting by deriving an explicit expression of the
correction term for the single-step model. We will also show that this
leading order contribution of the back mutations is exactly contained
in the estimate of $\bar{r}$ as derived from the maximum principle.

Let us assume, for notational simplicity, that the wildtype is also
the fittest type present in the equilibrium population, and remains so
if back mutations are switched off.  Suppose that the back mutation
rates $u_k^-$ are small compared to the fitness effects, but not
necessarily the deleterious mutation rates $u_k^+$.  We then obtain,
to linear order in $u_1^-$,
\begin{equation}
l \,  \simeq \,
u_0^+ - \frac{\partial \bar{r}}{\partial u_1^-}\bigg|_{u_1^-=0}
\cdot u_1^- = u_0^+ - \frac{p_1}{p_0}\bigg|_{u_1^-=0} \cdot u_1^- \,,
\end{equation}
where we have used \Eref{eq:linres2} for the derivative of $\bar{r}$
with respect to the mutation rates, and $z_0=1/p_0$, $z_1 = 0$ for
$u_1^- = 0$.  Calculating $p_1/p_0$ from the equilibrium condition for
the mutation--selection equation, we find
\begin{equation} \label{eq:loadu}
l = u^+_0 - \frac{u_0^+ u_1^-}{s_0^+ - u_0^+ + u_1^+} + O([u^-]^2)\,.
\end{equation}
This is in accordance with the result of \citet{BuHo94} if also
$u_0^+$, $u_1^+ \ll s_0^+$.

On the other hand, starting with a linear interpolation of the fitness
and mutation functions of the form $r(x) = r_0 + Nx(r_1-r_0)$, $u^+(x)
= u^+_0 + Nx(u^+_1-u^+_0)$, and $u^-(x) = Nx u^-_1$ for $0 \le x \le
1/N$, we find the load by using $\bar{r}$ from the maximum principle.
To linear order in $u^-$, a lengthy but elementary calculation yields
that $r(x) - g(x)$ is maximized at $Nx = u_0^+
u_1^-/(s_0^+-u_0^++u_1^+)^2$, and we again obtain \Eref{eq:loadu} for
the load. We can therefore conclude that the maximum principle, when
applied to finite $N$, still gives results that are correct to linear
order in the back mutation rates (cf \sref{sec:accuracy}).

In the preceding paragraphs, back mutations have merely played the
role of a small perturbation of the system with unidirectional
mutation.  Our main interest in this article, however, lies in the
case of sufficiently large mutation rates -- or sufficiently small
fitness effects of mutations (as in a nearly neutral landscape) such
that the equilibrium distribution is no longer dominated by one or a
few wildtype states, but is dispersed over many classes.  This is
exactly the situation in which one would assume back mutations to
become important, with effects beyond a second order correction term.
At the same time, this is the domain of validity of the mutation class
limit, in which the maximum principle is also exact.  We then obtain
\begin{equation}\label{eq:loadg}
l = \rmax - r(\hat{x}) + g(\hat{x}) \quad 
  \bigl[\le g(0) = u^+(0)\bigr] 
\end{equation}
as an estimate for the mutation load.  Clearly, the load is no longer
independent of the fitness function as soon as the ancestral mean
fitness $\hat{r} = r(\hat{x})$ differs from the wildtype fitness.
Note, however, that the only quantity that matters is the deviation of
the ancestor mean fitness from the wildtype fitness.

It is instructive to compare the load for different fitness functions.
Let $\rs$ and $r$ be fitness functions with $r(0)=\rs(0)=\rmax$, and
$\rs(x) \geq r(x)$ for all $x \in [0,1]$.  By the maximum principle,
the load with $\rs$ cannot be larger than with $r$. If $\rs(x) > r(x)$
at $x=\hat x$, the ancestral genotype under $r$, the load with $\rs$
is strictly smaller than with $r$.  In this sense, higher mutant
fitness tends to decrease the mutation load (and vice versa).

Let us now extend these thoughts to the evolution of mutational
effects.  To this end, we consider a general mutation--selection model
(i.e.\ not restricted to permutation invariant fitness or
single-step).  Assume there is an additional modifier locus, which is
tightly linked to the other loci and changes the fitness of one or
several of the original types or classes.  (In the biallelic model,
this may, for example, happen through epistatic interactions outside
our permutation invariant fitness scheme.)

Let now a modifier be introduced into the equilibrium population at
low frequency at time $t = 0$ (by mutation or migration), and consider
its fate for $t \to \infty$.  If there is no further mutation at the
modifier locus, the modifier will asymptotically fix (or get lost) in
terms of relative frequencies, $\bs{p}(t) = \bs{y}(t) / (\sum_i
y_i(t))$, if the modified system has a larger (smaller) leading
eigenvalue than the original one, in which case we write
$\delta\bar{r} > 0$ ($\delta\bar{r} < 0$).  If $\delta\bar{r} = 0$,
the modifier will equilibrate at an intermediate frequency, the exact
value of which depends on the initial conditions.

The above argument is analogous to the clonal competition mechanism as
described for mutation rate modifiers in asexual populations
\citep[for review, see][]{SGJS00}.  It requires slight modification if
mutation at the original loci is unidirectional.  Here, the fate of
the modifier also depends on the genetic background it is introduced
into at $t=0$.  If the fitness modifications are so small that the
fittest type present remains the same in equilibria with and without
modifier, the modifier will always get lost if it does not already
occur in individuals of that type at $t=0$.  This follows since all
other types asymptotically expect no offspring.

Note that the competition mechanism just described works \emph{within}
the population, the separation of genotypes with and without modifier
being due to tight linkage of the modifier to the primary loci.  In
particular, no group selection is implied.

What consequences, now, does this have for the possibility of
mutational effects to evolve?  Again, the answer involves the ancestor
distribution.  We have seen in \eref{eq:linresr} that changing the
fitness values $r_i$ to $r_i + \delta_i$ will change the equilibrium
mean fitness by
\begin{equation} \label{eq:deltar}
\delta\bar{r} \simeq \sum_i \delta_i a_i \,,
\end{equation} 
to first order in the $\delta_i$.  From this, we now obtain the
following intuitive picture: In order for a modification to prevail in
an equilibrium population, it has to invade the ancestors; otherwise,
it will be `washed away'.

Let us discuss this in some more detail.  According to our above
discussion, the fate of the modifier is entirely determined by
$\delta\bar{r}$ if we have back mutations.  Now, the right hand side
of \Eref{eq:deltar} may be interpreted as the selection coefficient of
the modifier with respect to the ancestor distribution -- assuming
that the modifier is statistically independent of the other loci. In
order to understand why this quantity governs the leading order of
$\delta\bar{r}$, consider infinitesimal small fitness changes
$\delta_i$, in which case \Eref{eq:deltar} becomes exact. Here,
mutation will indeed drive the modifier distribution towards
statistical independence in an initial period of time. In order to
eventually spread to fixation, the modifier now has to compete
successfully against those types whose descendents make up the
equilibrium population at an even later time.  These, however, follow
the ancestor distribution.  In this sense, $\delta\bar{r}$ may be
understood as measuring the modifier's growth within the ancestor
population.  In the same vein, the vector of the ancestor frequencies
can be seen as the gradient of the mean fitness, pointing into the
direction of the indirect (i.e.\ second order) selection pressure
exerted on the fitness values $r_i$.

This long-term picture is in sharp contrast to the initial growth of
the modifier in the population, which is determined by its selection
coefficient with respect to the \emph{equilibrium population} and of
course depends on the distribution of the modifier over the types at
$t=0$.  If $\delta\bar{r}$ is positive (negative), the modifier will
asymptotically fix (vanish) even if its initial selection coefficient
is negative (positive).  Note, however, that this process may be very
slow if $\delta \bar r$ is small.

If there is no back mutation to the fittest class (or type) $\hat{i}$
present, this is the absorbing state of the backward process, in which
all lineages end, and the ancestor distribution is entirely
concentrated there ($a_{\hat{i}} = 1$, $a_j = 0$, $j\neq \hat{i}$).
So \Eref{eq:deltar} leads back to the prediction of Haldane's
principle that the mean fitness is independent of the mutant fitness
values in this case.  In order to `invade the ancestors', a modifier
must be introduced into the fittest type in the first place, and
increase its fitness.

Assume now that the wildtype fitness is kept fixed but mutational
effects at the wildtype are modified by variations of the mutant
fitness values.  Such modifiers are \emph{canalizing} (or modifiers
for mutational robustness) if they increase the mutant fitnesses, and
\emph{decanalizing} (modifiers for antirobustness) if they decrease
the mutant fitnesses \citep[cf]{WBB97}.  It is now clear from
\Eref{eq:deltar} that only an \emph{increase} of mutant fitness values
may lead to an evolutionary advantage.  Independently of the fitness
landscape or of mutation patterns, we thus never find a potential for
the evolution of antirobustness in mutation--selection models;
however, mutational robustness may, indeed, evolve.  Here, modifiers
increasing the fitness of mutant classes with large ancestor
frequencies will be under particularly large (positive) selection
pressure. If modifiers have deleterious side-effects, these may even
be the only ones that persist and go to fixation.\footnote{Note,
  however, that no predictions are made here concerning invadability
  of modifier mutations, or fixation probabilities, if random drift
  becomes a weighty factor.}

Let us, for further analysis, consider two limiting cases of the
mutation scheme now. If mutation is unidirectional, neither
modifications for robustness nor for antirobustness will change the
mean fitness (at least under the usual assumption that the wildtype is
present in the original equilibrium). We may conclude that there is no
selection pressure on the mutant fitness values at all in this simple
setting, and hence no potential for these to evolve
either.\footnote{For very large, but finite populations (where
  Muller's ratchet does not operate but there is drift among classes
  of equal fitness) the fixation probability of clones with and
  without the modifier is ultimately determined only by the initial
  sizes of the respective wildtype classes \citep{GaBu00}.  Any
  modifier which enters the wildtype class at low frequency will
  therefore get lost from this class and, consequently, from the
  population with high probability.}  On the other hand, if the
mutation matrix is symmetric, $M_{ij} = M_{ji}$, the ancestor
frequencies are proportional to the square of the population
frequencies, $a_i \sim p_i^2$. Thus, the landscape is evolvable
exactly in those regions in which the equilibrium frequency of the
population distribution is high.

Note that we may come to different results here depending on whether
genotype classes or single genotypes are the relevant entities.  If
mutation between genotypes is symmetric (as in our biallelic model
with $\kappa = 0$), modifiers of single genotypes will be particularly
important if the corresponding equilibrium frequency is high.  For
modifiers of whole genotype classes, however, the asymmetric mutation
scheme with respect to the classes is relevant, and the maximum of the
ancestor distribution will in general deviate from the maximum of the
population distribution.

\begin{figure}
\centerline{%
\epsfig{file=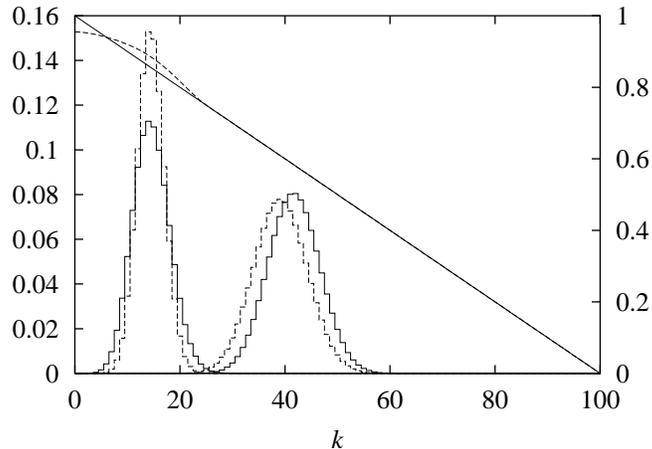,bbllx=67,bblly=52,bburx=308,bbury=236,clip=}}
\caption{Comparison of population frequencies $p_k$ (near $k=40$) and
  ancestor frequencies $a_k$ (near $k=15$) for the biallelic model
  with $\mu=0.365\gamma$ (where $\gamma$ is the loss in reproduction
  rate due to a single mutation as in \fref{fig:paz}),
  $\kappa=\frac12$, and $N=100$.  The right axis refers to the fitness
  functions used: additive fitness $r(x)/\gamma = 1-x$ (solid lines),
  and a modified version (dashed lines) that is favored with respect
  to the additive one.  The modified fitness is increased in regions
  of high ancestor frequencies.  In this particular example, it is
  slightly decreased at the wildtype and unchanged in other regions of
  vanishing ancestor frequencies, but note that the success of a
  modification is independent of the fitness values there.}
\label{fig:pafitevo}
\end{figure}
In order to see what happens between these limiting cases, let us
restrict our discussion again to the single-step mutation model.
Here, the ancestor distribution becomes sharply concentrated around
$\hat{x}$ with an increasing number of mutation classes (cf
\aref{app:proofinf}). Similar to the case of unidirectional mutation,
only a very minor part of the fitness function will thus experience
appreciable selection pressure. Note that this part need neither
extend to the types which contain the bulk of the equilibrium
distribution (concentrated around $\bar{r} < \hat{r}$), nor the
largest fitness values at $\rmax > \hat{r}$. If robustness modifiers
have deleterious side effects, only those which lead to buffering in
the ancestor region will prevail at all. Therefore, if robustness
evolves by the mechanism described, the strongly \emph{differential}
selection pressure might lead to the emergence of synergistic
epistasis at the same time. This is illustrated in
\fref{fig:pafitevo}, where modification of the fitness function leads
to a flattening near its `summit' at $x < \hat{x}$ relative to the
`slope' at $x > \hat{x}$. The example also shows that an increase in
fitness around $\hat{r}$ may compensate for a deleterious side-effect
of the modifier mutation which decreases the wildtype fitness.

\subsection{Accuracy of the approximation}
\label{sec:accuracy}

In this subsection we wish to illustrate the accuracy of the
analytical expressions for means and variances given in
\sref{sec:results}.  To pay respect to the invariance of the
equilibrium distributions under scaling of both reproduction and
mutation rates with the same factor, we introduce $\gamma$ as an
overall constant for the reproduction rates.  It should be chosen to
represent roughly the average effect of a single mutation on the
reproduction rate in a mutant genotype (with the maximum number of
mutations considered) as compared to the wildtype.  This does not
exclude the possibility that effects of single mutations may be quite
large.  In the figures, both reproduction and mutation rates are given
in units of this constant, i.e.\ as $r/\gamma$, respectively
$\mu/\gamma$.

\newcommand\rawvarthreegnuplot[1]{%
\epsfig{file=#1,bbllx=54.6,bblly=50,bburx=213.2,bbury=193,clip=}}
\begin{figure}
\centerline{\rawthreegnuplot{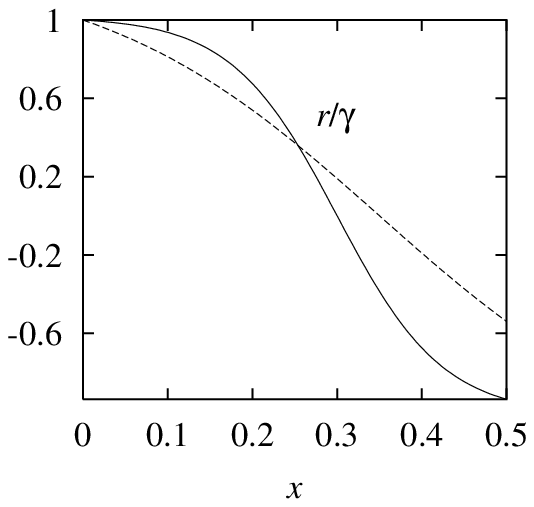}\hspace{0.5\columnsep}%
\rawthreegnuplot{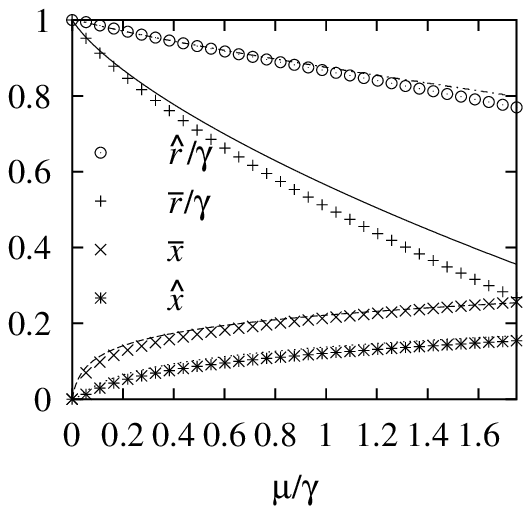}\hspace{0.5\columnsep}%
\rawvarthreegnuplot{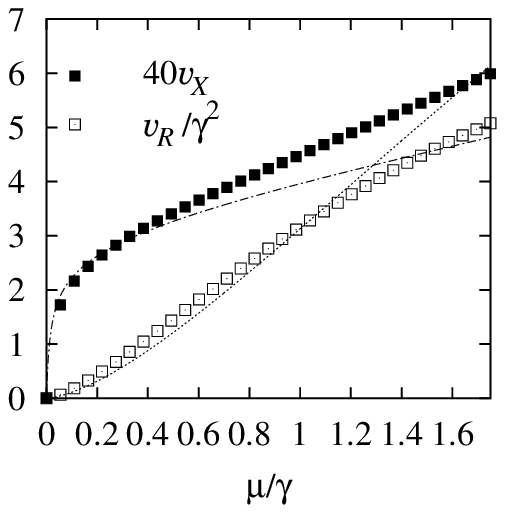}}
\threegnuplots{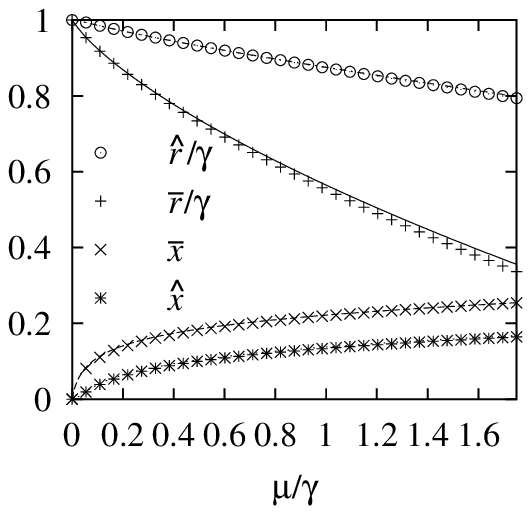}{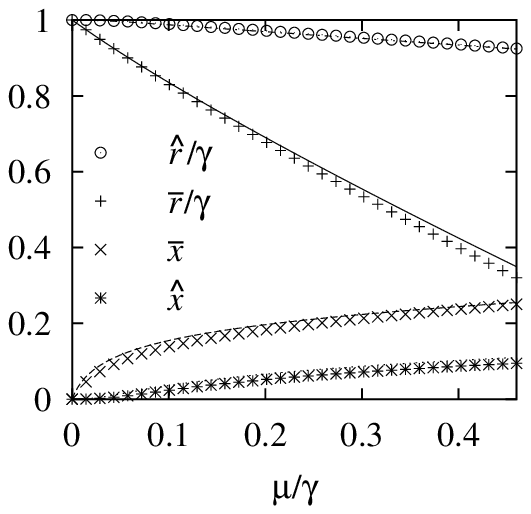}{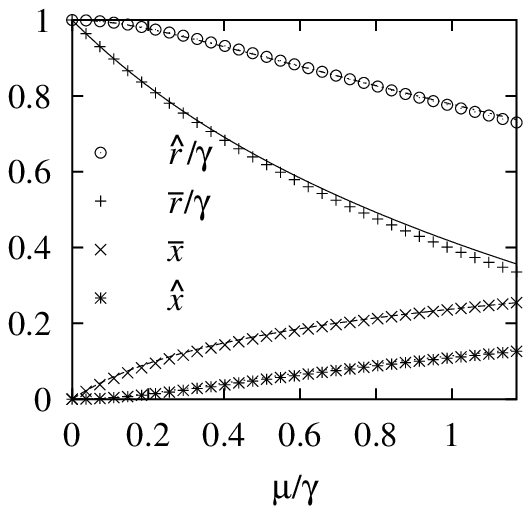}
\centerline{\rawvarthreegnuplot{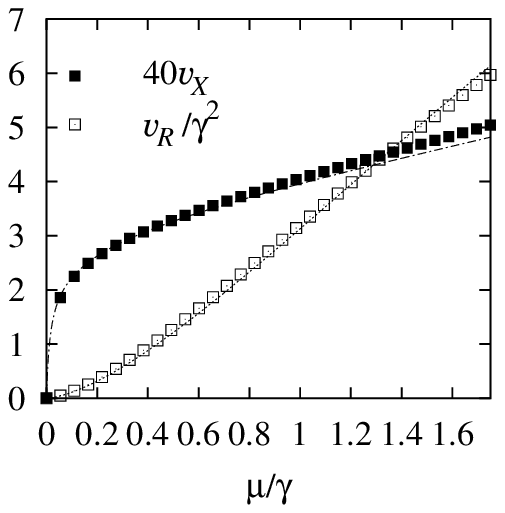}\hspace{0.5\columnsep}%
\rawthreegnuplot{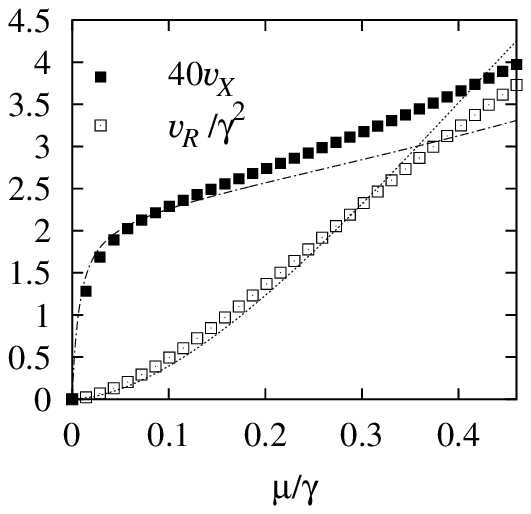}\hspace{0.5\columnsep}%
\rawthreegnuplot{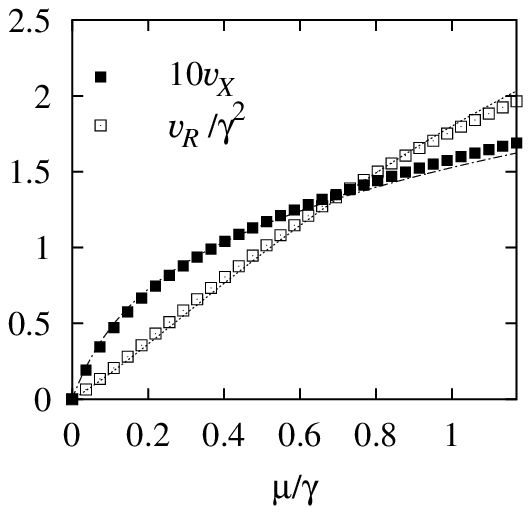}}
\caption{The top row refers to a biallelic model that deviates
  from all three exact limiting cases described in \sref{sec:limit} in
  having a strongly non-additive fitness function $r/\gamma$ (left,
  solid line), symmetric site mutation ($\kappa=0$), and small
  sequence length ($N=20$).  The mean values of the observables
  (middle) and corresponding variances (right) are shown as a function
  of the mutation rate $\mu/\gamma$, both for the model itself
  (symbols) and according to the expressions given in
  \sref{sec:results} (lines, sometimes hidden by symbols).  Even here,
  we find reasonable agreement.  Deviations, however, are visible for
  larger mutation rates.  As can be seen from the last two rows, going
  towards any of the three exact limits, i.e.\ increasing the number
  of mutation classes (left, $N=100$), going to more asymmetric
  mutation (middle, $\kappa=0.8$), or using a different fitness
  function with less curvature (right, $r/\gamma$: top left, dashed
  line), we find that these deviations vanish quickly.  In the case of
  increasingly asymmetric mutation, however, this is not true for the
  variances, since the approximation becomes only exact here in
  either of the other two limits (cf \sref{sec:derivrest}).}
\label{fig:tanh}
\end{figure}
\Fref{fig:tanh} displays an example of a biallelic model that deviates
from all three exact limiting cases described in \sref{sec:limit},
and, for comparison, three modifications that are closer to one of the
exact limits each.  All numerical values, also in the rest of this
article and in Figs.\ \ref{fig:paz} and \ref{fig:pafitevo}, are
virtually exact and, if not noted otherwise, obtained by the power
method \citep[][Ch.\ 9, also known as von Mises iteration]{Wilk65}
with the evolution matrix $\bd{H}$.  For continuous fitness functions,
the approximate expressions for the observable means agree with the
exact ones up to corrections of order $N^{-1}$ (as indicated by
numerical comparison, not shown) or of order $(u^-)^2$ (cf
\sref{sec:haldane}).  For fitness functions with jumps, the error
seems to be at most of order $N^{-1/2}$ (cf \fref{fig:truncsel}); for
a jump at $x=0$ such as in the sharply peaked landscape, however, the
corrections to $\bar{r}$ appear to be still of order $N^{-1}$ for the
biallelic model (cf \fref{fig:4types}).

Further examples, exhibiting more conspicuous features, are shown in
\sref{sec:thresholds}.  For most of them, one will also find good
agreement of numerical and analytical values for the means for
sequences of length $N=100$; for the variances, however, one sometimes
needs longer ones, like $N=1000$.  In the biallelic model, we
generally find stronger deviations for higher mutation rates, as in
this regime back mutations become more and more important, whereas for
small mutation rates, deviations are of linear order in $\mu$.

\section{More applications: threshold phenomena}
\label{sec:thresholds}

In this section, we will take a closer look at how the equilibrium
behavior of a mutation--selection system changes if the mutation rates
are allowed to vary relative to the corresponding mutational effects.
In order to keep the overall shapes of the fitness and mutation
functions constant, we vary all mutation rates by a common scalar
factor $\mu \ge 0$.  Concentrating on the single-step mutation model
in this section, we choose $\mu$ as the mean mutation rate over all
classes,
\begin{equation}
\label{eq:defmu}
\mu =(2N)^{-1} \sum_{k=0}^N (u_k^+ + u_k^-)
\end{equation}
(recall that $u_0^- = u_N^+ = 0$).  This is consistent with the
definition of $\mu$ as the mean point mutation rate for the biallelic
model, cf \Eref{eq:mutbiall} and \fref{fig:mutscheme}.  By slight
abuse of notation, we define the shape of the mutational loss function
as $g(1,x) = \mu^{-1} g(x)$ (which does not depend on $\mu$), and
introduce $\mu$ as a variable parameter via $g(\mu,x) = \mu \,
g(1,x)$.

\subsection{Mutation thresholds}
\label{sec:mutthres}

Consider a population in mutation--selection balance.  Usually, if
mutation rates change slightly, the population will move on to a new
equilibrium with the observables, like means and variances of traits
and fitness, close to the old ones.  At certain \emph{critical
  mutation rates}, however, threshold phenomena may occur, associated
with much larger effects on traits or fitness.  The prototype of this
kind of behavior is the so-called \emph{error threshold}, first
observed in a model of prebiotic evolution many years ago
\citep{Eig71} and discussed in numerous variants ever since \citep[for
review, see][]{EMcCS89,BaGa00}.

In the following, we will discuss and classify `error threshold like'
behavior in our model class.  We shall, however, avoid the term error
threshold as the collective name for all threshold effects that may be
observed, but rather, and more generally, speak of \emph{mutation
  thresholds}.  This is because the definition of the error
threshold is closely linked to the model in which it had been observed
originally, namely the quasispecies model with the sharply peaked
fitness landscape.  While many effects of the original error threshold
will turn out to generalize easily to the much larger class of models
considered here, the criterion of the \emph{loss of the wildtype},
which has frequently been taken as the defining property of the error
threshold, seems to be applicable only in special cases.

We want to be as general as possible as far as the fitness model and
mutation schemes are concerned, but specific about the responsible
evolutionary forces. Error thresholds have also been described as
driven by the joint action of mutation and segregation \citep{Higgs94}
or recombination \citep{BBN96}.  We will not consider these phenomena.

Let us now define the notion mutation threshold.  Ideally, a
characterization should give a precise mathematical definition in the
modeling framework which, at the same time, captures biologically
significant behavior. As may be seen from the varying and sometimes
incompatible definitions that have previously been suggested for the
error threshold \citep[see, e.g., the discussion in][]{BaGa00}, this
can be a complex problem. Let us therefore start with a verbal
description:
\begin{quote}
  A \emph{mutation threshold} for a particular trait or fitness is the
  pronounced change of the equilibrium distribution of the trait or
  fitness values within a narrow range of mutation rates.  Here, the
  threshold phenomenon is purely due to the interplay of mutation and
  selection.
\end{quote}
Note that we only consider effects on distributions, not on absolute
numbers. This demarcates mutation thresholds from mutational meltdown
effects \citep[cf][]{GLB93}.

In order to come to a stringent mathematical definition, a two-fold
limit must be considered for any general mutation--selection model
\eref{eq:paramuse}. These are the infinite population limit, which we
assumed right from the beginning, and the limit of an infinite number
of mutation classes.

Application of the \emph{infinite population limit} is a direct
consequence of the last condition in the verbal definition above.  As
mutation thresholds result from mutation and selection alone, they
must persist in the absence of genetic drift.  Hence, unlike drift
effects (like Muller's ratchet), these phenomena can not be avoided by
increasing population size.  For the purposes of analysis and
classification, therefore, deterministic models provide the right
framework.  Of course, aspects of thresholds should also persist in
(large) finite populations, if the phenomena are biologically
relevant.  For some models this has been confirmed in numerical
studies \citep{NoS89,BoSt93}: While certain properties of the
threshold (such as the critical mutation rate) may be altered by
finite population size, the threshold effect as such is not eliminated
by drift.

The \emph{infinite mutation class limit}, on the other hand, is needed
to give the vague notion of a `pronounced change' a more precise
meaning in mathematical terms. Our intention is to specify this notion
as a \emph{discontinuous} change of a biological observable (or, at
least, of one of its derivatives) as a function of $\mu$. In any
finite system with back mutations, however, this clearly conflicts
with the fact that the population frequencies are analytic functions
of the mutation rates.\footnote{This follows from the
  Perron--Frobenius theorem and the fact that the PF eigenvalue and
  eigenvector depend analytically on the matrix entries.  Since the PF
  eigenvalue is real and unique under the above conditions, it never
  crosses with the second largest eigenvalue as a function of any
  model parameter, such as mutation rates.}  The same problem also
arises for the definition of phase transitions in physics.  Phase
transitions, therefore, are defined as non-analyticity points of the
free energy in the \emph{thermodynamic limit} (i.e.\ for infinitely
large systems).  Since the infinite mutation class limit is just the
counterpart of the thermodynamic limit in our models (cf
\aref{app:physics}), we take this concept of theoretical physics as
our guideline and characterize different types of mutation thresholds
by discontinuities or kinks in the equilibrium mean and/or variance of
some trait or of fitness as a function of $\mu$ in the limit $N \to
\infty$.  (Therefore, we will omit the subscript $\infty$ throughout
this section.)

Let us add a few comments concerning this strategy:
\begin{enumerate}
\item Firstly, and most importantly, the proposed procedure is in
  accordance with the original definition of the error threshold: In
  the quasispecies model, a kink in the wildtype frequency (and thus
  the mean fitness) as a function of the total mutation rate was first
  established by an approximate formula for finite sequence length by
  \citet{Eig71}, which was later found to be exact in the limit $N \to
  \infty$ \citep{Swet82}. The finite system is thus effectively
  approximated by an infinite one.  In order to capture the behavior
  of the finite system in the limit, the total mutation rate and the
  selective advantage of the wildtype must scale with the number of
  classes $N$ (thus leaving the \emph{mean} mutational effect per
  class constant; cf \citeauthor{FrPe97}, \citeyear{FrPe97}). The
  equivalence of this phenomenon with a magnetic phase transition has
  first been established by \citet{Leut87a}, and was later used by
  \citet{Tara92} and many others.
\item Whereas we have introduced the mutation class limit mainly as an
  approximation for real systems with a finite number of classes, its
  use in the present context rather has a conceptual reason.
  Analogously to phase transitions in physics, the threshold should be
  considered as a property of the limit that manifests itself (as a
  `pronounced change') in finite systems as well (cf the numerical
  examples in Figs.\ \ref{fig:4types}, \ref{fig:fitnthr}, and
  \ref{fig:wtthr}--\ref{fig:unidir}).
\item Discontinuities in the biological observables can also arise in
  finite systems if the evolution matrix $\bd{H}$ is reducible (as for
  unidirectional mutation).  Then mutation thresholds can be directly
  defined for finite $N$.  This has previously been done by
  \citet{Wie97} and will be discussed in \sref{sec:unidir} below.
\end{enumerate}

\subsection{Description of threshold types}

\begin{figure}
\threegnuplots{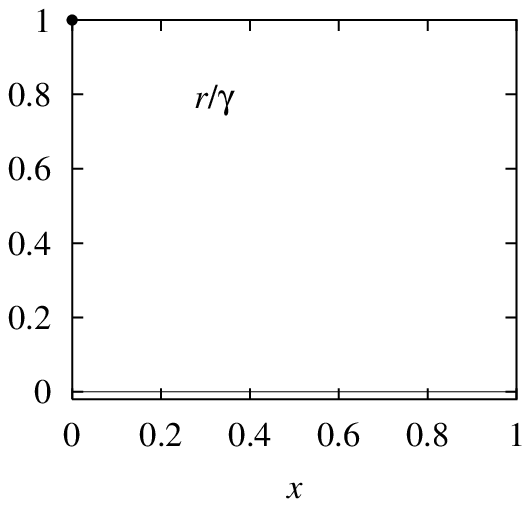}{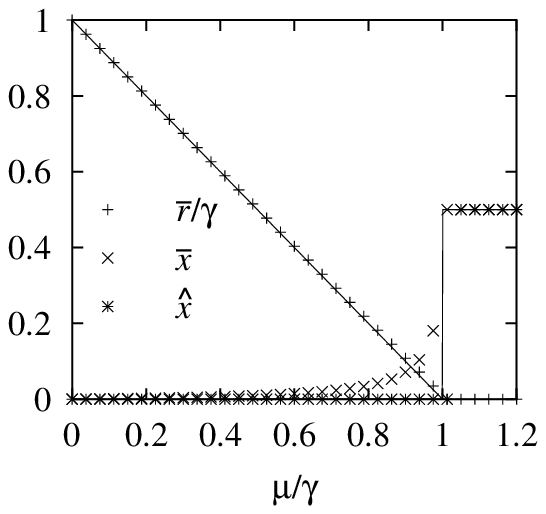}{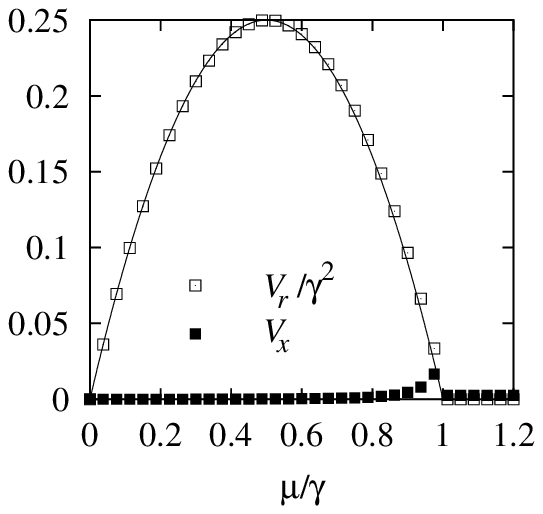}
\caption{The error threshold of the \textbf{sharply peaked landscape}
  (left) with $r(0) = \gamma$ (bullet) and $r(x) = 0$ for $x>0$
  (line), for the biallelic model with symmetric mutation
  ($\kappa=0$). The observable means are shown in the middle, the
  variances on the right.  Symbols correspond to $N=100$, lines to the
  expressions in \sref{sec:results}.  The ancestral fitness
  $\hat{r}(\mu)$ (not shown) jumps from $\gamma$ to 0 at $\mu=\gamma$.
  Note that $V_r$ follows the scaling described by \eref{eq:vr1} and
  is given by \eref{eq:varrjump} for $N\to\infty$.}
\label{fig:4types}
\end{figure}
Following the lines of the above reasoning, we now come to a
description of different types of mutation thresholds.  In our list we
will not include \emph{any} discontinuous change that might occur, but
rather concentrate on pronounced changes of potential evolutionary
significance.  To this end, we will take the original error threshold
of the sharply peaked landscape as our reference and analyze four of
its characteristic properties, namely (cf \fref{fig:4types}):
\begin{itemize}
\item A kink in the population mean fitness,
\item the loss of the wildtype from the population,
\item complete mutational degradation, and
\item a jump in the population mean of the mutational distance (or some
  additive trait).
\end{itemize}
For these threshold effects, we will check whether and how they extend
to the permutation-invariant class of mutation--selection models.  We
will discuss their origin, analyze how they are related, and formulate
criteria for the fitness function to exhibit each threshold effect, or
\emph{type of threshold}, separately.

\subsubsection{Fitness thresholds}
\label{sec:fitnthr}

As we will see below, the kink in the population mean fitness is, in
many respects, the most fundamental aspect to classify
mutation thresholds.  We therefore discuss it first.

\paragraph{Phenomenon.}
The most pronounced change that may happen to the \emph{fitness}
distribution at some critical mutation rate $\muc$ is characterized by
a kink in the mean fitness $\bar{r}$ as a function of $\mu$ (i.e.\ a
jump in its derivative). We will refer to this phenomenon as a
mutation threshold in fitness, or \emph{fitness threshold} for short.
Using \Eref{eq:loss} and the maximum principle, we see that an
alternative definition can be given in terms of the ancestor
distribution.  Here, a fitness threshold is defined by a jump in the
mutational loss (as a function of $\mu$), $g = g(\hat{x}) =
-\mu\,\partial\bar{r}/\partial\mu$, corresponding to jumps in
$\hat{x}$ and the ancestor mean fitness $\hat{r} = r(\hat{x})$.  As a
consequence of the kink in $\bar{r}$, the mean mutational distance
$\bar{x}$, and the variances $v_R$ and $v_X$, will typically show a
kink as well.

\paragraph{Interpretation and graphical representation.}

\begin{figure}
\centerline{\epsfig{file=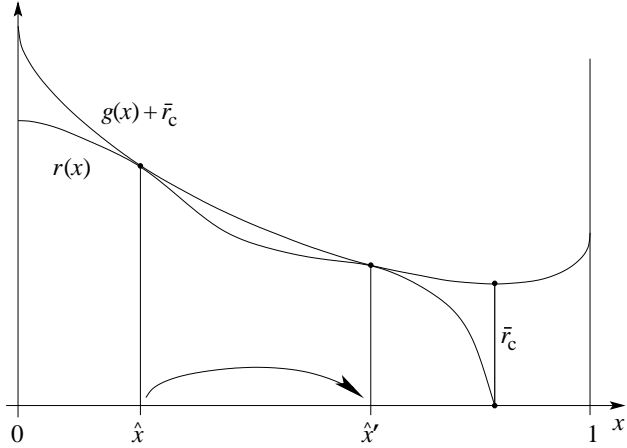}}
\caption{Graphical construction of the fitness threshold,
  following \fref{fig:comics}.  At the critical mutation rate $\muc$,
  the maximum of $r(x)-g(x)$ is not unique.  Thus, with $\mu$ being
  increased across $\muc$, the mean of the ancestor distribution jumps
  from a position of relatively high fitness and high mutational loss,
  $\hat{x}$, to lower fitness genotypes with less mutational loss at
  $\hat{x}'$.  The figure also shows how the population mean fitness
  is constructed at the threshold.}
\label{fig:comic5}
\end{figure}
The origin of a fitness threshold is easily understood from the
maximum principle. For a generic choice of $\mu$, the function $r(x) -
g(\mu,x)$ is maximized for a unique $x = \hat{x}$. For some fitness
functions, however, there are particular values of $\mu$ that lead to
multiple solutions. It is precisely this phenomenon of two distinct
ancestor distributions becoming degenerate with respect to the maximum
principle which marks the threshold. This may be illustrated
graphically as shown in \fref{fig:comic5}.

\begin{figure}
\threegnuplots{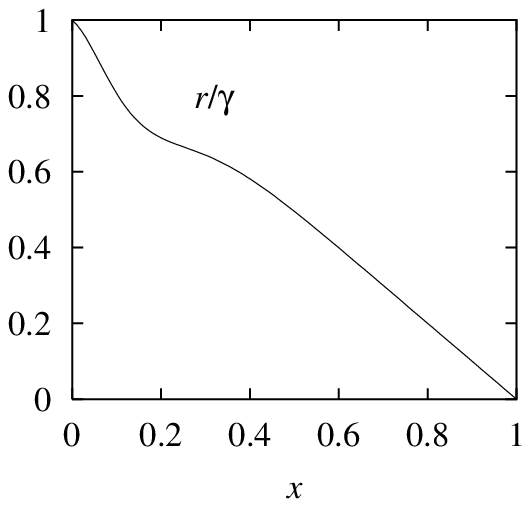}{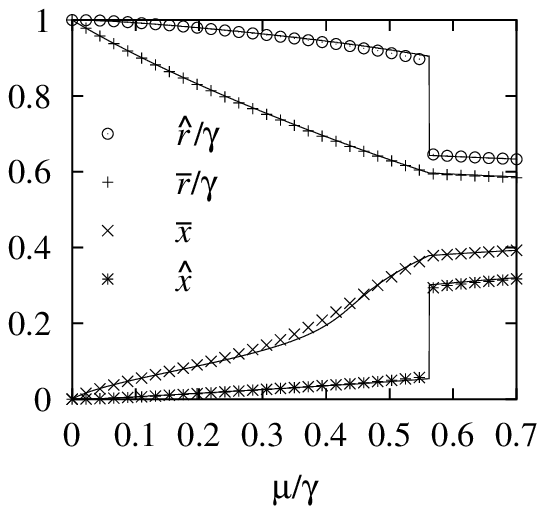}{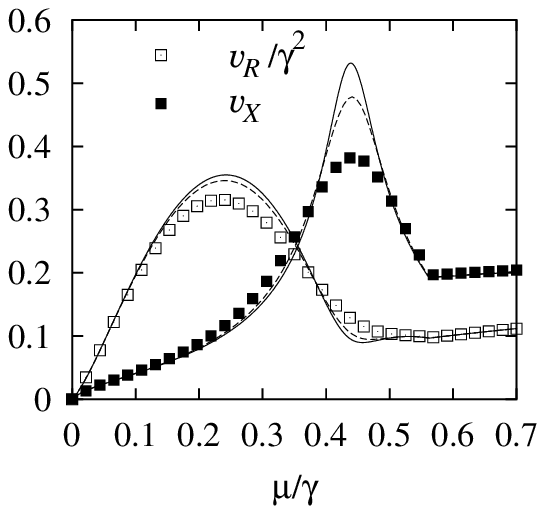}
\caption{Means (middle) and variances (right) for a biallelic
  model with asymmetric mutation ($\kappa = 0.4$), and
  a fitness function $r/\gamma$ (left) that
  displays strong positive epistasis near $x=0.15$.  One therefore
  observes a \textbf{fitness threshold} ($\muc/\gamma \simeq 0.562$).
  Symbols correspond to $N=100$, dashed lines to $N=500$, and solid
  lines to the expressions in \sref{sec:results}.}
\label{fig:fitnthr}
\end{figure}
Let us add a remark concerning the transferability of these notions to
the original `biological' model with fixed, finite $N$.  In defining
fitness thresholds in the mutation class limit, we have tacitly
assumed that the fitness function $r$ reasonably interpolates the
discrete fitness values of the original model. In order to avoid
`pseudo-thresholds' driven by purely local features of the fitness
function on a scale smaller than $1/N$, the effects should be stable
under different interpolations.  Note that one way to assure this is
to apply the maximum principle only to the discrete point set $\{x_k\}
= \{k/N\}$ and ask for a jump in $\hat{x}$ over more than one mutation
class.  In any case, the example in \fref{fig:fitnthr} and those in
Figs.\ \ref{fig:wtthr}--\ref{fig:unidir} show that the threshold
effects are usually clearly visible also for finite $N$.

\paragraph{Criterion.}

To derive a criterion for the existence of a fitness threshold for a
given fitness function $r$, we use the following argument.  According
to the above definition, a fitness threshold is signaled by a jump in
$\hat{x}$. Thus, in any fitness landscape \emph{without} a threshold,
$\hat{x}(\mu)$ varies continuously from the wildtype position $\xmin
:= \lim_{\mu\to0} \hat{x}(\mu)$ to the position of the mutation
equilibrium, $\xmax := \lim_{\mu\to\infty} \hat{x}(\mu)$, where
$g(\xmax) = 0$ (for the biallelic model, $x_{\text{max}} =
(1+\kappa)/2$).  Therefore, at each $x$ in the half-open interval
$\coint{\xmin,\xmax}$ the maximum in \eref{eq:maxprinc} is attained
for some finite $\mu$.  If $r$ and $u^\pm$ are twice continuously
differentiable in the closed interval $\ccint{\xmin,\xmax}$, then $g$
is twice continuously differentiable in $\ooint{\xmin,\xmax}$ and we
arrive at the following sufficient condition for the non-existence of
a fitness threshold:
\begin{equation}\label{eq:critfitx}
\forall x \in \ooint{\xmin,\xmax} \;\;\exists \mu > 0 \;:\;
  r'(x) = g'(\mu,x)  \quad \text{and} \quad r''(x) < g''(\mu,x) \,.
\end{equation}
Expressing $\mu = \mu(x)$ through the derivatives of $r$ and $g$, we
can state an existence condition in the following general form, cf
\aref{app:fitnthre}: \\
\emph{There is a fitness threshold in the mutation--selection
  equilibrium at some critical mutation rate $\muc$ if and only if}
\begin{equation}\label{eq:critfitn}
\sup_{x \in \ccint{\xmin,\xmax}} \left( r''(x) -
\frac{r'(x) g''(x)}{g'(x)}\right) \ge 0 \,.
\end{equation}
For the biallelic model, this reads:
\begin{equation} \label{eq:crit1}
\sup_{x \in \ccint{\xmin,\xmax}} 
\left( r''(x) - \frac{-r'(x)}{2x\left(1-x\right)\left(1-2x+
2\kappa\sqrt{\frac{x(1-x)}{1-\kappa^2}}\right)}\right) \ge 0 \,.
\end{equation}
In the special case that the supremum in \eref{eq:critfitn} is zero,
but is assumed only in a single point $x_0$, there is actually no
jump in $\hat{x}$.  Here, we obtain limiting cases of a threshold, in
the sense that a jump in $\hat{x}$ may be obtained by arbitrarily
small changes in the slope or curvature of $r$ or $g$.  Typically,
this limiting behavior is indicated by an infinite derivative of the
function $\hat{x}(\mu)$ at $\hat{x} = x_0$ (cf
\aref{app:fitnthre}).\footnote{In physics, this kind of behavior
  corresponds to the important class of continuous phase transitions,
  cf \aref{app:physics}. In the biological models, however, these 
non-generic limiting cases do not seem to
  justify a category of their own.}

Discontinuities in the fitness function or its derivatives can
formally be included in \eref{eq:critfitn} by considering left- and
right-sided limits separately. For a kink in $r$, we formally set $r''
= \infty$ or $r'' = - \infty$, respectively, if $r'$ increases or
decreases at this point (which makes \eref{eq:critfitn} true in the
former, but not in the latter case).  Finally, a jump in $r$ always
results in a fitness threshold.

Note that the criteria presented here do not indicate whether there
are one or multiple thresholds for a given combination of $r$ and
$u^\pm$. Neither do they provide direct information about the value of
$\bar{r}$ at the threshold, or about $\muc$. In fact,
\eref{eq:critfitn} and \eref{eq:crit1} are independent of the scalar
factor $\mu$, but only depend on the shapes of the mutation and
fitness function.  Answers to these questions, however, are easily
derived from the maximum principle for any specific $r$ and $u^\pm$,
and may also be obtained from the graphical construction, cf the
discussion in the preceding paragraph.

\paragraph{Discussion.}

\begin{figure}
\centerline{%
\epsfig{file=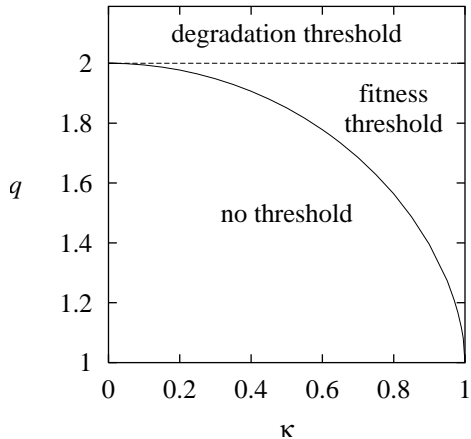,bbllx=58,bblly=52,bburx=231,bbury=216,clip=}}
\caption{The figure shows, as a solid line, the minimum exponent
  $q$, parametrizing epistasis of the fitness function $r(x) = (\xmax
  - x)^q$, that is needed to obtain a fitness threshold in the
  biallelic model as a function of the asymmetry parameter $\kappa$ of
  the site mutations rate.  The exponent varies continuously from
  quadratic (for symmetric site mutation, $\kappa=0$) to linear (for
  unidirectional mutation, $\kappa=1$).  For $q>2$ (dashed line), the
  fitness threshold is also a degradation threshold (see
  \sref{sec:degrthr}).  For this combination of fitness and mutation
  functions, a wildtype threshold only occurs for unidirectional
  mutation ($\kappa=1$).}
\label{fig:monom1}
\end{figure}
Under what conditions should we expect a fitness threshold to exist in
a mutation--selection system?  The above criterion \eref{eq:critfitn}
compares $r''$, which measures the epistasis of the fitness function,
with $g''$ weighted by a factor $r'/g'$. Under the reasonable
assumption that the rate of back mutations $u^-$ increases with the
distance to the wildtype, whereas the rate of deleterious mutations
$u^+$ decreases, we have $g'<0$ and $r'/g' >0$ for decreasing fitness
functions. Typically, if the curvature of $u^+$ and $u^-$ is not too
large, we also find $g'' > 0$. The criterion then shows that a finite
minimum strength of positive epistasis ($r''>0$, cf the end of
\sref{sec:mutflow}) is required for a fitness threshold.  For the
biallelic model with $r(x) = (\xmax - x)^q$, this is shown in
\fref{fig:monom1}.  Vanishing curvature or even concavity of the
mutational loss function, $g'' \le 0$, on the other hand, may even
lead to thresholds for fitness functions with negative epistasis.

As will become apparent in \aref{app:physics}, the fitness threshold
as defined above is the biological counterpart of a first order phase
transition in physics. Since $\hat{x}$, which translates into the
magnetization, plays the role of the order parameter, the phase
transition is generically first order, and continuous only in the
limiting case mentioned above.  Note that positive epistasis with
quadratic exponent $q=2$ in a biallelic model with symmetric site
mutation ($\kappa=0$), as has been discussed by \citet{BaWa01}, is
just such a limiting case. The physical analogy shows that a fitness
threshold is indeed a true collective phenomenon on the level of the
sites or loci. The essential self-enhancing effect simply is that in
regions of positive epistasis the selection pressure decreases with
any new deleterious mutation.

\subsubsection{Wildtype thresholds}
\label{sec:wtthr}

The \emph{loss of the wildtype} is the classic criterion for the
original error threshold as defined by \citet{Eig71}: For the sharply
peaked landscape, the frequency $p_0$ of the wildtype (or master
sequence) remains finite for small mutation rates even for $N \to
\infty$, but vanishes above the critical mutation rate.  The same
effect may be observed for any fitness function with a jump at the
wildtype position $x_{\text{min}}$.\footnote{As the mean fitness
  varies continuously, the wildtype frequency in the limit decreases
  linearly with the mutation rate, until the mean fitness reaches the
  lower value at the jump.  For larger mutation rates, the wildtype
  frequency in the limit is zero due to the sharpness of the
  population distribution for $N\to\infty$ (cf \sref{sec:derivrest}).}
Note that this does not depend on whether we assume the wildtype class
to contain only a single or a large number of genotypes (the latter
case has sometimes been called the \emph{phenotypic error threshold},
cf \citec{HSF96}).

If $r$ is continuous at $x_{\text{min}}$, however, the population
distribution spreads over a large number of mutation classes with
similar fitness for any finite mutation rate.  While for finite $N$
the frequency in any class remains positive for arbitrary $\mu$ (as
long as there are back mutations), the frequency of any single
mutation class (including the wildtype class) vanishes for $N \to
\infty$.  According to the original definition, error thresholds
therefore depend on strongly decanalized wildtypes in the sense that
deleterious mutations with small mutational effects are virtually
absent. While such a model was found to be adequate in certain cases,
such as the evolution of coliphage Q$\beta$ and certain viruses
\citep[cf][]{EiBi88}, and could be favored by pleiotropy
\citep{WaPe98}, slightly deleterious mutations are generally assumed
to occur in most biologically relevant situations (\citec{Kim83}, Ch.\ 
8.7; \citec{Ohta98}).

Still, one may ask for some related phenomenon that goes together with
the loss of the wildtype in all models in which this effect is
observed,\footnote{i.e.\ basically for fitness functions with a jump
  at the wildtype, and for certain models with unidirectional
  mutation, see the discussion in \sref{sec:unidir}} but defines a
threshold also in a broader model class.  The fitness threshold as
defined above does not meet this requirement, since fitness functions
with a jump at the wildtype may well have multiple fitness thresholds,
but only lose their wildtype once.  Instead, we will give a definition
which is based on the ancestor distribution.

\paragraph{Phenomenon.}
We define the wildtype threshold as the largest mutation rate $\muc^-
> 0$ below which the ancestral mean fitness coincides with the fitness
of the wildtype:
\begin{equation}
 \hat{r}(\mu) = \hat{r}(0) = \rmax, \quad \mu < \muc^-\,.
\end{equation}
The threshold may equivalently be defined as the largest $\muc^-$
below which $\hat{x}(\mu) = x_{\text{min}}$.  As a consequence, the
population mean fitness $\bar{r}$ responds linearly to an increase of
the wildtype fitness if $ \mu < \muc^-$, but becomes independent of
(sufficiently small) changes in the wildtype fitness above the
threshold.

Note that for unidirectional mutation, the ancestral average $\hat{x}$
(in general) also denotes the fittest class with non-vanishing
equilibrium frequency for any finite $N$, cf \Eref{eq:unimaxfit}.  In
this special case, the wildtype thus indeed vanishes from the
population at $\muc^-$. Threshold criteria in models with special
unidirectional mutation schemes have been derived previously, see the
discussion below in \sref{sec:unidir}.

\paragraph{Criterion.} 

For a wildtype threshold to occur, $r(x) - g(\mu,x)$ must be maximized
at $x=\xmin$ for some $\mu > 0$.  Assuming $r$ and $g$ to be
continuously differentiable for $x>x_{\text{min}}$, we arrive at the
criterion
\begin{equation} \label{eq:critwt}
\lim_{x{\scriptscriptstyle\searrow} x_{\text{min}}} 
  \frac{g(1,x) - g(1,x_{\text{min}})}{r(x) - r(x_{\text{min}})} =
\lim_{x{\scriptscriptstyle\searrow} x_{\text{min}}} 
  \frac{g'(1,x)}{r'(x)} < \infty \,,
\end{equation}
see \aref{app:wtthre} for a proof.  Fitness functions with a jump at
the wildtype position lead to a threshold for any continuous $g$.

\paragraph{Discussion.}

Note first that a wildtype threshold will always lead to
non-analytic behavior of $\hat{x}(\mu)$ and $\bar{r}(\mu)$ in
$\mu_c^-$ and is therefore closely related to a fitness threshold.  In
general, however, it need not show up as a prominent feature with a
jump in means or variances as functions of the mutation rate.  If we
have a fitness threshold with a jump in $\hat{x}(\mu)$ at $\hat{x} =
x_{\text{min}}$, however, this will also be a wildtype threshold. In a
system with a series of thresholds, the wildtype threshold (if it
exists) is always the one with the smallest $\muc$.

\begin{figure}
\threegnuplots{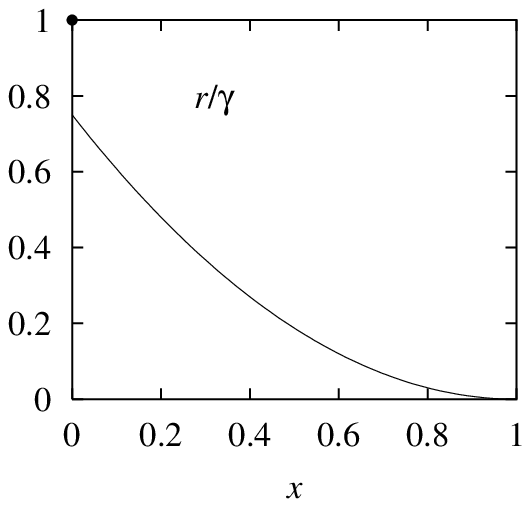}{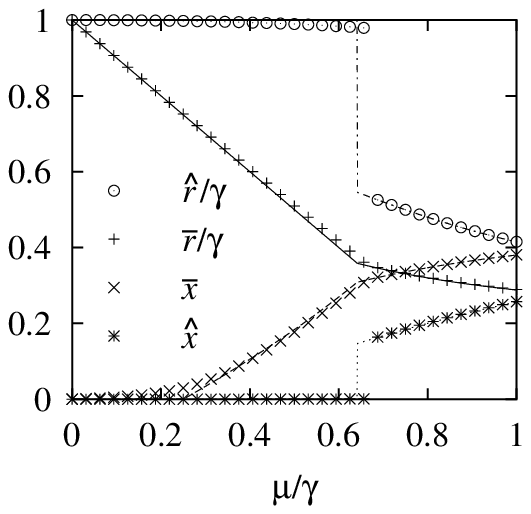}{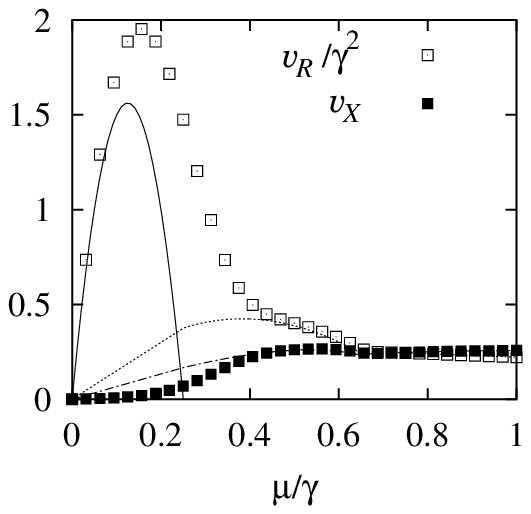}
\caption{Means (middle) and variances (right) for a model with
  symmetric mutation ($\kappa = 0$), $N=100$ (symbols), and the
  fitness function $r(x) = \frac34 \gamma\, (1-x)^2$ with an
  additional single peak of height $\gamma$ at $x=0$ (left).  Due to
  the latter, one finds a \textbf{wildtype threshold} ($\muc^-/\gamma
  \simeq 0.641$), which is also a fitness threshold. Lines correspond
  to the expressions in \sref{sec:results}.  For $1 \le \bar{r}/\gamma
  < \frac34$, i.e.\ $0 \le \mu/\gamma < \frac14$, the variance in
  fitness no longer follows \Eref{eq:varbiall}, but scales differently
  and is given by \eref{eq:varrjump} for $N\to\infty$ (see the
  discussion in \sref{sec:derivrest}). For finite $N$, we can
  approximate $v_R$ by a combination of both relations, where
  \eref{eq:varrjump} and \eref{eq:varbiall} dominate for small and
  large $\mu$, respectively. Note that $\bar{r}$ is analytic at
  $\mu/\gamma = \frac14$; we thus have no fitness threshold at this
  point.}
\label{fig:wtthr}
\end{figure}
The existence of a wildtype threshold, and also the `loss of the
wildtype' where applicable, depends on the strength of the deleterious
mutational effect at the wildtype, measured by $r'(x_{\text{min}})$.
The degree to which the wildtype requires a fitness advantage to avoid
the threshold depends on the mutational loss function.  If $g$ has a
finite derivative at $x_{\text{min}}$, we always obtain a threshold if
the mutational effects do not tend to zero.  In many important
situations, like the biallelic model and $x_{\text{min}} =0$, however,
a wildtype threshold requires fitness functions with a rather sharp
peak, like $r(x) \sim -x^p$ with $p \le 1/2$ or the one used in the
example in \fref{fig:wtthr}.  Note that this result depends on back
mutations, which make the slope of $g$ diverge at $x=0$.  For $u^-
\equiv 0$, however, the situation changes drastically, and we obtain a
threshold if only $r'(0) < 0$, as described above.

Since $g(0) = u^+(0)$, we see from Eqs.\ \eref{eq:maxprinc} and
\eref{eq:hatx} that $\bar{r}$ and $\bar{x}$ (but not necessarily the
variances) are unaffected by back mutations for mutation rates below
the wildtype threshold.  Further, the mutation load coincides with the
mutational loss, $l = \rmax - \bar{r} = \hat{r} - \bar{r} = u^+(0)$,
and therefore provides a meaningful measure for changes in $\bar{r}$
if the mutation rate is varied.  In this sense $\muc^-$ may be seen as
a point up to which back mutations can be safely ignored.

\subsubsection{Degradation thresholds}
\label{sec:degrthr}

\paragraph{Phenomenon.}
A far reaching effect of the error threshold is that selection
altogether ceases to operate.  We define a degradation threshold as
the smallest mutation rate $\muc^+$ above which the population mean
fitness is insensitive to any further increase of the mutation rate:
\begin{equation}
\frac{\partial\bar{r}}{\partial\mu} = -g(1,\hat{x}(\mu)) = 0, 
\quad \mu > \muc^+ \,.
\end{equation}
This is equivalent to the condition $\hat{x}(\mu) = \xmax$ for $\mu >
\muc^+$.  Also, the other means and variances then coincide with their
values in mutation equilibrium, and the population is degenerate.

\paragraph{Criterion.}
Selection ceases to operate according to the above definition if and
only if $r(x)-g(\mu,x)$ is maximal at $x=\xmax$ (where $g(\mu,\xmax) =
0$) for any finite $\mu > \muc^+$. Since $g$ is continuous and
strictly positive for $x < x_{\text{max}}$ and $\mu > 0$, it is
sufficient to compare the asymptotic behavior of $r$ and $g$ in the
neighborhood of $x_{\text{max}}$, cf \aref{app:degrthre}:
\begin{equation} \label{eq:critdegr}
\lim_{x {\scriptscriptstyle\nearrow} x_{\text{max}}} 
  \frac{r(x) - r(x_{\text{max}})}{g(1,x)} 
= \lim_{x {\scriptscriptstyle\nearrow} x_{\text{max}}} 
  \frac{r'(x)}{g'(1,x)} < \infty \,.
\end{equation}

\paragraph{Discussion.}

The degradation threshold is related to the fitness threshold in an
analogous way as the wildtype threshold above. In particular, we
always find non-analytic behavior of $\hat{x}(\mu)$ and $\bar{r}(\mu)$
at $\mu_c^+$, but not necessarily a jump or a kink.  However, a
fitness threshold with a jump of $\hat{x}(\mu)$ onto $x_{\text{max}}$
is necessarily a degradation threshold.  If there is a series of
thresholds connected with a system fulfilling \eref{eq:critdegr}, the
degradation threshold obviously is the last one as $\mu$ increases.

The criterion \eref{eq:critdegr} implies an important necessary
condition for a degradation threshold, namely $r(x_{\text{max}}) >
-\infty$; i.e.\ genotypes should not be lethal at this point.  This
parallels a well-known sufficient condition for the existence of a
normalizable limit distribution for arbitrary mutation rates in models
with non-compact state space \citep[][p.\ 128]{Mor77, Bue00}.  From a
biological point of view, a finite value of $r(x_{\text{max}})$ means
that not the whole genome, but only the part relevant for a specific
function or phenotypic property is included in the model, and the
genetic background is under sufficiently strong selection to be stable
under the mutation rates considered and guarantees survival of the
population.  We may then obtain mutational degradation \emph{w.r.t.\ 
  the function under consideration} if this function is less robust
under mutation than the background, and fitness thus levels out at a
finite value. Essentially, this is the threshold criterion previously
given by \citet{WaKr93} in their treatment of single-step models with
unidirectional mutation (see the discussion in \sref{sec:unidir}).
 
\begin{figure}
\threegnuplots{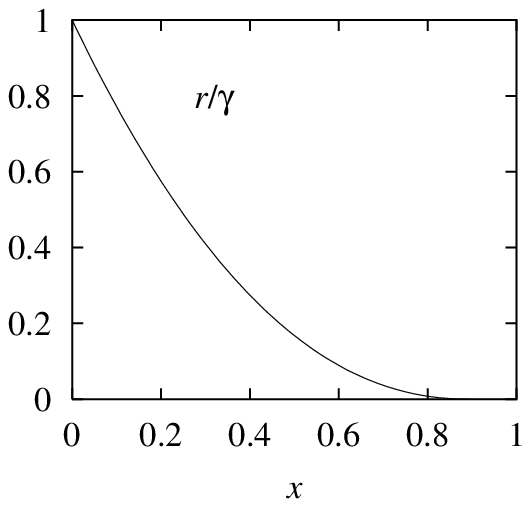}{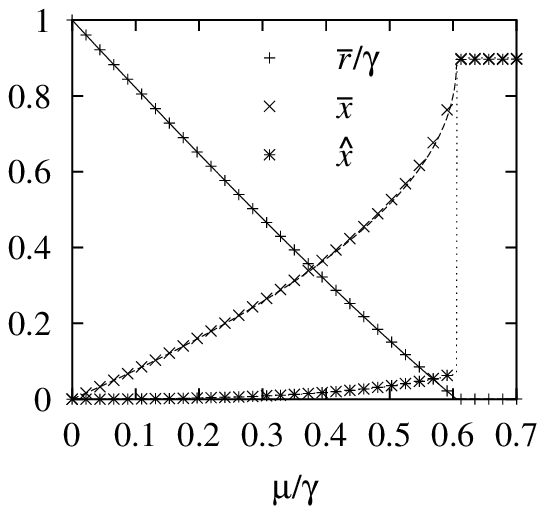}{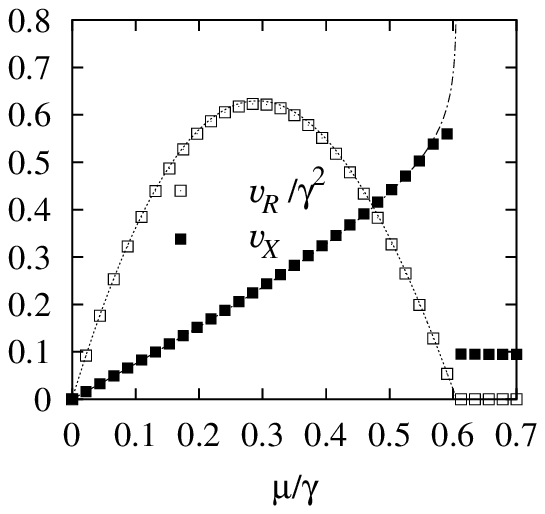}
\caption{Means (middle) and variances (right) for a model with
  asymmetric mutation ($\kappa = 0.8$), $N=100$ (symbols), and the
  fitness function $r(x) = \gamma\,(\xmax - x)^q/(\xmax)^q$ with
  $\xmax = (1+\kappa)/2 = 0.9$ and $q = 2.2$ (left).  As $q>2$, one
  finds a \textbf{degradation threshold} ($\muc^+/\gamma \simeq
  0.606$), which is also a fitness threshold, cf \fref{fig:monom1}.
  As $\hat{r}$ behaves just like $r(\hat{x})$ with a similar accuracy
  of the approximation, it is not shown here.}
\label{fig:degrthr}
\end{figure}
For the more general model with back mutations, we see that $r(x)$
must approach the fitness level at $r(x_{\text{max}})$ sufficiently
fast in order to fulfill \eref{eq:critdegr}.  For the biallelic model,
it is easy to show that we need positive epistasis with at least a
quadratic exponent, i.e.\ $r \sim r(x_{\text{max}}) +
\alpha\,(x_{\text{max}}-x)^2$.  Clearly, we always obtain mutational
degradation if $r(x) = r(\xmax)$ already for $x < \xmax$,
corresponding to the reasonable assumption that a minimum of
non-random coding region is needed for the gene or function considered
to show a fitness effect at all. An example for a degradation
threshold is given in \fref{fig:degrthr}.

\begin{figure}
\threegnuplots{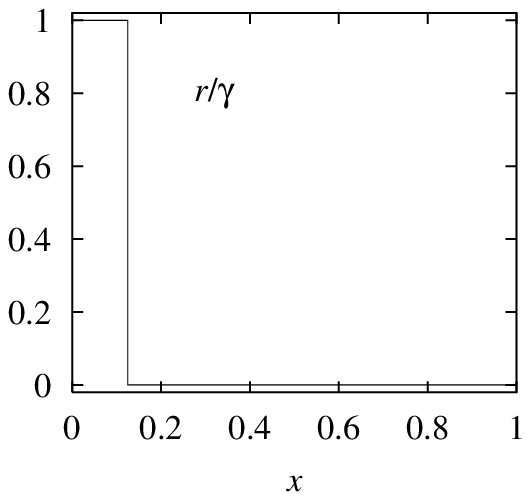}{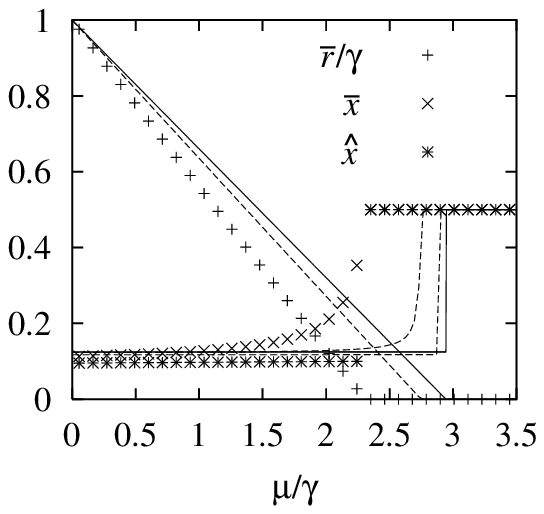}{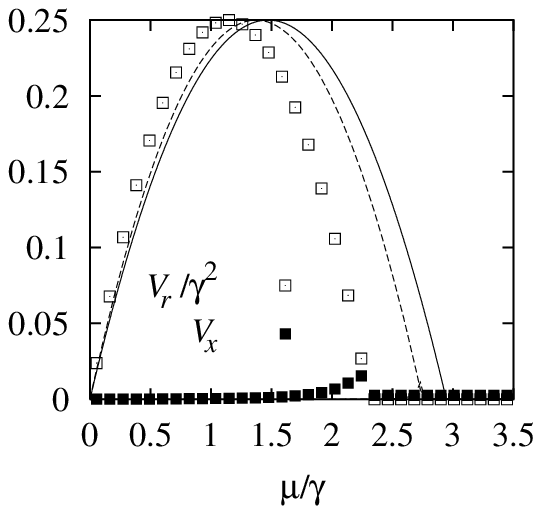}
\caption{Means (middle) and variances (right) for a model with
  symmetric mutation ($\kappa = 0$) and \textbf{truncation selection},
  i.e.\ $r(x) = \gamma$ for $x\le\tfrac18$ and $r(x)=0$ otherwise
  (left).  As in the sharply peaked landscape, cf \fref{fig:4types},
  one finds a combined fitness, wildtype, degradation, and trait
  threshold ($\muc/\gamma \simeq 2.94$).  Also, the variance in
  fitness follows the different kind of scaling as described by
  \eref{eq:vr1} and is given by \eref{eq:varrjump} for $N\to\infty$.
  Symbols correspond to $N=100$, dashed lines to $N=1000$, and solid
  lines to the expressions in \sref{sec:results}.  As $\hat{r}$
  behaves just like $r(\hat{x})$ with similar accuracy, it is not
  shown here.  Note that the deviations of the approximate expressions
  are somewhat stronger (of order $N^{-1/2}$) for fitness functions
  with jumps, cf \sref{sec:accuracy}.}
\label{fig:truncsel}
\end{figure}
Note finally that we obtain a degradation threshold that at the same
time is a wildtype threshold (and a fitness threshold with a jump of
$\hat{x}$ from $x_{\text{min}}$ to $x_{\text{max}}$) if and only if
\begin{equation} \label{eq:critwtdegr}
\sup_{x\in\ccint{\xmin,\xmax}} \left( r(x) -
r(x_{\text{max}}) - g(x) \frac{r(x_{\text{min}}) -
r(x_{\text{max}})}{g(x_{\text{min}})} \right) \le 0\,,
\end{equation}
as is most easily seen with the help of the graphical representation,
cf \fref{fig:comic5}.  Clearly, \Eref{eq:critwtdegr} is fulfilled for
the sharply peaked landscape used in \fref{fig:4types}, but also for
truncation selection, see \fref{fig:truncsel}.

\subsubsection{Trait thresholds}
\label{sec:traitthr}

\paragraph{Phenomenon.}
As stated above, there is usually a kink in the population mean of the
mutational distance $\bar{x}(\mu)$ (or some other trait) at a fitness
threshold.  The most pronounced change in the equilibrium distribution
of $x$, however, is a jump of $\bar{x}$ at some mutation rate
$\muc^x$, referred to as a \emph{trait threshold}.  Since a
discontinuous change in $\bar{x}$ is usually accompanied by a jump in
the \emph{local} mutation rates $u^\pm(\bar{x})$ as well as
$r'(\bar{x})$, it typically also leads to jumps in $v_X$ and $v_R$.
The mean fitness, however, is not at all affected at such points (if
they do not coincide with a fitness threshold as defined above).

\paragraph{Criterion.}
Since the equilibrium mean fitness $\bar{r}(\mu)$ as a function of the
mutation rate is always continuous, we easily conclude from $\bar{r} =
r(\bar{x})$ that a jump in $\bar{x}$ occurs if and only if the fitness
function is \emph{not} strictly decreasing from $x_{\text{min}}$ to
$x_{\text{max}}$.

\begin{figure}
\threegnuplots{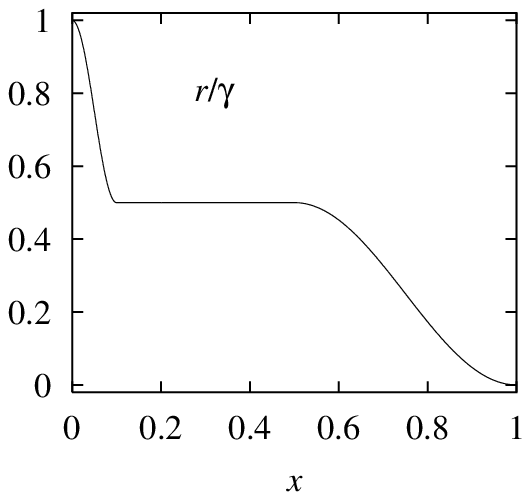}{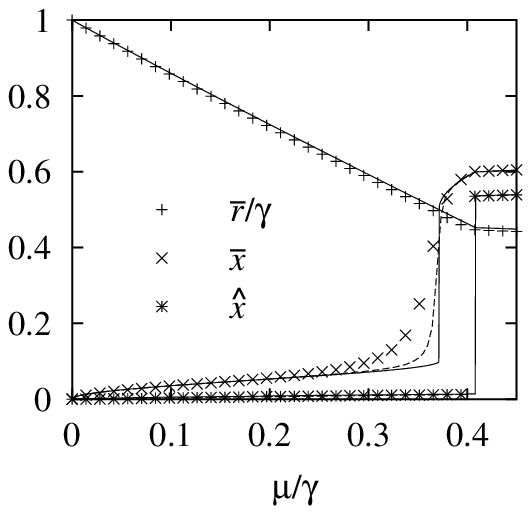}{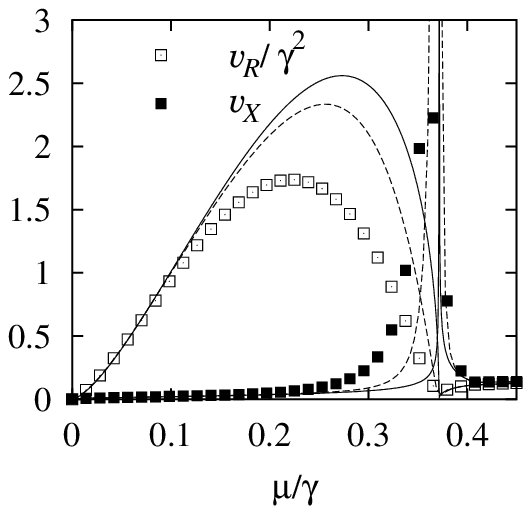}
\caption{Means (middle) and variances (right) for a model with
  asymmetric mutation ($\kappa = 0.5$) and a fitness function
  $r/\gamma$ (left) with an ambiguity for $r(x)/\gamma = 0.5$.  Thus,
  one finds a \textbf{trait threshold} ($\muc^x/\gamma \simeq 0.372$),
  which precedes a fitness threshold ($\muc/\gamma \simeq 0.408$), cf
  \sref{sec:traitthr}.  Symbols correspond to $N=100$, dashed lines to
  $N=500$, and solid lines to the expressions in \sref{sec:results}.
  As $\hat{r}$ behaves just like $r(\hat{x})$ with similar accuracy,
  it is not shown here.}
\label{fig:traitthr}
\end{figure}
\paragraph{Discussion.}
Obviously, any fitness landscape with a trait threshold also fulfills
\eref{eq:critfitn} and thus also has a fitness threshold, but not vice
versa.  We have $\muc \ge \muc^x$ (i.e.\ the jump in $\bar{x}$ in
general precedes the fitness transition with the jump in $\hat{x}$);
see the example in \fref{fig:traitthr}. This shows, in particular,
that with varying mutation rate there may be large changes in the
phenotype that may be accompanied by changes in the fitness variance,
but have virtually no effect on mean fitness.  Trait and fitness
thresholds should, therefore, be clearly distinguished.  In contrast
to the fitness threshold or a phase transition in physics, the trait
threshold is not driven by collective (self-enhancing) action, but
simply mirrors a local feature of the fitness function.

\subsection{Unidirectional mutation}
\label{sec:unidir}

\begin{figure}
\threegnuplots{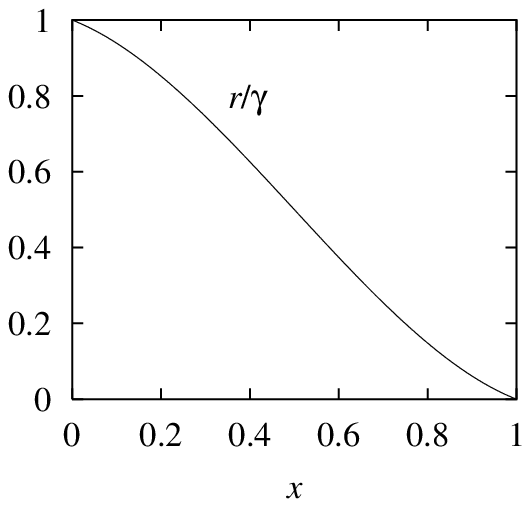}{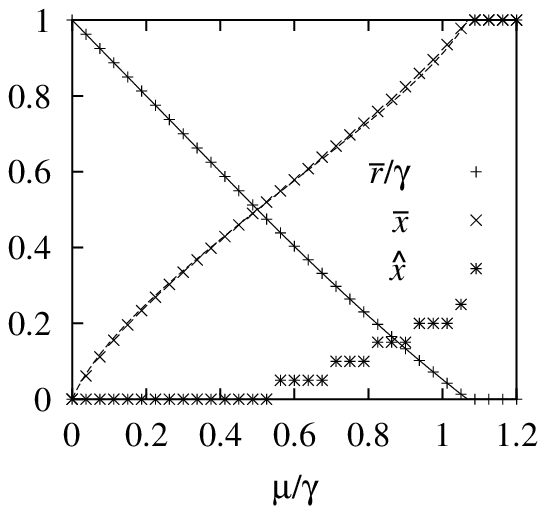}{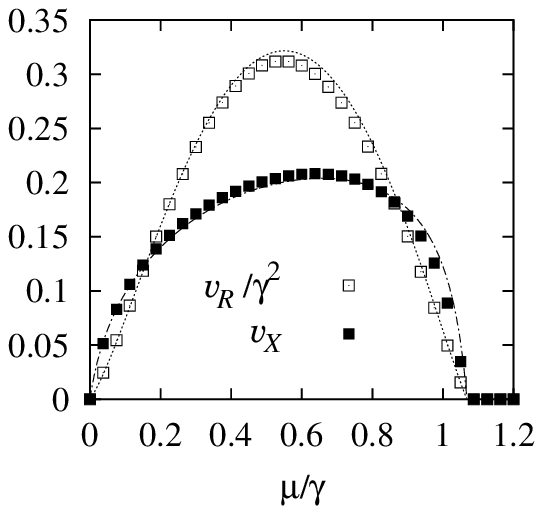}
\caption{Means (middle) and variances (right) for a model with
  \textbf{unidirectional mutation} ($\kappa = 1$), $N=20$ (symbols),
  and the fitness function $r/\gamma$ shown on the left.  The means
  $\bar{r}$ and $\hat{x}$ were calculated via the discrete maximum
  principle \eref{eq:max1}.  For $\bar{x}$ and the variances the
  population distribution was calculated explicitly using the
  recursion following from \eref{eq:step} for $U^-_k \equiv 0$; solid
  lines refer to the expressions from \sref{sec:results}.  One
  observes both a wildtype and a degradation threshold.  As $\hat{r}$
  is exactly $r(\hat{x})$, it is not shown here.}
\label{fig:unidir}
\end{figure}
In this section, we briefly discuss how the definitions of mutation
thresholds specialize for unidirectional mutation ($\kappa = 1$ for
the biallelic model).  An example is given in \fref{fig:unidir}.  We
shall also take the chance to make contact with previous results on
threshold criteria by \citet{WaKr93} and by \citet{Wie97}, where
related models were studied.

For the above definition of thresholds, the maximum principle in the
mutation class limit has played a central role.  Since, for vanishing
back mutations, it reduces to Haldane's principle and is also exact
for finite $N$, many of the notions above can be formulated directly,
avoiding the limit.  As has been described in \sref{sec:derivuni}, the
ancestral mean of the mutational distance (or trait) agrees with the
minimal $x$ in the equilibrium population.  Since this minimum can
only assume discrete values for finite $N$, jumps in $\hat{x}$ will
necessarily occur for some $\mu$. For a system with a large number of
mutation classes (which we consider here), this should, however, not
be regarded relevant. In line with the above reasoning on the
applicability of the fitness threshold definition for finite systems
and previous definitions of the error threshold for unidirectional
mutation, it seems reasonable to restrict the term threshold to the
first and the last jump, i.e.\ the loss of the wildtype
\citep{WaKr93,Wie97} and the point of complete mutational degradation
\citep{Wie97}, and to jumps of $\hat{x}$ over more than one class.

Threshold criteria are easily found as analogs of the above relations
(note that $g$ reduces to $u^+$ if back mutations vanish).  As
condition for the existence of a fitness threshold with a jump over
more than one class, for example, we obtain for monotonic $u^+_k$:
\begin{equation} \label{eq:crituni}
\max_k \left[ \frac{s_k^+}{s_{k+1}^+} - \frac{u_k^+ -
u_{k+1}^+}{u_{k+1}^+ - u_{k+2}^+} \right] \ge 0\,.
\end{equation}
We \emph{always} find a wildtype threshold with loss of the fittest
class if the total range of fitness values is \emph{finite}. If there
are lethal genotypes ($r_k = - \infty$), we obtain \emph{no} such
threshold if (and only if) the mutation rate at all non-lethal types
is larger or equal to the mutation rate at the wildtype. For the
special case of a constant mutation rate ($u^+_k = \text{const}$,
$k<N$, $u^+_N = 0$), this reproduces the criterion by \citet{WaKr93}.
A degradation threshold is found if and only if there are no lethals.

For the special case of the biallelic model with linearly decreasing
mutation rates, \Eref{eq:crituni} reduces to $\max_k s_k^+/s_{k+1}^+
\ge 1$ and we obtain a threshold for any degree of positive epistasis,
but also for linear parts of the fitness function (with any three
fitness values on a straight line).  For fitness functions of the form
$r(x) = r_0 (1-x)^\alpha$, finally, we can confirm the result by
\citet{Wie97} that wildtype and degradation thresholds coincide if and
only if $\alpha \le 0$ (no or negative epistasis).  Note that the
result by \citet{Wie97} was derived for a different mutation scheme
(with mutations coupled to reproduction).

\subsection{Variation of fitness values and sequence lengths} 

Up to this point, we have discussed mutation thresholds as effects
that may occur as the mutation rate varies, while the fitness function
and the number of mutation classes are kept fixed (note that the
mutation class limit is always understood as an approximation to a
given finite system).  Here are two alternative points of view.

Firstly, we can consider threshold effects as the \emph{fitness
  values} vary, while the mutation rates remain constant.  As already
mentioned in the discussion of Haldane's principle
(\sref{sec:haldane}), mean fitness is largely independent of local
variations in the fitness function, but only depends on the shape of
$r$ in regions with substantial weight in the ancestral distribution.
For most values of the mutation rate, this has a unique peak, and
therefore only the neighborhood of the mean ancestral mutational
distance, $\hat{x}$, matters.  At fitness thresholds, however, we
find, in general, two peaks at which variations in $r$ can change the
mean fitness.

Secondly, in line with the original work of \citet{Eig71}, we can
increase \emph{sequence length} while leaving the mutation rate
\emph{per site} fixed, thus altering the \emph{total} mutation rate.
The question of interest then is: Given a certain (fixed) fitness
advantage of some function, and fixed mutation rates per site in a
molecular model, how long can the coding region for the function
become and still be maintained intact by selection?  In this case,
with $u^\pm/\gamma \sim N$ (where $\gamma$ denotes the range of
fitness values under consideration), we obtain thresholds which are
inversely related to sequence length, $\muc \sim 1/N$, in all
situations above.  Note that this is in accordance with the original
findings for the sharply peaked landscape \citep[cf][]{EiBi88}, but at
variance with results by \citet{Wie97}.  The latter are artificial
effects caused by the use of a different scaling of the fitness
functions, in which $\gamma$ is not kept fixed, but increases with $N$
in just those cases where conflicting results have been found.

\subsection{Implications of mutation thresholds}
\label{sec:threshdisc}

At mutation thresholds, mutation--selection balance is unstable with
respect to small changes in the model parameters.  There is no real
lower limit on the mutation rates at which these phenomena may happen,
but for fitness and degradation thresholds mutation rates must be
comparable to the average mutational effect $\gamma$ to obtain effects
of significant magnitude (cf our examples in Figs.\ \ref{fig:fitnthr}
and \ref{fig:wtthr}--\ref{fig:unidir}).  In this case, the
\emph{average} effect of the mutations considered will be very
slightly deleterious (or almost neutral) for realistic values of the
mutation rate.  The model then pays respect to the rationale that
these mutations are the relevant ones for the discontinuous behavior.
Since they may be numerous \citep[cf, e.g.,][]{Kond95}, their
collective effect may nevertheless be quite large.  Mutations with
much stronger effects, on the other hand, will only occur at very low
frequency in the population and contribute smooth changes to the
system observables if the mutation rate is varied.  They may therefore
be excluded from these considerations.

An important consequence of the original error threshold of the
sharply peaked landscape (and, more generally, of any degradation
threshold in our typology), which has been stressed in particular by
\citet{Eig71} as well as by \citet[][Ch.\ 4.3]{MaySS95}, is its
potential importance for the evolution of mutation rates.  Since the
total mutation rate increases with the sequence length (see the
previous subsection), site mutation rates must evolve below the
threshold value to allow functions to prevail that need a certain
minimum length of the coding region as their genetic basis.  This
might have been a severe problem for early replicators since the
mutational repair mechanisms required to reduce the mutation rate
depend on enzymes with relatively large coding regions.  Since we find
degradation thresholds for a rather broad class of fitness functions,
this is also a plausible hypothesis with respect to our more general
model class.\footnote{Note that the fitness effect of mutational
  repair is always an indirect one caused by an increase in the
  copying accuracy in parts of the sequence that are directly related
  to fitness.}

A closer look at the effect of thresholds on the mutational loss
reveals yet another mechanism by which degradation thresholds, and
fitness thresholds as well, may be important for the evolution of
mutation rates, even if mutational repair itself is not the function
endangered.  Assume that the mutation rate may be reduced by
modifications of the replication accuracy.  Recall further that the
mutational loss $g = \hat{r} - \bar{r}$ provides a measure for the
indirect fitness advantage $\delta\bar{r}$ gained by the decrease of
$\mu$.  Therefore, a system beyond a degradation threshold (where
$g=0$) will never experience any selection pressure for decreasing
mutation rates, and thus cannot evolve in this direction.  But even a
fitness threshold (with a jump in $g$, but $g>0$ for $\mu>\muc$) may
have a similar effect. This is because modifiers for reduced mutation
usually have deleterious physiological side-effects, dubbed the `cost
of fidelity' \citep[see][for a recent review]{SGJS00}.  Clearly, for
the modifier to prevail, the indirect fitness advantage
$\delta\bar{r}$ gained by the decrease of $\mu$ must be at least as
high.  Therefore, a jump in $g$ separates two different evolutionary
regimes: for $\mu < \muc$, much larger costs can be counteracted than
for $\mu > \muc$.

In a second line of interpretation, the critical mutation rate of an
error threshold has often been argued to provide a strict upper limit
that must be avoided in all real organisms.  Certain kinds of viruses
are perceived as thriving just below that value as to maximize their
adaptability in a changing environment \citep{EiBi88}. While it is
certainly true that wildtype sequences or certain functions can get
lost at threshold points, it is, however, much more difficult to argue
why evolution should care about them.  After all, $g$ drops at the
threshold, thus making a further increase in adaptability less costly.
Further, the equilibrium mean fitness changes continuously with the
mutation rate in arbitrary deterministic mutation--selection systems,
even at threshold points.  Mutation thresholds, therefore, can not be
seen as strict limits constraining the evolution of mutation rates.
This may be different if further evolutionary forces are relevant,
most importantly drift. Indeed, numerical studies show that the mean
fitness (averaged over time) may drop discontinuously at critical
mutation rates in a finite population \citep{NoS89}.  A jump in the
mean fitness has also been found for sexually reproducing populations
with dominance \citep{Higgs94}. This, however, is outside our model
class and, according to our definition, no longer a property of a
\emph{mutation} threshold but essentially a drift (or segregation)
phenomenon.

Let us finally turn to yet another effect that has previously been
described as characteristic of the error threshold
\citep[e.g.][]{BoSt93}. Assume that mutation classes increase in size
with the distance from the wildtype (as is the case for the biallelic
model for $k$ up to $N/2$), which is reflected by asymmetric mutation
rates between neighboring classes.  Then, a jump in $\bar{x}$, as at
the critical mutation rate $\muc^x$ of a trait threshold, entails a
\emph{delocalization effect}.  It should be stressed, however, that
this effect has \emph{no} direct consequences for the evolution of
mutation rates, which are entirely connected to the population mean
fitness and thus only to fitness, wildtype, and degradation
thresholds.

\section{Summary and outlook}
\label{sec:summary} 

The findings of this article, and the future directions they might
lead to, fall into three parts, which we would like to discuss in
turn.

\paragraph{Ancestors.}
As a crucial concept for the study of (asexual) mutation--selection
models, we have identified the ancestor distribution of genotypes, or
genotype classes, which, in mutation--selection balance, is the
equilibrium distribution of the time-reversed evolution process.  The
ancestor frequency of the $i$-th genotype (or class) is given as $a_i
= z_i p_i$, where $z_i$, the relative reproductive success, and
$p_i$,the equilibrium frequency, are the $i$-th entries of the left
and right leading (PF) eigenvectors of the evolution matrix. In the
biology--physics analogy laid down in \Aref{app:physics}, the ancestor
distribution corresponds to the distribution of the bulk magnetization
in spin models.  Biologically, measurements of ancestor frequencies in
real populations should in principle be possible by marker techniques.
In the equilibrium dynamics, the ancestors permanently feed the swarm
of mutants that is observed at any instant of time.  Significant
evolutionary change is indicated by modification of this ancestor
population.  We have shown this in a couple of instances.

If the fitness values $R_i$ are subject to change, the $a_i$ measure
the sensitivity of the equilibrium mean fitness $\bar{R}$ to these
changes.  The net total change in $\bar{R}$ is given (to linear order)
by \Eref{eq:linresr}. If the fitness changes are due to a modifier
mutation, \Eref{eq:deltar} can be read as the selection coefficient of
this modification with respect to the ancestors.  Such a modifier will
asymptotically fix if and only if it increases the fitness of the
ancestors.  The vector of ancestor frequencies can therefore be seen
as the gradient which points into the direction of the effective
selection pressure on the fitness function and determines the course
of evolution -- given that the appropriate modifier mutations are
available.

Since the $a_i$ are non-negative, selection will always favor an
increase of fitness values.  In the case of modifiers that change the
mutant fitness values, we thus find a tendency for the evolution of
robustness, or canalization, in systems with back mutations, whereas
anti-robustness, or decanalization, cannot result in this simple
setup.  As the selection pressure is strongly differential, one can
even speculate that this mechanism is a cause for negative
(synergistic) epistasis (as in the example in \fref{fig:paz}), which
is considered a rather general phenomenon by many \citep[][and
references therein]{CrSi83,POW00}.  As always with indirect selection,
however, selective forces are weak and probably of relevance only in
large populations and for rather high mutation rates.  In the limiting
case of unidirectional mutation, the ancestor distribution is
concentrated at the wildtype (if present in the equilibrium
population).  Then, only modifiers that increase the wildtype fitness
will go to fixation, whereas modifications of mutant fitness values
have no effect on the equilibrium mean fitness -- in line with the
predictions of the Haldane--Muller principle.

We have defined the \emph{mutational loss} $G$ as the difference
between ancestor and population mean fitness, which equals the
long-term loss in progeny that the equilibrium system suffers due to
mutation.  \Eref{eq:linres4} shows that the loss determines the change
in the equilibrium mean fitness if the mutation rate is subject to
change.  Again, it is thus the ancestor distribution that provides the
link between external variations of model parameters and the
equilibrium response.  We always have $G \le L$, with $L$ the mutation
load, and equality only in systems without back mutation to the
fittest type.  Measurements of the mutational loss should be possible
by fitness measurements in mutator strains or by direct determination
of the ancestor fitness distribution using genetic markers.

The ancestor concept, as introduced in this article, is independent of
modeling assumptions on fitness landscapes and mutation schemes.  We
have derived a few basic results that hold for this general case, and
extend the Haldane--Muller principle.  Under additional assumptions
much stronger results may be obtained, as we have seen for the
single-step mutation model.  We expect that, of the many successful
approximation methods that are routinely applied to the population
distribution, some could also be applicable for the ancestor
distribution and yield further interesting results.  However, in order
to apply this approach to more general situations, namely including
genetic drift, the concept will have to be extended.  The question is
whether it is possible to characterize the distribution of genotypes
on a single lineage backward in time, and to relate this to the
mutation--selection--drift equilibrium.

\paragraph{The maximum principle.}
The reformulation of the equilibrium condition in terms of ancestor
variables leads to a maximum principle for the equilibrium mean
fitness, which we have exploited for the single-step mutation model.
In this model, fitness is an arbitrary function of the number of
mutations (or some other additive trait).  Mutation proceeds stepwise
on the mutation classes, but mutation rates (as well as back mutation
rates) may vary from class to class.  Here, the maximum principle may
be recast into a particularly simple form, which yields the mean
fitness as the maximum of the difference between the fitness function
and the mutational loss function (see Eqs.\ \eref{eq:maxprinc} and
\eref{eq:g}).  The position of the maximum determines the mean
ancestral genotype and the corresponding value of the mutational loss
function yields the mutational loss $G$ (\Eref{eq:hatx}).  The
simplicity of the maximum principle results from the fact that
maximization is over one single scalar variable only, and may be
performed explicitly, or with the help of a simple graphical
construction (\fref{fig:comics}).  A different maximum principle has
been suggested previously for mutation--selection models
\citep{Deme83}.  It relies on general variational principles in the
framework of ergodic theory, in which maximization is over all
possible genealogies, and therefore not constructive.
 
Our maximum principle is exact in three independent limiting cases,
namely unidirectional mutation, models with a linear dependence of
both mutation rates and fitness on an underlying trait (including
multilocus wildtype--mutant models without epistasis), and in the
limit of an infinite number of mutation classes.  For small back
mutation rates, $u^- \ll (u^+ + s)$, the resulting estimate for the
equilibrium mean fitness is exact to linear order in $u^-$. In
general, the maximum principle holds as an approximation that leads to
quantitatively reasonable results for a wide range of parameters and
quickly becomes accurate if one of the exact limits is approached.

Starting from the mean fitness, we have explicitly calculated the
fitness variance and the mean and variance of the trait.  All formulas
are collected in \Sref{sec:results}.  The fitness variance is both
proportional to the mean mutational effect and the mean difference of
deleterious and back mutation rates; the trait variance has the same
dependence on the mutation rates, but is inversely proportional to the
mean mutational effect (\Eref{eq:var}).  These formulas give the
amount of genetic variability that is maintained by the balance
between mutation and selection.

Extensions of the maximum principle to a larger model class is
possible in various ways. Following the lines of this paper, it is
relatively straightforward to include double or multiple mutations in
the theory.  Poisson distributed mutations (which emerge naturally in
the biallelic model if mutation is coupled to reproduction) can also
be treated. A necessary ingredient is that the evolution matrix can
still be symmetrized by transformation to the ancestor frequencies.

The models discussed here all assume fitness to depend only on the
distance to a reference class (the Hamming distance to the reference
type in the biallelic case).  Especially in a molecular context, this
is, of course, a severe oversimplification.  But also in classical
population genetics, the importance of variance of additive and
epistatic effects has often been highlighted \citep[see,
e.g.,][]{BuGi99,POW00}.  Progress in this direction can be made by
applying methods of inhomogeneous mean-field theory from statistical
physics to the biallelic model.  Here, it is possible to derive a
simple maximum principle for models in which groups of sites or loci
have different weights assigned that scale their respective direct and
epistatic fitness effects (H.W., unpublished results).  With similar
techniques, fitness landscapes with more than one trait, such as the
multiple quantitative trait model \citep{TaHi00}, can also be treated.
Here, the equilibrium mean fitness is derived from a maximum principle
over an $n$-dimensional space, if $n$ is the number of traits.
Finally, multilocus models with more than two alleles per locus (or
states per site) may be considered.  In the molecular context, an
explicit treatment of the four-letter case with Kimura 3ST mutation
scheme \citep[cf][]{SOWH95} has already been given by \citet{HWB01}.

\paragraph{Mutation thresholds.}
Inspired by the definition of phase transitions in statistical
physics, we have used the concept of the mutation class limit to
define threshold behavior in mutation--selection models as the
discontinuous change of statistical observables (such as the mean
fitness or the mean number of mutations) with the mutation rate $\mu$.
Four different types of thresholds have been singled out, which all
coincide in Eigen's original \emph{error threshold} for the model with
the sharply peaked landscape, but should be distinguished for general
fitness functions. With the help of the maximum principle, 
criteria have been given to characterize the fitness functions and
mutation schemes that lead to each type of threshold.

\emph{Fitness thresholds} are characterized by a kink in the
population mean fitness and a jump in the mutational loss $G$.  They
precisely occur at mutation rates for which the equilibrium ancestor
distribution that solves the maximum principle is non-unique in the
mutation class limit.  The evolutionary significance of a fitness
threshold lies in its potential impact on the evolution of mutation
rates. Since the mutational loss jumps and may take much smaller
values for $\mu$ exceeding the critical mutation rate, the gain in
mean fitness by reduction of $\mu$ may be very small until the
threshold is reached. If this gain in fitness must (over)compensate
costs connected with mutational repair, evolution for lower mutation
rates might be slowed down in the presence of a threshold.  For the
existence of fitness thresholds, positive (antagonistic) epistasis is
required for many mutation schemes.  Small convex parts of the fitness
function, however, may already be sufficient.  Fitness thresholds are
collective phenomena and correspond to phase transitions in physics.

Whereas the loss of the wildtype from the population is not a
well-defined notion for most of the models treated here, we consider
the ancestor mean in the mutation class limit instead.  A
\emph{wildtype threshold} is then characterized by a critical mutation
rate $\muc^- >0$ below which the ancestor mean fitness coincides with
the wildtype fitness, and the ancestor distribution is concentrated at
the wildtype.  Below a wildtype threshold, the system behaves, in many
respects, as a system with unidirectional mutation.  For the biallelic
model, wildtype thresholds occur only for fitness functions with very
sharp peaks at the wildtype position.

A \emph{degradation threshold} is characterized by the fact that
selection altogether ceases to operate and the mean fitness does not
change any further for mutation rates exceeding a critical value
$\muc^+$.  A necessary condition for a degradation threshold is that
the fitness function does not diverge to minus infinity. This is
reminiscent of a threshold criterion derived for a model with
unidirectional mutation by \citet{WaKr93}. Degradation thresholds have
similar implications for the evolution of mutation rates as fitness
thresholds.

A \emph{trait threshold}, finally, is characterized by a jump in the
trait or mean number of mutations $\bar{X}$.  In the sequence space
picture, a trait threshold is connected with (partial) delocalization
of the equilibrium population in genotype space.  It is important to
note that a trait threshold is not a collective phenomenon but is
simply caused by non-monotonic parts of the fitness function. The
delocalization effect is not connected with any significant change in
the mean fitness (unless the trait threshold goes together with a
fitness threshold), and thus has no direct impact on the selection
pressure on the mutation rate.

The types of thresholds found here should also be observable in
mutation--selection models with more general fitness landscapes and
mutation schemes. Explicit threshold criteria can be obtained at least
in some cases, such as the four-state model treated by \citet{HWB01}
(J.H., unpublished result).

\appendix
\section{The connection to physics}
\label{app:physics}

For a number of models from statistical physics, a relation to
mutation--selection models has been demonstrated, see \citet{BaGa00}
for an overview.  In the present investigation, too, concepts and
techniques from theoretical physics have served as a guideline for
analysis.  Most importantly, the maximum principle \eref{eq:maxprinc}
derives from the physical principle by which a system seeks to
minimize its free energy.  In our definitions of mutation thresholds,
we also exploited the correspondence between thresholds and physical
phase transitions, which has first been pointed out by
\citet{Leut87a}.

Whereas such correspondencies can be very fruitful, they require a
detour through the physical world, which remains unsatisfactory from
the biological point of view.  Therefore, our intention in the body of
the article has been to develop and discuss concepts entirely within
the biological framework.  Nevertheless, for readers with a physical
background, as well as for biologists who are familiar with the
interface to statistical mechanics, we will briefly sketch the
relationship between both approaches. This may, on the one hand,
facilitate further transfer of methods; on the other hand, limitations
of certain `imported' concepts in the biological context become
obvious.  Last but not least, it is exactly this connection which
resolves a few issues that had remained enigmatic so far.

Concentrating on the biallelic model with types $\bs{s}$ in this
appendix, we can rely on a connection to a model of quantum
statistical mechanics that was previously established by \citet{BBW97}
\citep[see also][]{WBG98}. More precisely, the evolution operator of
the biallelic model with symmetric mutation was shown to be exactly
equivalent to the Hamilton operator of an Ising quantum chain (up to a
minus sign). Generalizing this slightly to include asymmetric mutation
rates, and assuming a suitable ordering of genotypes, we may represent
the quantum chain Hamiltonian as
\begin{equation}\label{eq:qc}
{\bd H}  = \mu \left[  \sum_n (\sigma^x_n  -\Id) -
 \kappa \sum_n (i\sigma^y_n + \sigma^z_n)\right] +
   \sum_I \eta^{}_I \prod_{n \in I} \sigma^z_n
    = {\bd M} + {\bd R}\,.
\end{equation}
Here,  
\begin{equation}
  \sigma^a_n := 
   \underbrace{\Id \otimes \ldots \otimes \Id}_{n-1 \; \text{copies}}
   \otimes \, \sigma_{}^a \otimes
   \underbrace{ \Id \otimes \ldots \otimes \Id}_{N-n \; \text{copies}},
\end{equation}
where $a$ equals $x$, $y$ or $z$, and $\sigma_{}^{x,y,z}$ are
\emph{Pauli's matrices}\/, i.e.
\begin{equation}
\sigma_{}^x := \begin{pmatrix} 0 & 1 \\ 1 & 0 \end{pmatrix} ,\quad
\sigma_{}^y := \begin{pmatrix} 0 & -i \\ i & 0 \end{pmatrix} , 
\quad \text{and} \quad
\sigma_{}^z := \begin{pmatrix} 1 & 0 \\ 0 & -1 \end{pmatrix} \,.
\end{equation}
The last sum in \eref{eq:qc} runs over all index sets $I \subset
\{1,\dots, N\}$, and $\eta^{}_I$ are the interaction coefficients of
the spins within the chain, or with a longitudinal field. The
collection of the $\eta_I^{}$ determines the fitness function. Further,
there are transversal fields in $x$ and $y$ direction that account for
mutation. Note that the Hamiltonian is non-Hermitian for asymmetric
mutation.

This equivalence was used by \citet{WBG98} and \citet{BaWa01} to solve
the model for a couple of fitness functions and symmetric mutation
($\kappa = 0$) with the help of methods from quantum statistical
mechanics.  In the current investigation, we have chosen an equivalent
formulation, which remains entirely within classical probability, to
analyze more general mutation and fitness schemes.  In order to
briefly sketch the connection between the approaches, we first
symmetrize $\bd{H}$ by means of a similarity transform, i.e.\ 
$\tilde{\bd{H}}=\bd{S} \bd{H} \bd{S}^{-1}$ with $\bd{S} :=
\prod_{n=1}^{N} \bigl( \cosh(\theta/2) \Id + \sinh(\theta/2)
\sigma^z_n\bigr)$ and $\theta= \operatorname{artanh}(\kappa)$.  (Note
that this transformation  relies on the sequence space
representation of $\bd{H}$ \eref{eq:qc}, in contrast to the
symmetrization in \Sref{sec:branching}, which starts out from a
mutation class representation.)

The central concept now required is the \emph{quantum mechanical
  expectation} $\scal{\bd{O}}$ of an operator $\bd{O}$,
defined by
\begin{equation}\label{eq:qmexp}
  \scal{\bd{O}}(t) := \tr \bigl( \bd{O} \bs{\rho}(t) \bigr) \,,
\end{equation}
where $\bs{\rho}(t)$ is the so-called \emph{density operator}
\begin{equation}
  \bs{\rho}(t) := \exp(t \tilde{\bd{H}}) /   
    \tr \bigl(\exp(t \tilde{\bd{H}}) \bigr) \,,
\end{equation}
and $t$ corresponds to the inverse temperature.  For the choice
$\bd{O} := \frac{1}{N} \sum_{n=1}^{N} \sigma^z_n$, one obtains the
\emph{quantum mechanical magnetization}, which is the crucial
\emph{order parameter} for the quantum chain.

We will concentrate on the limit $t \to \infty$ (the ground state),
where $\bs{\rho}(t)$ becomes identical with the time evolution
operator of the critical branching process we have met in
\Sref{sec:branching}.  That is, $\bs{\rho}=\lim_{t \to \infty}
\bs{\rho}(t) = \lim_{t \to \infty} \exp \bigl(t(\tilde{\bd{H}} -
\lambdamax \Id) \bigr) = \tilde {\bs{p}} \tilde {\bs{p}}^T /
\scal{\tilde{\bs{p}},\tilde{\bs{p}}}$, where $\scal{\cdot,\cdot}$
denotes the scalar product, and $T$ means transposition.  In this
limit, the quantum mechanical expectation of any diagonal operator
$\bd{O}$ (defined by the elements $O_{\bs{s}\bs{s}}$) therefore turns
out to coincide with the corresponding \emph{ancestral average} (cf
\sref{sec:averages}):
\begin{equation}\label{eq:ancav} 
\scal{\bd{O}} = \tr ( \bd{O} \bs{\rho}) = 
  \frac{\scal{\tilde {\bs{p}}, \bd{O} \tilde{\bs{p}}}}{\scal{\tilde {\bs{p}}, 
\tilde{\bs{p}}}}
  = \sum^{}_{\bs{s}} O_{\bs{s}\bs{s}} \frac{\tilde p_{\bs{s}}^2}{\sum_{\bs{s'}} 
  \tilde{p}_{\bs{s'}}^2} = 
   \sum_{\bs{s}} O_{\bs{s}\bs{s}}a_{\bs{s}} = \hat O,
\end{equation} 
where we have used that $a_{\bs{s}} = \tilde
p_{\bs{s}}^2/\sum_{\bs{s'}}\tilde{p}_{\bs{s'}}^2$ for symmetric
$\tilde {\bd{H}}$, in line with \sref{sec:branching}.  In particular,
the quantum mechanical magnetization (given by $O_{\bs{s}\bs{s}} =
(N-2 \dH(\bs{s},\bs{s}^+)/N$)), which has, so far, appeared as a
crucial but technical quantity unrelated to any biological observable,
now emerges as the mean ancestral genotype $\hat x$ (up to a factor
and an additive constant).  In contrast, the classical magnetization
$\sum_{\bs{s}} O_{\bs{s}\bs{s}} p_{\bs{s}}$, which we had termed
surplus previously, translates into the population average $\bar x$.
Let us note in passing that the change in normalization performed in
\citet[Eqs.\ (7), (11)]{BBW98} and \citet[][Eqs.\ (55), (56)]{BaWa01}
to formulate Rayleigh's principle for the PF eigenvalue (i.e.\ 
$\lambdamax = \sup \scal{\bs{x}, \tilde{\bd{H}} \bs{x}}$ where the
supremum is taken over all $\bs{x}$ with
$\scal{\bs{x},\bs{x}}=\|\bs{x}\|_2=1$) is equivalent to our ancestral
transformation in \eref{eq:evsymm}.  This way, we may take advantage
of $L^2$ theory although the original problem is inherently in the
realm of $L^1$.  Finally, the expectation of the \emph{non-diagonal}
operator $\bd M$ is $\scal{\bd{M}} = \sum_{\bs{s},\bs{s'}}
\tilde{p}_{\bs{s}} \tilde{M}_{\bs{s},\bs{s'}} \tilde{p}_{\bs{s'}} =
\sum_{\bs{s},\bs{s'}} z_{\bs{s}} M_{\bs{s},\bs{s'}} p_{\bs{s'}}$ (with
$\tilde{\bd{M}} := \bd{S} \bd{M} \bd{S}^{-1}$), which we have
identified with the loss $G$ in offspring due to mutation (cf
\sref{sec:interpret}).
 
The concept of ancestral distributions is very general and does not
rely on our special dynamical system.  It also applies to discrete
dynamical systems, as long as they are linear (or may be transformed
to a linear system) and admit a unique stable stationary state. This
is true if a system is defined by an iteration matrix $\bd{T}$ for
which the Perron--Frobenius theorem holds \citep[hints in this
direction may be found in][]{Deme83}.  In particular, this applies to
a discrete-time version of the quasispecies model defined by
$T_{\bs{s} \bs{s'}}=v_{\bs{s}\bs{s'}} w_{\bs{s'}}$, where $v_{\bs{s}
  \bs{s'}}$ is the mutation probability from $\bs{s'}$ to $\bs{s}$,
and $w_{\bs{s'}}$ is Wrightian fitness of $\bs{s'}$. As observed by
\cite{Leut86,Leut87a} and reviewed by \citet[][Appendix II]{BaWa01},
this model is equivalent to a classical two-dimensional Ising model
with row transfer matrix $\bd{T}$, where the rows correspond to
genotypes, and the columns to generations.  Hence, every 2D
configuration corresponds to one line of descent, \emph{conditional on
  nonextinction at present}.

Here, considerable confusion has arisen in the literature as to the
distinction and meaning of \emph{surface} and \emph{bulk}
magnetization \citep{Leut87a,Tara92,FrPe97}.  \emph{Surface}
quantities correspond to the last row (in the time direction) of a
configuration with open boundary conditions, i.e.\ the current
population; therefore, surface averages are population averages.  In
contrast, \emph{bulk} quantities are averages over the entire 2D
configuration.  In the limit of an infinite number of rows, they
become identical with averages over a single row `in the middle' of
the configuration (i.e.\ at infinite distance from both the first and
the last row), as given by $\lim_{n \to \infty} \tr (\bd{T}^{n} \bd{O}
\bd{T}^{n}) / \tr (\bd{T}^{2n})$.  Therefore, the bulk average is,
again, the ancestral average (also compare with \eref{eq:qmexp} and
\eref{eq:ancav}).

Everything we have said so far holds for arbitrary, finite $N$.
Clearly, the infinite mutation class limit $N \to \infty$ is the
thermodynamic limit of the statistical mechanics system with its
extensive scaling of energy and magnetic field terms. Technical
aspects related to this scaling in the biological system are covered
by \citet{BaWa01}.  While the thermodynamic limit may be taken as a
matter of course in most classical situations in solid state physics,
however, the adequacy of the corresponding limit as an approximation
in biological applications must be thoroughly considered.  In
particular, the mutation class limit should be clearly distinguished
from the infinite-sites limit, which is widely used in theoretical
population genetics; see the discussion in \sref{sec:limit}, and
\citet{BaWa01}.

Clearly, the fitness thresholds described in \sref{sec:thresholds}
correspond to the phase transitions of the physical system, in the
sense of a non-analytic point of the free energy of the classical
Ising system or the ground state energy of the quantum chain (the mean
fitness in the biological model). Most importantly, the idea to use
the thermodynamic limit for the mathematical definition of the concept
is taken from physics. As we have pointed out (\sref{sec:mutthres}),
this is in accordance with the original definition of the error
threshold for the sharply peaked landscape. It should be noted,
however, that the fitness functions of the biological system typically
lack the symmetries inherent in physics.  As a consequence, the usual
classification of phase transitions in physics according to orders of
the non-analyticity as well as the consideration of critical exponents
does not seem to be particularly meaningful in the biological context.
Fitness thresholds are typically first order and exhibit a jump in the
ancestral mean $\hat{x}$, which parallels the physical magnetization.
Note at this point that neither the population mean $\bar{x}$
\citep[as suggested by][]{Tara92,FrPe97} nor the mean fitness itself
\citep[as implicitly in][]{Higgs94} should be mistaken as an order
parameter, in the sense that jumps in these quantities do not
characterize first order phase transitions.

\section{Proofs from \sref{sec:deriv}}
\label{app:proofsderiv}

\subsection{The additive case}
\label{app:prooflin}

Let us prove here that, if fitness and mutation rates depend linearly
on some trait $y_k = y(x_k)$ as described in \eref{eq:lincond}, the
system \eref{eq:symm} reduces to just two equations, one corresponding
to the necessary extremum condition following from \eref{eq:maxprinc},
the other being the defining equation \eref{eq:hatx} for the ancestral
mean $\hat{y}$ (for $y(x)=x$).  For the sake of simplicity, we write
$x_k$ instead of $y_k$ here.  Taking the difference of two arbitrary
equations of the linear system \eref{eq:symm}, say for $k$ and $\ell$,
divided by $\sqrt{a_k}$ and $\sqrt{a_\ell}$, respectively, we get
\begin{multline}\label{eq:diff}
(\beta^+ - \beta^- - \alpha) (x_k - x_\ell) + \sqrt{\beta^+\beta^-} \bigg(
\sqrt{x_k(1-x_{k-1})}\sqrt{\frac{a_{k-1}}{a_k}} - 
\sqrt{x_\ell(1-x_{\ell-1})}\sqrt{\frac{a_{\ell-1}}{a_\ell}}
\\
+ \sqrt{x_{k+1}(1-x_k)}\sqrt{\frac{a_{k+1}}{a_k}} -
\sqrt{x_{\ell+1}(1-x_\ell)}\sqrt{\frac{a_{\ell+1}}{a_\ell}}\bigg) = 0\,.
\end{multline}
With the \emph{ansatz}
\begin{equation} \label{eq:ansatz}
\frac{a_{k-1}}{a_k} = C \frac{x_k}{1-x_{k-1}} 
\quad \Leftrightarrow \quad 
\frac{a_{k+1}}{a_k} = C^{-1} \frac{1-x_k}{x_{k+1}} \qquad 
\forall \, k \in \{ 1, \dots, N \}
\end{equation}
\Eref{eq:diff} can be divided by $(x_k-x_\ell)$ and becomes
independent of $k$ and $\ell$. Note that \eref{eq:ansatz} also takes
care of the boundary conditions $a_{-1} = a_{N+1} = 0$ if $x_0 = 0$
and $x_N = 1$.  Summing both sides of $(1-x_{k-1})a_{k-1} = C x_k a_k$
over $k$, we obtain $C = (1-\hat{x})/\hat{x}$ and thus from
\eref{eq:diff}
\begin{equation} \label{eq:A1}
\beta^+ - \beta^- - \alpha + \sqrt{\beta^+\beta^-}\frac{1-
2\hat{x}}{\sqrt{\hat{x}(1-\hat{x})}} = 0 \,,
\end{equation}
which is exactly the extremum condition $r'(\hat{x}) = g'(\hat{x})$
following from \eref{eq:maxprinc}.  Together with the negative second
derivative this implies the maximum principle.

On the other hand, we can use \eref{eq:ansatz} to eliminate
$a_{k\pm1}$ from \eref{eq:symm}.  After multiplication by $\sqrt{a_k}$
this reads
\begin{equation}
\bigg[r_0 - \alpha x_k - \bar{r} - \beta^+(1-x_k) - \beta^- x_k +
\sqrt{\beta^+\beta^-} \bigg(x_k \sqrt{\frac{1-\hat{x}}{\hat{x}}} + 
(1-x_k)\sqrt{\frac{\hat{x}}{1-\hat{x}}}\bigg)\bigg]a_k = 0
\end{equation} 
and we obtain by summation over $k$
\begin{equation}\label{eq:A2}
\bar{r} = r_0 - \alpha\hat{x} - \beta^+(1-\hat{x})-\beta^-\hat{x} 
+ 2\sqrt{\beta^+\beta^-\hat{x}(1-\hat{x})} \,=\, r(\hat{x}) -
g(\hat{x}) \,,
\end{equation}
which corresponds to \eref{eq:hatx}.  Since fitness is assumed
linear in the trait, the mean values with respect to 
the population distribution are also related via $\bar{r} = r(\bar{x})$.

\subsection{The case $N\to\infty$}
\label{app:proofinf}

Let $u^\pm: [0,1] \to \RR_{\geq 0}$ be continuous and positive, but
fulfill $u^+(1) = u^-(0) = 0$.  Let $r: [0,1] \to \RR$ have at most
finitely many discontinuities, being either left or right continuous
at each discontinuity in $\openunitint$.  Then, with the scaling
described at the end of \sref{sec:scaling}, the maximum principle
\eref{eq:maxprinc} holds in the limit $N\to\infty$.

For a proof, we follow the arguments and notation introduced in
\sref{sec:discinf}.  First note that the lower bound for $\bar{r}_N$
in \eref{eq:lowerfin} is itself greater than or equal to
\begin{equation}
\label{eq:lkmn}
\rho_{k,m,n}^{} :=
  \inf_{y \in I_{k,m,n}} \bigl(r(y) - g(y)\bigr) -
  \sup_{y \in I_{k,m,n}} \bigl|g(y) - g_N(y)\bigr| -
  \frac{\sqrt{u^+_{k-m-1}u^-_{k-m}} +
    \sqrt{u^+_{k+n}u^-_{k+n+1}}}{m+n+1} \,,
\end{equation}
where $I_{k,m,n} = [\frac{k-m}{N},\frac{k+n}{N}]$ and the rules for
inf/sup have been applied.  We will now construct a sequence
$\rho_N(x) := \rho_{k_N(x),m_N(x),n_N(x)}$ for each $x\in[0,1]$, using
suitable sequences for the indices, such that
\begin{equation}
\label{eq:rhon}
\rho_N(x) \to r(x) - g(x) \,.
\end{equation}
Since, by definition, $\lim_{N\to\infty} \bar{r}_N = \bar{r}_\infty$,
Eqs.\ \eref{eq:lowerfin}, \eref{eq:lkmn}, and \eref{eq:rhon} will then
establish $\bar{r}_\infty \ge \sup_{x\in[0,1]} (r(x)-g(x))$, from
which, together with the upper bound in \eref{eq:inequal}, the claim
will follow.

Note first that, for $x=0$ or $x=1$, $\rho_N(x) = \rho_{xN,0,0} = r(x)
- g(x)$ holds for arbitrary $N$.  Now, fix $x\in\openunitint$.  If $r$
is continuous in $[x-d,x]$ for a suitable $d>0$, let $k_N(x) :=
\lfloor xN \rfloor$, $m_N(x) = \lfloor d\sqrt{N} \rfloor$, and $n_N(x)
\equiv 0$.  Otherwise $r$ is continuous in $[x,x+d]$ for some $d>0$,
and we define $k_N(x) := \lceil xN \rceil$, $m_N(x) \equiv 0$, and
$n_N(x) = \lfloor d\sqrt{N} \rfloor$.  With these choices, the last
term in \eref{eq:lkmn} vanishes for $N\to\infty$ since $m_N(x)+n_N(x)
\to \infty$, and the enumerator is bounded.  So does the supremum term
because of the uniform convergence $g_N \to g$: $\sup_{y \in
  I_{k_N,m_N,n_N}} |g(x)-g_N(x)| \le \sup_{y\in[0,1]} |g(x)-g_N(x)|
\to 0$.  The latter follows from the uniform continuity of
$\sqrt{u^\pm}$ since, in
\begin{multline}
|g(x)-g_N(x)| = \\
  \left| \left(\sqrt{u^+(x-\tfrac1N)}-\sqrt{u^+(x)}\right)\sqrt{u^-(x)} +
   \sqrt{u^+(x)}\left(\sqrt{u^-(x+\tfrac1N)}-\sqrt{u^-(x)}\right)
  \right| \,,
\end{multline}
the terms in parentheses vanish uniformly in $x$ as $N\to\infty$ and
$\sqrt{u^\pm(x)}$ is bounded.  Finally, the infimum term in
\eref{eq:lkmn}, and thus $\rho_N(x)$, converges to $r(x)-g(x)$ since
$x_{k_N(x)} \to x$, $r$ is continuous in all $I_N \ni x$, and $|I_N| =
(m_N(x)+n_N(x))/N \to 0$.  This was to be shown.

Now, let us prove that the ancestor distribution is concentrated
around those $x$ for which $r(x)-g(x)$ is maximal, from which
\Eref{eq:hatx} follows if the maximum is unique.  Multiplying the
evolution equation in ancestor form \eref{eq:symm} by $\sqrt{a_k}$, we
obtain, after summation over $k$:
\begin{multline}
\bar{r}_N=\sum_{k=0}^N \Big[\big(r(x_k) - u^+(x_k) - u^-(x_k)\big) a_k \,+\\
\sqrt{u^+(x_{k-1}) u^-(x_k)} \sqrt{a_ka_{k-1}} + \sqrt{u^+(x_k) u^-(x_{k+1})} 
\sqrt{a_{k+1}a_k} \Big] \,.
\end{multline}
Using $\sqrt{a_ka_{k\pm1}} \le \frac{1}{2}(a_k + a_{k\pm1})$, we get
\begin{equation}
\bar{r}_N \le \sum_{k=0}^N \left[r(x_k) - g_N(x_k) \right]a_k = 
\hat{r}_N-\widehat{(g_N)}_N^{}
\end{equation} 
with $g_N(x_k)$ as defined in \sref{sec:discinf}.  Since $\bar{r}_N
\to \bar{r}$ and $g_N(x) \to g(x)$ uniformly in $x \in [0,1]$, we can
find for any given $\varepsilon > 0$ some $N_\varepsilon$, such that
for all $N > N_\varepsilon$:
\begin{equation}
\sum_{k=0}^N \big[ r(x_k) - g(x_k) \big] a_k > \bar{r} - \varepsilon^2 \,.
\end{equation}
We now divide this sum into two parts, $\sum_k := \sum_{k_>} +
\sum_{k_{\le}}$. The first part, $\sum_{k_>}$, collects all $k$ with 
$r(x_k) - g(x_k) > \bar{r} - \varepsilon$, the second part contains the
rest. We then obtain
\begin{equation}
\bar{r} - \varepsilon^2 < \sum_{k=0}^N \big[ r(x_k) - g(x_k) \big] a_k \le
\bar{r} \sum_{k_>} a_k + (\bar{r}-\varepsilon)\sum_{k_{\le}} a_k = \bar{r} -
\varepsilon \sum_{k_{\le}} a_k
\end{equation}
and thus $\sum_{k_{\le}} a_k < \varepsilon$.  We conclude that for $N$
sufficiently large, the ancestor distribution is concentrated in those
mutation classes for which $r(x) - g(x)$ is arbitrarily close to its
maximum, $\bar{r}$.

\section{Proofs from \sref{sec:thresholds}}
\label{app:proofsthr}

\subsection{Proof of \eref{eq:critfitn}}
\label{app:fitnthre}

We first prove that the negation of \eref{eq:critfitn},
\begin{equation} \label{eq:negcrit}
r''(x) - \frac{r'(x)g''(x)}{g'(x)} < 0 \,,\quad \forall x 
\in [x_{\text{min}},x_{\text{max}}] \,,
\end{equation}
implies \eref{eq:critfitx} and is therefore a sufficient condition for
the absence of a fitness threshold.  We start by showing that both $r$
and $g$ are strictly decreasing in $\ooint{\xmin,\xmax}$.  To see
this, suppose there exists an $x>\xmin$ with $r'(x) = 0$, and let
$x_r$ be the smallest such $x$.  Then either $g'(1,x_r) = 0$ and
$\lim_{x \to x_r} \left( r''(x) - \frac{r'(x) g''(x)}{g'(x)}\right) =
r''(x_r) - r''(x_r) = 0$ in contradiction to \eref{eq:negcrit}, or
$g'(1,x_r) \neq 0$, in which case we obtain $r''(x_r) <0$ in
contradiction to $r'(x) < 0$ for $x \in \ooint{\xmin,x_r}$.  On the
other hand, imagine $g'(x)=0$ for some $x \in \ooint{\xmin,\xmax}$,
and let $x_g$ be the largest such $x$.  Then, since $g'(x)<0$ for $x
\in \ooint{x_g,\xmax}$, we have $g''(x_g) \le 0$ and thus $\lim_{x \to
  x_g} g''(x)/g'(x) = + \infty$ for the right-sided limit, which again
contradicts \eref{eq:negcrit} since $r'(x_g) < 0$.  Therefore,
$\mu(x):= r'(x)/g'(1,x)$ is well-defined everywhere in
$\ooint{\xmin,\xmax}$, it guarantees $r'(x) = g'(\mu(x),x)$, and
\eref{eq:negcrit} yields $r''(x) < g''(\mu(x),x)$, which completes the
first part of the proof.

We now prove that \eref{eq:critfitn} implies a threshold.  Assume
first that the supremum in \eref{eq:critfitn} is larger than zero. Due
to the continuity of $r$, $g$, and their derivatives, we then find an
$x_0$ in $\ooint{\xmin,\xmax}$ with $r''(x_0) -
r'(x_0)g''(x_0)/g'(x_0) > 0$. This, however, implies $r''(x_0) -
g''(x_0) > 0$ whenever $r'(x_0) - g'(x_0) = 0$. Therefore, the maximum
of $r(x) - g(x)$ is never attained at $x_0$ and we must have a jump in
$\hat{x}(\mu$). If the supremum in \eref{eq:critfitn} is exactly zero,
we argue as follows.  For the absence of a threshold, we need a
continuous function $\hat{x}(\mu)$ whose inverse, by the maximum
principle, is $\mu(\hat{x}) = r'(\hat{x})/g'(1,\hat{x}) >0$ for
$\hat{x} \in \ooint{\xmin,\xmax}$.  For the derivative of
$\mu(\hat{x})$, we find $\mu'(\hat{x}) = [r''(\hat{x}) - r'(\hat{x})
g''(\hat{x})/g'(\hat{x})]/g'(1,\hat{x})$, which must be non-negative
in the absence of a threshold. Consider now those $\hat{x}$ at which
the supremum in \eref{eq:critfitn} is attained. For $g'(1,\hat{x})
\neq 0$, we have $\mu'(\hat{x})=0$. Since $\hat{x}'(\mu) =
1/\mu'(\hat{x})$, $\hat{x}(\mu)$ has a diverging derivative at these
points, and a jump if the supremum is attained (and thus
$\mu'(\hat{x}) = 0$) on a whole interval.  Finally, we also obtain a
jump if the supremum is attained on an interval where also $g'(1,x) =
0$ as then the whole interval is degenerate as a maximum.  We exclude
the special case that $g'(1,x) = 0$ at an isolated $x$ to avoid
lengthy technicalities.

\subsection{Proof of \eref{eq:critwt}}
\label{app:wtthre}

Note first that existence of a wildtype threshold obviously implies a
lower bound of $1/\muc^-$ on the left hand side of \eref{eq:critwt}.
Assume, on the other hand, that there are sequences $x_i$ in
$[x_{\text{min}},x_{\text{max}}]$ and $\mu_i>0$ with $\mu_i \to 0$ and
$r(x_i) - \mu_i g(1,x_i) > r(x_{\text{min}}) - \mu_i
g(1,x_{\text{min}})$ for all $i$. Let then $x_j \to x_\infty$ be a
convergent subsequence. Since $r$ and $g$ are assumed to be
continuous, we have $r(x_\infty) \ge r(x_{\text{min}})$ and hence
$x_\infty = x_{\text{min}}$, since $r(x_{\text{min}})$ is the unique
maximum of $r$ in $[x_{\text{min}},x_{\text{max}}]$. Thus, we find
\begin{equation}
\frac{g(1,x_j) - g(1,x_{\text{min}})}{r(x_j) - r(x_{\text{min}})} > 
\frac{1}{\mu_j} \to \infty \,,
\end{equation}
contradicting \eref{eq:critwt} and proving the criterion.

\subsection{Proof of \eref{eq:critdegr}}
\label{app:degrthre}

The proof is analogous to the case of the wildtype threshold.  On the
one hand, existence of the threshold implies the criterion with a
bound $\muc^+$. On the other hand, if we have sequences $x_i$ in
$[x_{\text{min}},x_{\text{max}}]$ and $\mu_i$ with $r(x_i) - \mu_i
g(1,x_i) > r(x_{\text{max}})$ for $\mu_i \to \infty$, we can again 
choose a convergent subsequence $x_j \to x_\infty$. Since $g(1,x)$ is
continuous and $x_{\text{max}}$ is the only zero of $g$ in
$[x_{\text{min}},x_{\text{max}}]$, we have $x_\infty =
x_{\text{max}}$. As in the wildtype case above, this contradicts
\eref{eq:critdegr} and proves the criterion.

\section*{Acknowledgments}

It is our pleasure to thank Michael Baake and G\"unter P. Wagner for
many fruitful discussions, and for critically reading the manuscript.
Many clarifications were produced by Reinhard B\"urger's careful
scrutiny.  We uniformly thank an anonymous reviewer for hints towards
simplification of the proof in \aref{app:proofinf}, and Tini Garske
for useful comments.  J.H. gratefully acknowledges funding by an Emmy
Noether Fellowship from the German Science Foundation (DFG).

\bibliography{eb}

\end{document}